\documentclass{iopart}

\usepackage[dvipsnames]{xcolor}
\usepackage{stackrel,bbold}
\usepackage{hyperref,amssymb}
\usepackage{graphicx,slashed}
\usepackage{slashed}
\newcommand{\dv}[1]{\frac{\mathrm{d}}{\mathrm{d} #1}}
\newcommand{\pdv}[1]{\frac{\partial}{\partial #1}}

\newcommand{\dd}{\mathrm{d}}
\newcommand{\bd}{\mathbb{d}}

\newcommand{\uV}{\mkern 0mu\underline{\mkern-0mu V \mkern-4.5mu}\mkern 2.5mu}
\newcommand{\ubeta}{\beta \!\!\!\! \underline{\phantom{\alpha}}}

\newcommand{\utrchi}{\tr \!\!\!\! \! \underline{\phantom{\chi}} \underline{\chi}}

\newcommand{\uD}{\mkern-2mu \underline{\mkern 0mu D \mkern-2mu} \mkern 1mu}
\newcommand{\uK}{\underline{K}}

\newcommand{\ueta}{\underline{\eta}}

\newcommand{\uh}{\underline{h}}

\newcommand{\oh}{\overline{h}\phantom{h}\mkern-10mu}
\newcommand{\ioh}{(\overline{h} \mkern-10mu\phantom{h}^{-1})}

\newcommand{\cGp}{\mathcal{G}_+}
\newcommand{\cGm}{\mathcal{G}_-}

\newcommand{\cH}{\mathcal{H}}

\newcommand{\twoS}{\tilde{\Sigma}}
\newcommand{\symS}{\mathcal{H}}

\newcommand{\andd}{\; \; \mbox{and} \; \;}

\begin{document}

\title[]{Exotic marginally outer trapped surfaces \\
in rotating spacetimes of any dimension}

\author{Ivan Booth$^1$,  Kam To Billy Chan$^2$, Robie A.~Hennigar$^{3}$, Hari Kunduri$^4$, Sarah Muth$^1$}

\address{$^1$ Department of Mathematics and Statistics, Memorial University of Newfoundland, St. John's, Newfoundland and Labrador, A1C 5S7, Canada}

\address{$^2$ Department of Physics and Physical Oceanography, Memorial University of Newfoundland, St. John's, Newfoundland and Labrador, A1B 3X7, Canada}
   
\address{$^3$ Departament de F{\'\i}sica Qu\`antica i Astrof\'{\i}sica, Institut de
Ci\`encies del Cosmos,
 Universitat de
Barcelona, Mart\'{\i} i Franqu\`es 1, E-08028 Barcelona, Spain}   

\address{$^4$ Department of Mathematics and Statistics and Department of Physics and Astronomy, McMaster University, Hamilton, Ontario, L8S 4M1, Canada}

\begin{abstract}
The recently developed MOTSodesic method for locating marginally outer trapped
surfaces was effectively restricted to non-rotating spacetimes. In this paper we extend the method to include  (multi-)axisymmetric time slices of (multi-)axisymmetric spacetimes 
of any dimension. We then apply this method to study marginally outer
trapped surfaces (MOTSs) in the BTZ, Kerr and Myers-Perry black holes. While there are many similarities between the
MOTSs observed in these spacetimes and those seen in Schwarzschild and Reissner-Nordstr\"om, details of
the more complicated geometries also introduce some new, previously unseen, behaviours.

\end{abstract}

\maketitle

%===================================================================

\section{Introduction}

Trapped surfaces play an indispensable role in the theory of black holes. The existence of a trapped surface is related to the inevitable formation of a spacetime singularity through the Hawking-Penrose singularity theorems~\cite{Penrose:1964wq, hawking_ellis_1973}. Trapped surfaces lead to a characterization of black holes and their properties that is quasi-local in space and local in time, providing an alternative to the teleological event horizon~\cite{Booth:2005qc}. As such, these methods are routinely used in numerical relativity, where the ability to locate the black hole horizon at a given moment of time has obvious practical utility~\cite{Smarr:1976qy, Thornburg:2006zb, Schnetter:2006yt,Baumgarte:2010ndz}. 

A key object in the quasi-local approach to black hole boundaries is the \emph{marginally outer trapped surface} (MOTS). In a $D$-dimensional spacetime, an \emph{outer trapped surface} is a closed, space-like $(D\!\!-\!\!2)$-dimensional surface such that light rays emitted in the future outward-pointing null normal directions are converging. A (fully) trapped surface is one for which both inward and outward light rays converge. A MOTS is a limiting case of an outer trapped surface: a closed, space-like $(D\!\!-\! \!2)$-dimensional surface for which the light rays emitted in the future-pointing outward direction are neither converging nor expanding, while no constraint is placed on the light rays emitted in the inward direction. In the simplest black hole spacetimes, for example Schwarzschild,  within the event horizon the spheres of constant radius and time are trapped surfaces, while cross-sections of the event horizon are MOTSs.  

Many important results from the mathematical theory of black holes can be formulated with reference to either event horizons or MOTSs. For example, MOTSs can be assigned physical quantities, such as mass and angular momentum, and variations in these quantities are constrained, through the Einstein equations, into variational identities analogous to the laws of black hole mechanics~\cite{Hayward:1993wb, Ashtekar:1998sp, Ashtekar:1999yj,  Ashtekar:2002ag, Ashtekar:2003hk}. However, theoretical developments in this area have tended to be more difficult than comparable analyses based on the event horizon. Progress toward addressing these questions has partly been illuminated through a better understanding of MOTSs in simpler, exact spacetimes.   %However, theoretical developments in this area have tended to be somewhat more difficult (or at least more conceptually complicated) than comparable analyses based on the event horizon. Progress toward addressing these questions has partly been illuminated through a better understanding of MOTSs in simpler, exact spacetimes.  

An example of this concerns the notion of the ``second law'', or area monotonicity theorem, for MOTSs.  A number of examples of marginally trapped tubes were studied in~\cite{Booth:2005ng}, where the full hypersurface consisted of both time-like and space-like segments. It was remarked in~\cite{Booth:2005ng} that the area evolution of the marginally trapped surfaces foliating the tube was monotonic, if one takes the time-like and space-like segments together to form a single entity. It was subsequently proven, with a few additional assumptions,\footnote{We now understand that some of these additional assumptions do not apply to certain MOTSs that are present, for example, in the interior of a binary merger~\cite{Pook-Kolb:2020zhm}.} that the area of MOTSs along a marginal tube is indeed monotonic in general~\cite{Bousso:2015mqa}. Another example is the ``apparent horizon'' Penrose inequality, for which a counter-example was found by studying MOTSs in time-dependent spherically symmetric backgrounds~\cite{Ben-Dov:2004lmn}. A number of further examples can be found in~\cite{Bengtsson:2010tj}. 

%\ivan{Maybe some AMS papers?} \robie{Please, go ahead and add any other papers you have in mind. I don't know which ones you mean for sure.}

A more recent case where there has been positive back-and-forth interaction between exact solution analyses and more general dynamical situations has been in understanding the final fate problem in a black hole collision. That is:  what happens to the original black hole horizons (event or apparent) when two black holes merge to become one? In the case of the event horizon, this problem has been well-understood for over half a century. The result is nicely summarized in the well-known ``pair of pants'' diagram~\cite{hawking_ellis_1973}. While some understanding of this problem in the quasi-local paradigm dates back to the same time, much of the progress in resolving this question happened only in the last few years~\cite{Mosta:2015sga, Pook-Kolb:2018igu, Pook-Kolb:2019iao, Pook-Kolb:2019ssg, Booth:2021sow, Pook-Kolb:2021jpd} --- see~\cite{Pook-Kolb:2021gsh} for a succinct summary of the current understanding. A number of results that have emerged from this analysis have been important for revealing novel properties of MOTSs, and in clarifying which MOTSs can be appropriately called black hole boundaries.  

It was observed in~\cite{Pook-Kolb:2019iao, Pook-Kolb:2019ssg} that MOTSs with self-intersections play a role in a binary black hole merger. More specifically, it was found that the inner common MOTS, which forms in a bifurcation with the apparent horizon when the two original black holes become sufficiently close, develops self-intersections after penetrating the MOTSs describing the apparent horizons of the original black holes. This was the first observation of such surfaces, in part because previously existing  numerical MOTS finders were built so that they could only find ``star-shaped'' surfaces. Hence they were blind to surfaces with self-intersections. The key to the work of~\cite{Pook-Kolb:2019iao, Pook-Kolb:2019ssg} was the development of a reference-shape based MOTS finder. Subsequently it was shown that self-intersecting MOTSs are, in fact, extremely common~\cite{Booth:2020qhb}, when it was found that even within the event horizon of the Schwarzschild space-time there exist an apparently infinite family of self-intersecting MOTSs. 

As a short-hand, we will refer to the (now large) class of ``new'' MOTSs as \emph{exotic}. This is inspired by their generally complicated geometries, as opposed to the relatively sedate geometries of black hole horizons.

The observation of an infinite number of self-intersecting MOTSs was not the original motivation behind the work~\cite{Booth:2020qhb}. Instead, the aim was to describe the evolution of the apparent horizon during an extreme mass ratio binary black hole merger. This was based on the insights of~\cite{Emparan:2016ylg, Emparan:2017vyp, Emparan:2020uvt}, where the event horizon evolution during a merger was understood exactly in the (strict) extreme mass ratio limit. In this framework, one tracks the paths traced by a sheet of null rays --- corresponding to the event horizon of the large black hole ---  in the spacetime of the small black hole. While a complete analysis of the apparent horizon was not achieved in~\cite{Booth:2020qhb}, it was found that a number of structures exist within a single black hole that mimic those seen in a full numerical simulation of colliding black holes. These observations, and the methods used, provided the impetus for the analysis carried out and numerical methods developed in~\cite{Pook-Kolb:2019ssg, Booth:2021sow, Pook-Kolb:2021jpd}.

%Thus, exploring the properties of MOTSs in exact (explicit) spacetimes can inform more general cases. 

\begin{figure}
    \centering
    \includegraphics{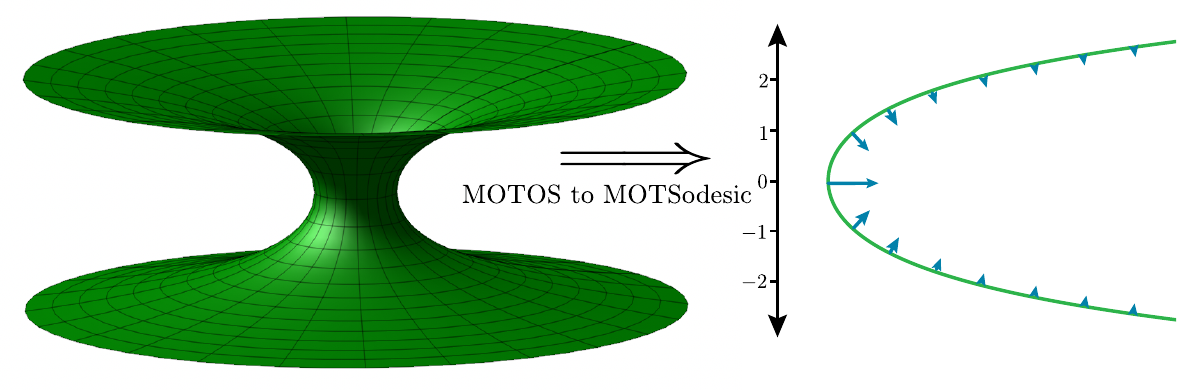}
    \caption{MOT(O)Ss are closely related to minimal surfaces. This figure shows a planar slice of Minkowski space in which MOTOSs (there are no MOTSs) \emph{are} minimal surfaces. 
    Then the shown catenoid is a MOTOS. Taking a slice of this rotationally symmetric surface (such as the half-plane $x>0, y = 0, z \in \mathbb{R}$) we see this recast as a MOTSodesic. 
    On the three-dimensional side, the surface is minimal as the curvature associated with the rotational symmetry (towards the centre) is equal and opposite to that of the catenoid 
    curving outwards. As a MOTSodesic that curvature from the rotational symmetry manifests as an acceleration (the blue arrows) that can be intuitively thought of as ``repelling'' the
    curve from the $z$-axis. }
    \label{fig:catenoid}
\end{figure}

The work done in~\cite{Booth:2020qhb} found MOTSs using a simple numerical algorithm that made use of the axisymmetry of the spacetime and a shooting method. 
The problem of finding MOTSs was reduced, via axisymmetry, to the problem of finding curves in the 
two-dimensional orbit space, or in concrete terms, the $(r,\theta)$-plane (see Figure \ref{fig:catenoid}).
As was shown more elegantly in later papers~\cite{Booth:2021sow,Hennigar:2021ogw}, those curves are
defined according to the equation
\begin{equation} 
T^a \bar{D}_a T^b = \kappa N^b \, ,
\end{equation}
where $T^a$ is the tangent to the curve,  $N^b$ is its normal, $\bar{D}$ is the  two-dimensional covariant derivative compatible with the metric
induced in the plane, and the acceleration $\kappa$ is that necessary to generate a curve that will rotate into a surface of vanishing outward null expansion. 
It encodes all the information about how the plane is embedded in the full spacetime and almost 
all of the real work in the method involves finding $\kappa$. Most such surfaces are in fact marginally outer trapped open surfaces (MOTOSs), in that they don't close. 
The shooting method is used to search through the space of MOTOS to find the surfaces that close and so are MOTSs. 
The resemblance to the geodesic equation (which would have $\kappa = 0$) and the fact that they
rotated into MOT(O)S led to these curves being called `MOTSodesics'. 

MOTSodesics, while applicable only in the case of axisymmetry, have proven to be a rather powerful device. Indeed, the techniques of~\cite{Booth:2020qhb} were generalized in~\cite{Booth:2021sow} to allow for the MOTSodesic shooting method to be applied directly to data obtained from a full numerical simulation~\cite{pook_kolb_daniel_2021_4687700}. The benefit of this is a fully shape-agnostic MOTS finder, without including additional assumptions of star-shaped surfaces, or requiring a reference surface or initial guess. This method was applied to initial data in~\cite{Booth:2021sow} and to the results of a numerical simulation of axisymmetric non-spinning black holes in~\cite{Pook-Kolb:2021jpd}, finding a large (possibly infinite) number of MOTSs which evolve and interact within the spacetime of a binary black hole merger. Like the self-intersecting MOTSs found in~\cite{Pook-Kolb:2019iao, Pook-Kolb:2019ssg}, the new MOTSs uncovered in~\cite{Pook-Kolb:2021jpd} possessed exotic geometric features. However, it was shown that these surfaces play an important role in the merger, being responsible for the ultimate annihilation of the original black hole horizons. 

One may worry that the existence of an infinite number of MOTSs dooms the quasi-local paradigm to irrelevance. However, as discussed in~\cite{Booth:2021sow, Pook-Kolb:2021jpd}, there emerges a natural way to understand whether or not a MOTS should be considered a black hole boundary. This is based on the stability operator first defined in~\cite{Andersson:2005gq,Andersson:2007fh}. A MOTS is said to be strictly stable if the principal (smallest real) eigenvalue of the stability operator is positive, while if it is negative they are unstable. Strictly stable MOTSs possess a \textit{barrier property}, which means they serve as boundaries between trapped/untrapped surfaces\footnote{In more detail, `inward' perturbation result in outer trapped surfaces while `outward' pertubations give rise to outer untrapped surfaces.}. In the results of~\cite{Booth:2021sow, Pook-Kolb:2021jpd} only three strictly stable MOTSs were found: the common apparent horizon, and the apparent horizons of the two original black holes. It is precisely these surfaces that should be considered as bonafide black hole boundaries. Moreover, it was found that the stability operator imposes some semblance of order on the multitude of MOTSs present in the merger: every time a MOTS was formed or was annihilated, this coincided with the vanishing of an eigenvalue of the stability operator. While mathematical results indicated this to be the case for the principal eigenvalue, no such proofs exist for higher eigenvalues of the stability operator. 
% Got to think about the phrasing here.
Additional evidence for this result has recently been found through a study of MOTSs in static, spherically symmetric black hole spacetimes with inner horizons~\cite{Hennigar:2021ogw}. 

In this manuscript, we are concerned with identifying MOTSs in the spacetimes of rotating black holes of arbitrary dimension. This turns out to be significantly more
complicated than for non-rotating black holes, and we focus on cases where the $D-$dimensional spacetime metric has $D\!\!-\!\!3$ spatial commuting Killing vector fields, so that the orbit space is two-dimensional (we will refer such spacetimes simply as `axisymmetric' for simplicity). We develop the formalism in two steps. 
First, in Section \ref{ArbDim}, we generalize the existing formalism from $(3+1)$ to arbitrary dimensions. This is essentially
a straightforward extension of the derivation found in \cite{Booth:2021sow} but is useful to establish notation and ideas that are used in later sections. 
We also consider non-rotating $(2+1)$-dimensional BTZ black holes and $(4+1)$-dimensional Schwarzschild to establish baselines for comparison with rotating 
black holes. 
Then, in Section \ref{RotBH}, we lift the mathematical assumption that restricted earlier works to studying non-rotating spacetimes and tackle arbitrary 
axisymmetric spaces. This is a non-trivial extension but having established the formalism, 
we then apply it in Sections \ref{(2+1)}, \ref{(3+1)} and \ref{(4+1)} to study MOTSs in a variety of exact rotating black hole solutions: the BTZ black hole, the Kerr black hole solution, and a particular subset of Myers-Perry black holes. Our exploration here is focused on identifying general properties and novel features of the MOTSs in these spacetimes, rather than the relatively complete catalogues of behaviours attempted in earlier works \cite{Booth:2020qhb,Booth:2021sow, Pook-Kolb:2021jpd,Hennigar:2021ogw}. 

As mentioned, previous experience suggests that the MOTSs in exact solutions provide clues as to the kinds of structures which may appear in a full merger. Therefore, our results in this context may be useful in considering numerical simulations of rotating black holes.

\section{MOTSs in non-rotating spacetimes of arbitrary dimension}
\label{ArbDim}

%This method works for $n$-dimensional slices of an $(n+1)$-dimensional spacetime whose induced metric and extrinsic curvature depend
%only on two of the spatial coordinates. That is, there are $(n-2)$ coordinate symmetries. These are the cases where it makes sense to 
%consider a similarly symmetric MOTS with all the ``rotational'' symmetries factored out, so just a curve in a two-dimensional plane remains. 
%
%Then the MOTS which are  $(n-1)$-dimensional hypersurfaces in a Riemannian $n$-manifold are analogues of axisymmetric surfaces 
%in a rotationally symmetric Riemannian three-manifold. I think that it is quite likely that this could be reformulated in a coordinate free way 
%(so that we only require Killing vector fields rather than coordinate Killing vector fields) but since we are applying only to exact solutions and 
%for those solutions we can put them in a form with the coordinate symmetries it's probably not worth it. We can do the cases that we are 
%interested in with these methods. 
%
%
%\robie{One thing we should maybe emphasize in the introduction is that the interior MOTSs of Schwarzschild ended up having a lot of the same type of structure as appears in a non-spinning merger. Maybe the rotating cases will shed some light on a merger of spinning black holes?}

The spaces described in the following sections are geometrically complicated and described by a correspondingly complex formalism. To motivate
and hopefully illuminate that formalism, we start with a much simpler case 
which covers non-rotating spacetimes. This is 
essentially the formalism used in \cite{Booth:2021sow} 
but now developed for arbitrary dimensions. 

The notation may appear to be overly complicated for the cases considered here, but it is chosen to be consistent with later sections. 
Throughout we are working with definite coordinate systems and decompositions of the spacetime with respect to those coordinates. Hence
indices in this paper should be understood as concrete rather than abstract. 

%that may be 
%hard to follow on first reading. Hence we begin with an outline of our procedures including a summary of the simpler cases considered in 
%previous works (though for arbitrary dimension). 

\subsection{Background space}
\label{Background}

Let   $(\Sigma, h_{ij}, D_i)$ be an  $n$-dimensional Riemannian manifold. 
It is a time-slice of an $D=(n+1)$-dimensional Lorentzian spacetime and has extrinsic curvature $K_{ij}$. We will assume that the initial data admits an Abelian isometry group whose principal orbits are $(n-2)$ tori $T^{n-2}$ (the isometries leave both $h_{ij}$ and $K_{ij}$ invariant).  The torus action is degenerate (have fixed point sets) on the symmetry axes.  Away from these fixed point sets, $\Sigma \cong \tilde{\Sigma} \times \mathcal{H}$  where $\tilde{\Sigma}$ is a two-dimensional manifold and $\mathcal{H}$ is an $(n-2)$-torus\footnote{We expect that this could be extended to other topologies. However the current formalism covers all examples in which we are interested and so we will be content with this version. }. More precisely, the orbit space $\tilde{\Sigma}:=\Sigma \setminus T^{n-2}$ is a manifold with boundary. 

$\Sigma$ is symmetric over the $\mathcal{H}$-component with the geometry depending only on $\tilde{\Sigma}$. 
In coordinate terms, we assume that there exists a coordinate system $y^i$, $i \in \{1, \dots, n\}$ on $\Sigma$ 
with $x^a := y^a$, $a \in \{1, 2\}$ parameterizing an open subset of $\tilde{\Sigma}$. For the examples we consider later in this section, 
$\Sigma$ will be a half-plane and the $x^a$ will be be polar-type coordinates 
$x^a = (r, \theta)$ (though they could also be Cartesian as they were in \cite{Booth:2021sow}). The symmetry then manifests so that
\begin{eqnarray}
h_{ij} = h_{ij} (r, \theta) \; \; \mbox{and} \; \; K_{ij} = K_{ij} (r, \theta) \; . 
\end{eqnarray}
We label the remaining $(n-2)$ toroidal coordinates $\phi^A := y^A$, with 
$A \in \{3, \dots, n \}$. 

To simplify later expressions, we introduce shorter forms for the coordinate vectors and one-forms:
\begin{equation}
\left\{\xi_a = \pdv{x^a},  \xi_A = \pdv{\phi^A} \right\}  \; \; \mbox{and} \; \; \left\{ \Phi^a = \dd x^a, \Phi^A = \dd \phi^A \right\} \; . 
\end{equation}
These have components
\begin{equation}
\xi_a^i = \delta_a^i \; , \; \;   \xi^i_A = \delta_A^i  \; ,  \; \; \Phi_i^a = \delta_i^a \; \mbox{and} \; \; \Phi_i^A = \delta_i^A \label{constcomp}
\end{equation}
where $\delta$ is the Kronecker delta. They obey the usual relations for coordinate vectors and one-forms which in our notation are
\begin{equation}
\Phi^a_i \xi^i_b = \delta^a_b \; , \; \; \Phi^A_i \xi^i_b = 0 \; , \; \;  \Phi^a_i \xi^i_B = 0 \; \mbox{and} \; \; \Phi^A_i \xi^i_B = \delta^A_B \; . 
\end{equation}

It is often useful to think of the coordinates as foliating  $\Sigma$ into two-dimensional surfaces 
$\twoS_{(\phi)}$ of constant $\phi^A$ and 
$(n\!-\!2)$-dimensional surfaces $\symS_{(x)}$ of constant $x^a$. Here the $(\phi)$ and $(x)$ subscripts indicate which quantities are held constant on the surface. Then each $\twoS_{(\phi)}$ has two tangent vectors
$\xi_a$ and $(n \! - \! 2)$ normal one-forms $\Phi^A$ while each $\symS_{(x)}$ has $(n \! - \! 2)$  tangent vectors $\xi_A$ and two normal one-forms $\Phi^a$.

By the symmetry assumption, the $\xi_A$ (but not the $\xi_a$) are Killing vector fields and also generate symmetries of the extrinsic curvature:
\begin{eqnarray}
\mathcal{L}_{\xi_A} h_{ij}  = \mathcal{L}_{\xi_A} K _{ij} = 0 \; . 
\end{eqnarray}
Given this symmetry, the $\twoS_{(\phi)}$ are geometrically identical and we can understand the geometry of the full system 
by focusing on 
a single surface $\twoS_{(\phi)}$ of constant $\phi^A_o$. Since they are identical, it doesn't matter which we choose to work on and so we drop the $(\phi)$ labels from those surfaces and from now on just refer to $\twoS$.

\subsection{Warped product metrics}
\label{Warp}

The preceeding holds for all spacetimes studied in this paper. However in this section we further specialize 
to those spacetimes whose spatial metric is a warped product of $\tilde{\Sigma}$ and $\mathcal{H}$, and so block-diagonal:
\begin{eqnarray}
h_{ij} \dd y^i \dd y^j = \underline{h}_{ab} \dd x^a \dd x^b   + \uh_{AB} \dd \phi^A \dd \phi^B  \label{linelBD}
\end{eqnarray}
which is equivalent to the expansion
\begin{eqnarray}
h_{ij} = \uh_{ab} \Phi^a_i \Phi^b_j  + \uh_{AB} \Phi^A_i \Phi^B_j \, , 
\end{eqnarray}
for
\begin{eqnarray}
\uh_{ab} := h_{ij} \xi^i_a \xi^j_b = h_{ab}   \; \; \mbox{and} \; \; 
\uh_{AB}  := h_{ij} \xi^i_A \xi^j_B = h_{AB}.
\end{eqnarray}
In these expressions, the underline indicates that the base object $h$ has been broken into components by contracting with 
coordinate tangent vectors, while the index labels indicate which vectors were used. 

Of course, contracting the metric with $\xi_a^i$ is equivalent to pulling the metric back into $\twoS$  and 
contracting with $\xi_A^i$ is equivalent to pulling back to $\symS_{(x)}$. 
Hence $\uh_{ab}$ is the induced metric on $\twoS$ and $\uh_{AB}$ is the 
induced metric on $\symS_{(x)}$. As is clear from (\ref{linelBD}) the components of these metrics are 
identical to the coordinate-restricted components of the full metric. This equality of the components in the metrics
in the different spaces  follows from the fact that we are working 
with constant coordinate surfaces and so the $\xi_a \in T \twoS$ and $\xi_A \in T \symS_{(x)}$.

%The underlines are here intended to indicate that these quantities are the pull-backs of the full $h_{ij}$ into either $\tilde{\Sigma}$ or
%$\mathcal{H}$. That is
%\begin{eqnarray}
%h_{ab} = h_{ij} \xi^i_a \xi^i_b \; \; \mbox{and} \; \; 
%h_{AB} = h_{ij} \xi^i_A \xi^i_B
%\end{eqnarray}
%respectively. Of course since we are working with coordinate surfaces,  
%the components of the pull-backs are the same as the corresponding components of the full metric. 

The inverse  metric is similarly block-diagonal:
\begin{eqnarray}
h^{ij} \pdv{y^i}   \pdv{y^j} & = \oh^{ab} \pdv{x^a} \pdv{x^b} +  \oh^{AB} \pdv{\phi^A}  \pdv{\phi^B}  \label{invhBD}
\end{eqnarray}
which is equivalent to the expansion
\begin{eqnarray}
h^{ij} = \oh^{ab}  \xi_a^i \xi_b^j   + \oh^{AB} \xi_A^i \xi_B^j
\end{eqnarray}
for
\begin{eqnarray}
\oh^{ab} := h^{ij} \Phi^a_i \Phi^b_j \; \; \mbox{and} \; \;   \oh^{AB} := h^{ij} \Phi^A_i \Phi^B_j \; . 
\end{eqnarray}
Here the overline indicates that the base object $h$ has been broken into components by contracting with
coordinate one-forms while the index labels indicate which one-forms were used. However, due to the block-diagonal form of the metric,  it is also true that
\begin{eqnarray}
\overline{h}^{ab} = \left(  \underline{h}_{ab} \right)^{-1} \; \; \mbox{and} \; \; \overline{h}^{AB} = \left(  \underline{h}_{AB} \right)^{-1} \; . \label{inverseBD}
\end{eqnarray}
In this case, the full inverse metric is also block diagonal with the blocks being the inverses of the two metric components. Hence we don't need to distinguish 
between $\underline{h}^{AB}$ and $\overline{h}^{AB}$. Nonetheless, we will continue to do so for compatibility with Section \ref{RotBH}, where they
are not equivalent. 

We also use the same underline/overline and index notation for the decomposition of other tensors. For example
\begin{eqnarray}
K_{ij} =  \underline{K}_{ab} \Phi^a_i \Phi^b_j +  \underline{K}_{Ab} (\Phi^A_i \Phi^b_j + \Phi^A_j \Phi^b_i) +  \underline{K}_{AB} \Phi^A_i \Phi^B_j \;  \label{KXX}
\end{eqnarray}
for 
\begin{eqnarray}
\underline{K}_{ab} = K_{ij} \xi^i_a \xi^j_b \; , \; \; \underline{K}_{Ab} = K_{ij} \xi^i_A \xi^j_b \; , \; \;  \underline{K}_{AB} = K_{ij} \xi^i_A \xi^j_B  \; . 
\label{uK}
\end{eqnarray}
We do not assume that $K_{ij}$ is block-diagonal. 

Similarly for a general vector $v^i$ we can expand
\begin{equation}
v^i = \overline{v}^a \xi_a^i + \overline{v}^A \xi_A^i
\end{equation}
for 
\begin{equation}
\overline{v}^a := v^i \Phi_i^a \andd \overline{v}^A := v^i \Phi_i^A \; . 
\end{equation}
while a general one-form $w_j$ expands as 
\begin{equation}
w_i = \underline{w}_a \Phi^a_i + \underline{w}_A \Phi^A_i
\end{equation}
for 
\begin{equation}
\underline{w}_a : = w_i \xi^i_a \andd \underline{w}_A := w_i \xi^i_A \; . 
\end{equation}
Apart from the notation, there is nothing new here: expansion with respect to coordinate bases is the most common way to do calculations in general relativity.

%\begin{eqnarray}
%h^{ij} = \overline{h}^{ab}(r,\theta) \xi^i_a \xi^j_b +   \overline{h}^{AB}(r,\theta) \xi^i_A \xi^j_B:
%\end{eqnarray}
%where  $\overline{h}^{ab} = \left(  \underline{h}_{ab} \right)^{-1}$ and $\overline{h}^{AB} = \left(  \underline{h}_{AB} \right)^{-1}$ 
%(for now, this notation is overkill but it is adopted for later consistency). The overlines are here to indicate that these are the 
%inverses of the corresponding underlined metrics. For metrics of the form (\ref{BDmetric}) the components of these inverse-metrics are the 
%same as the corresponding full inverse metric: $h^{ab} = \overline{h}^{ab}$ and $h^{AB} = \overline{h}^{AB}$. 
%
%We extend the same notational conventions to the extrinsic curvature
%
%\begin{eqnarray}
%\underline{K}_{AB} = K_{ij} \xi^i_A \xi^j_B
%\end{eqnarray}

\subsection{Curves in $\twoS$}
\label{curves1}

\begin{figure}
\centering
\includegraphics[scale=1]{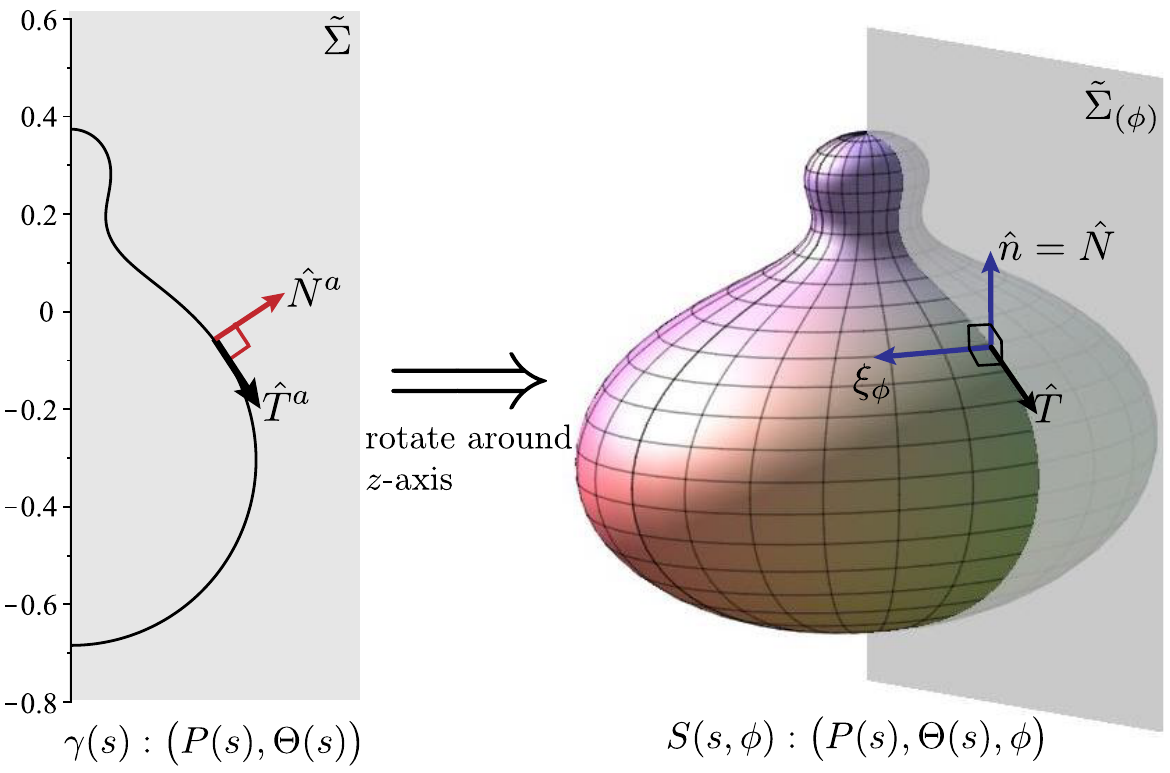}
\caption{The rotation of a curve $\gamma$ into a surface $S$ for a three-dimensional $\Sigma$ with the simple warp-product metric
 (\ref{linelBD}) along with the resulting surface-associated triad: $\{ \hat{T}, \hat{N}, \xi_\phi \}$. 
 } 
 \label{NonRotFig}
\end{figure}

We now consider the geometry of a general curve $\gamma$ in $(\tilde{\Sigma}, \underline{h}_{ab}, \uD_a)$. 
We parameterize by arclength so that 
$\gamma(s): x^a = X^a(s)$ in $\tilde{\Sigma}$. This has unit tangent vector
\begin{equation}
\hat{T} = \dv{s} = \dot{X}^a \pdv{x^a} = \dot{X^1} \pdv{x^1} + \dot{X^2} \pdv{x^2} \, \; . 
\end{equation}
Thanks to the parameterization and the fact that $\twoS$ is two-dimensional, for any such curve we must have
\begin{equation}
\hat{T}^a \uD_a \hat{T}^b = \kappa \hat{N}^b \label{curveEq}
\end{equation}
for some function $\kappa$ and $\hat{N}$ the  unit normal vector to $\hat{T}$ which we orient via
\begin{equation}
\hat{N}_a= \underline{\epsilon}_{ab} \hat{T}^b =   \sqrt{\det(\underline{h}_{cd})}  \left(\dot{X^2} [\dd x^1]_a - \dot{X^1} [\dd x^2]_a  \right) \, ,  \label{hN}
\end{equation}
where 
\begin{equation}
    \epsilon_{ab} = \sqrt{\det (\uh_{cd})} 
    \left[
      \begin{array}{cc}
          0 & 1 \\ -1 & 0 
      \end{array}
    \right] 
    \label{eq:LCdn}
\end{equation}
%$\underline{\epsilon} = \sqrt{\det(\underline{h}_{cd})} \dd x^1 \wedge \dd x^2$ 
is the Levi-Civita tensor on $\tilde\Sigma$ (the area form). This
orientation is chosen to match the four-dimensional Schwarzschild solution with $x^1 = r$ and $x^2=\theta$, so that moving away from the north pole in the $\pdv{\theta}$ direction, $\hat{N}$ (as a vector) is outward pointing. This construction is shown in Figure \ref{NonRotFig}. Our goal is to find a $\kappa$ that will 
generate curves that rotate into MOTSs. 
%
%
%
%Given a specific curve we can, of course, calculate $\kappa$ by contracting with $\hat{N}$:
%\begin{equation}
%\kappa =  \hat{N}_b \hat{T}^a \uD_a \hat{T}^b \; . 
%\end{equation}

Along $\gamma$, $\hat{T}$ and $\hat{N}$ provide an orthonormal basis for $T \twoS$ and so we can expand
\begin{equation}
\underline{h}_{ab} = \hat{T}_a \hat{T}_b + \hat{N}_a \hat{N}_b 
\; \; \mbox{and} \; \; 
\underline{\epsilon}_{ab} = \hat{N}_a \hat{T}_b - \hat{T}_a \hat{N}_b \; . 
\end{equation}
%Both of these quantities are, of course, actually independent of the dyad. 

\subsection{Surfaces in $\Sigma$}
\label{sec:2.4}

%\subsubsection{Rotating $\gamma$ into $S$}

Using the symmetry vector fields $\xi_A$, we   `rotate' $\gamma$ into an $(n\!-\!1)$-dimensional hypersurface $S$ in $\Sigma$. This is parameterized by the  coordinates $\vartheta^{\alpha} = (s, \phi^A)$,  where we use Greek indices for quantities that are intrinsic to this surface. $TS$ is then spanned by the coordinate vectors
\begin{equation} 
 \left\{  \pdv{\vartheta^\alpha} \right\}  = \left\{ \hat{T}, \xi_A   \right\}  \; . \label{spanning1}
\end{equation}
%Note that, if we don't include indices, there is an ambiguity in our notation. $\hat{T}$ could be any of $\hat{T}^a$, $\hat{T}^\alpha$ or $\hat{T}^i$ and $\xi_I$ could be either $\xi_I^i \in T\Sigma$ or $\xi_I^\alpha \in TS$. However since the higher dimensional versions of these vectors are just push-forwards of the lower-dimensional versions,  this should not cause any confusion. 
and $T^\star \!  S$ is spanned by the coordinate one-forms
\begin{equation}
%\left\{  \dd \vartheta^\alpha \right\} = \left\{ \tilde{\Phi}^s , \tilde{\Phi}^A \right\} = \left\{ \overline{\dd s}, \overline{\dd \phi}^A \right\} \; 
\left\{  \bd \vartheta^\alpha \right\} = \left\{ \bd s, \bd \phi^A \right\} \, , \label{spanning2}
\end{equation}
where the black-board $\bd$ is used as a reminder that  these are one-forms in $T^\star S$ rather than the full $T^\star \Sigma$. As one-forms they don't 
push-forward into the full space and so it is useful to have a visual reminder that $\dd \phi^A = \Phi^A  \neq \bd \phi^A$. %{\bf (SAME $\Phi$?)}
%Note that in the absence of upper indices there is an ambiguity in our notation: $\xi_I$ could mean either $\xi_I^i \in T\Sigma$ or $\xi_I^A \in TS$. 
%However this should not cause any confusion. 

 Then  the induced metric $q_{\alpha \beta}$ on $S$ is
\begin{equation}
q_{\alpha \beta} \bd \vartheta^\alpha \bd \vartheta^\beta  
=  \bd s^2 + \uh_{AB} \bd \phi^A \bd \phi^B \\
%
%= \tilde{\Phi}^s_\alpha \tilde{\Phi}^s_\beta + \beta_A \left(\tilde{\Phi}^s_\alpha \tilde{\Phi}^A_\beta + \tilde{\Phi}^A_\alpha \tilde{\Phi}^s_\beta  \right) + \uh^{(x)}_{AB} \tilde{\Phi}^A_\alpha \tilde{\Phi}^B_\alpha
\end{equation}
which has inverse
\begin{eqnarray}
q^{\alpha \beta} &=\hat{T}^\alpha \hat{T}^\beta + \oh^{AB} \xi_A^\alpha \xi_B^\beta \nonumber 
\end{eqnarray}
and pushes-forward from $S$ into $\Sigma$ in a trivial way:
\begin{equation}
q^{ij} =  \hat{T}^i \hat{T}^j +  \oh^{AB} \xi^i_A \xi^j_B \; . \label{qijBD}
\end{equation}

As a simple example consider a three-dimensional $\Sigma$. Then, as shown in Figure \ref{NonRotFig}, 
there is a single rotational Killing vector field $\xi_\phi$ and so 
\begin{equation}
q^{ij} =  \hat{T}^i \hat{T}^j +  \oh^{\phi \phi} \xi^i_\phi \xi^j_\phi =  \hat{T}^i \hat{T}^j + \hat{\phi}^i \hat{\phi}^j \, , \label{qBD3}
\end{equation}
where 
\begin{equation}
\hat{\phi} = \frac{\xi_\phi}{\| \xi_\phi \|} \label{hp}
\end{equation}
is both the unit normal vector pointing in the rotational direction and also the unit normal to the $\twoS$ in $\Sigma$. 

For a four-dimensional
$\Sigma$ this will be more complicated. Then, for rotational Killing vector fields $\xi_\phi$ and $\xi_\psi$, 
\begin{equation}
q^{ij} =  \hat{T}^i \hat{T}^j +  \oh^{\phi \phi} \xi^i_\phi \xi^j_\phi + \oh^{\phi \psi} (\xi^i_\phi \xi^j_\psi + \xi^i_\psi \xi^j_\phi)  + \oh^{\psi \psi} \xi^i_\psi \xi^j_\psi  \; . \label{qij5}
\end{equation}
If the Killing vectors are orthogonal to each other  ($h_{\phi \psi} = 0$), then one can again use (\ref{hp}) and the equivalent expression for $\hat{\psi}$ to write
\begin{equation}
q^{ij} =   \hat{T}^i \hat{T}^j + \hat{\phi}^i \hat{\phi}^j  + \hat{\psi}^i \hat{\psi}^j \; . \label{qBD4}
\end{equation}
If they are not orthogonal to each other ($h_{\phi \psi} \neq 0$) then one can still construct an orthonormal dyad, however the two basis vectors cannot both be
parallel to Killing directions. In such cases (which will mainly show up in later sections) we stay with the more general (\ref{qij5}).

Finally,  consider the unit normal $\hat{n}_i$ to $S$ in $\Sigma$ and its relation to that of the curve normal $\hat{N}$. 
The push-forward $\hat{N}^a \in T\tilde{\Sigma}$ into $\Sigma$ is
\begin{equation}
\hat{N}^i = \hat{N}^a \xi^i_a \;  \label{Nn}
\end{equation}
and it remains unit length and orthogonal to $\hat{T}$. Further, thanks to the block-diagonal structure of (\ref{linelBD}) it is also orthgonal to the 
$\xi_A$  (the other tangent vectors to $S$):
\begin{equation}
h_{ij} \hat{N}^i \xi_A^j = h_{ij} (\hat{N}^a \xi^i_a) \xi_A^j = h_{aA} \hat{N}^a = 0 \, , 
\end{equation}
since $h_{aA} = 0$. Hence, $\hat{n}^i = \hat{N}^i$.
%\begin{equation}
%\hat{n}^i = \hat{N}^i \; . 
%\end{equation}
This is not very surprising, but it is important to emphasize that the equality only holds because the `off-diagonal' $h_{aA} = 0$. This is 
another condition that
will not hold in the more complicated spacetimes considered in later sections. 

\subsection{MOTSs and MOTSodesics}
%
%and the full inverse metric can be expanded as
%\begin{equation}
%h^{ij} = (\hat{T}^a \hat{T}^b + \hat{N}^a \hat{N}^b) \xi_a^i \xi_b^j + \oh^{ij} \, , \label{metricBD}
%\end{equation}
%where $\oh^{ij} = \oh^{ab} \xi_a^i \xi_b^j$. 

We are now ready to derive the MOTSodesic equations. 

$S$ is a marginally outer trapped surface if the outward (in the $\hat{n}$ direction) null expansion vanishes:
\begin{equation}
\theta_{\ell} = 0  \; \; \Longrightarrow \; \;  k_{u} + k_{n} = 0 \label{eq:Meq}
\end{equation}
where 
\begin{equation}
k_{u} = q^{ij} K_{ij} \label{eq:ku} \, ,  
\end{equation}
is the extrinsic curvature of $S$ in the time-like direction perpendicular to $\Sigma$ and  
\begin{equation}
k_n = q^{ij} D_i \hat{n}_j  \label{eq:kn}
\end{equation}
is the  extrinsic curvature of $S$ in $\Sigma$.

By (\ref{qijBD}) and (\ref{Nn}) we can expand
\begin{eqnarray}
k_n  & = \left(\hat{T}^i \hat{T}^j + \oh^{ij} \right) D_i \hat{N}_j  \\ 
& = - \kappa_{\mbox{\tiny{MOTS}}}^\perp +  \oh^{ij} D_i \hat{N}_j  \nonumber
\end{eqnarray}
where 
\begin{equation}
\kappa_{\mbox{\tiny{MOTS}}}^\perp  = \hat{N}_j \hat{T}^i D_i \hat{T}^j = \hat{N}_b \hat{T}^a \uD_a \hat{T}^b \label{kappa} 
\end{equation}
with $\uD_a$ the $\uh_{ab}$-compatible derivative on $\tilde{\Sigma}$ and the $\perp$ superscript indicating 
that this is the special case where $T \tilde{\Sigma} \perp T \mathcal{H}$. Further we define
\begin{eqnarray}
\overline{h}^{ij} := \xi^i_A \xi^j_B \overline{h}^{AB} = \xi^i_A \xi^j_B \Phi^A_k \Phi^B_l h^{kl} \; .  \label{ohij}
\end{eqnarray}
This notation is ambiguous: based on previous uses of the notation, it could equally well mean
$\xi^i_a \xi^j_b \overline{h}^{ab}$. 
However, it is only the form (\ref{ohij}) that appears in our calculations.

Then the expansion vanishes for 
\begin{eqnarray}
\kappa_{\mbox{\tiny{MOTS}}}^\perp  = q^{ij} K_{ij} + \oh^{AB} \xi_A^i \xi_B^j D_i \hat{N}_j \; . 
\end{eqnarray}
Using (\ref{qBD3}) this can be rearranged into the form
\begin{equation}
\kappa_{\mbox{\tiny{MOTS}}}^\perp  = \mathcal{K} + \mathcal{K}_{\hat{N}}  + \mathcal{K}_{\hat{T} \hat{T}} \label{kM0}
\end{equation}
where
\begin{eqnarray}
\mathcal{K} &:=  \underline{h}^{AB} \uK_{AB} \label{kM1} \\
\mathcal{K}_{\hat{N}}  & := (\uD_a \ln \sqrt{\underline{h}}) \hat{N}^a = (\uD_a \ln \sqrt{\underline{h}}) \underline{\epsilon}^a_{\phantom{a} b} \hat{T}^b \label{kM2} \\
\mathcal{K}_{\hat{T} \hat{T}} & := (\uK_{ab} )\hat{T}^a \hat{T}^b \; . \label{kM3}
\end{eqnarray}
The origin of the first and third terms is obvious with $\uK_{AB}$ and 
$\uK_{ab}$ defined in (\ref{uK}).

The second is a little more complicated. Note that
\begin{eqnarray}
\oh^{AB} \xi_A^i \xi_B^j D_i \hat{N}_j & = \frac{1}{2} \uh^{AB} \xi_A^i \xi_B^j \mathcal{L}_{\hat{N}} h_{ij} \label{LNh} \\
& = \frac{1}{2} \uh^{AB}  \mathcal{L}_{\hat{N}} \uh_{AB} \nonumber \\
& = \frac{1}{2 \underline{h}} \mathcal{L}_{\hat{N}} \uh \nonumber \\
& = \mathcal{L}_{\hat{N}} \ln \sqrt{\underline{h}} \nonumber \; . 
\end{eqnarray}
In this derivation $\underline{h} = \det (\underline{h}_{AB})$. 
In the first line we apply the definition of the Lie derivative, while in the second we apply $\mathcal{L}_{\xi_A} \hat{N}^i = 0$ (since the 
$\xi_A$ are Killing vector fields). The third applies the Jacobi identity for the derivative of a determinant and the last line elementary algebra. 
Finally, (\ref{hN}) can be used to rewrite the $\hat{N}^a$ dependence as a $\hat{T}^a$ dependence.\footnote{
There is  an alternate `acceleration'  form of $\mathcal{K}_{\hat{N}}$ that has appeared in previous papers. We can equivalently write
\begin{eqnarray}
\mathcal{K}_{\hat{N}} = (-\xi_a^j   \xi_A^i \oh^{AB}  D_i  \xi_{Bj}) \hat{N}^a \; . 
\end{eqnarray}
This is simpler to derive but generally more complicated to calculate than (\ref{kM2}). One exception is when $\Sigma$ is three-dimensional (the case
for all previous calculations) where this becomes the geometrically appealing
\begin{equation}
\mathcal{K}_{\hat{N}} = ( - \xi_a^j \hat{\phi}^i D_i \hat{\phi}_j) \hat{N}^a \; . 
\end{equation}
That is, the acceleration of the unit normals to $\twoS$ in $\Sigma$ in the $\hat{N}^a$ direction. This is the form that appeared in \cite{Booth:2021sow}. 
In four-dimensions in cases where (\ref{qBD4}) holds, one could similarly write:
\begin{equation}
\mathcal{K}_{\hat{N}} = ( - \xi_a^j \hat{\phi}^i D_i \hat{\phi}_j -   \xi_a^j \hat{\psi}^i D_i \hat{\psi}_j) \hat{N}^a \; , 
\end{equation}
and this would extend in the obvious way to arbitrary dimensions for a similarly broken-up metric. 
However beyond three-dimensions, (\ref{kM2}) will usually be the simplest form to use in calculations.  }

This $\mathcal{K}_N$ term generates
the `repulsion' from the $z$-axis mentioned in caption to Figure \ref{fig:catenoid}: the `inward' curvatures associated with the symmetries can balance
an outward curvature. As a curve gets closer to the symmetry axis, this inward curvature approaches infinity and so the corresponding outward curvature also 
dramatically increases (though the amount will be modulated by $\mathcal{K}$ and $\mathcal{K}_{\hat{T}}$). For MOTSodesics, this looks like a repulsion from
the $z$-axis. See Section \ref{sec:3p1GC} for further discussion of this term in $(3\!+\!1)$-dimensions.

Writing $\kappa_{\mbox{\tiny{MOTS}}}$ in the form (\ref{kM0}) has the advantage that all of the geometry except that intrinsic to $\twoS$ 
has been subsumed into one scalar, one one-form and one two-index tensor field, with each defined over $\twoS$ and only contracted with $\hat{T}^a$s. Hence $\kappa_{\mbox{\tiny{MOTS}}}$ depends quadratically on $\hat{T}^a$. 

Then the solutions of 
\begin{equation}
\hat{T}^a \uD_a \hat{T}^b = \kappa_{\mbox{\tiny{MOTS}}}^\perp  \hat{N}^b
\end{equation}
are our MOTSodesics: curves in $\tilde{\Sigma}$ that rotate into full MOTSs. This is a pair of coupled non-linear, second-order ODES for 
$X^a(s)$ which can easily be solved by standard mathematical packages. Explicitly
\begin{equation}
\frac{\mathrm{d}^2 X^a}{\mathrm{d}s^2}  =-  \underline{\Gamma}_{bc}^{a} \frac{\mathrm{d} X^b}{\mathrm{d} s} \frac{\mathrm{d}X^c}{\mathrm{d}s} +  {\kappa}_{\mbox{\tiny{MOTS}}}^\perp  \underline{\epsilon}^a_{\phantom{a} b} \frac{\mathrm{d} X^b}{\mathrm{d} s}
 \label{MOTSperp2}
\end{equation}
where the $ \underline{\Gamma}_{bc}^{a}$ are the two-dimensional Christoffel symbols associated with $\uh_{ab}$ on $\tilde{\Sigma}$:
\begin{equation}
\underline{\Gamma}_{ab}^d =   \frac{1}{2} \underline{h}^{dc} \left(\partial_a \underline{h}_{bc} +  \partial_b \underline{h}_{ca} - \partial_c \underline{h}_{ab} \right) = \uh^{dc} \xi_a^i \xi_b^j \xi_c^k \Gamma_{kij}  \; , \label{C2}
\end{equation}
where the second equality relates it to the full $h_{ij}$-connection. This can be derived  by expanding the full Christoffel
symbol and keeping in mind that by (\ref{constcomp}), $\partial_a \xi^i_b = \partial_a \delta_b^j = 0$.

To get a feeling for the effects of lower and higher dimensions on 
the properties of MOTSs, we apply the formalism to studying 
three-dimensional non-rotating BTZ and five-dimensional Schwarzschild spacetimes. 

\subsection{Non-rotating BTZ}

\subsubsection{General considerations} 

For a $(2+1)$-dimensional spacetime the preceding formalism greatly  simplifies. The
hypersurfaces $\Sigma$ are two-dimensional and thus the MOTSs are simply curves. 
Further, there are no $\phi^A$ coordinates and so neither do there need to be any symmetry assumptions.
The set of all MOTSodesics is exactly the same as the set of all MOTSs.

Further, the  MOTSodesic equations are  very simple. All terms involving capital Latin indices vanish. %and the normals are equal ($\hat{n} = \hat{N} \Leftrightarrow  \alpha =0 $ ). 
Then
\begin{equation}
{\kappa}_{\mbox{\tiny{MOTS}}}^{\mbox{\tiny{2D}}}  =  {K}_{ab} \hat{T}^a \hat{T}^b 
\end{equation} 
and so the MOTSodesic equations are simply
\begin{equation}
\hat{T}^a \uD_a \hat{T}^b = ({K}_{cd} \hat{T}^c \hat{T}^d ) \hat{N}^b  \label{MOTS1nr} \; . 
\end{equation} 
Note in particular that the $\mathcal{K}_{\hat{N}}$ vanishes: there are no symmetry axes providing extra sources of curvature. Hence in $(2+1)$ we can no longer expect MOTSodesics to 
be `repelled' by any axes. 

In the special case where $\Sigma$ is extrinsically flat ($K_{ij} = 0$) things become even simpler: the MOTSodesic equations are the (space-like) geodesic equations. Hence in this case, the set of MOTSs and the set of Riemannian geodesics in $\Sigma$ 
are identical. 

\subsubsection{BTZ black hole spacetime: stationary coordinates}
\label{sec:BTZnonrot}

The non-rotating BTZ black hole spacetime is a $(2+1)$-dimensional solution of Einstein's equations with a negative cosmological constant\cite{Banados:1992gq,Banados:1992wn} . It takes the form
\begin{equation}
 \dd s^2 = -F \dd t^2 + \frac{\dd r^2}{F} + r^2 \dd\phi ^2
 \end{equation} where
 \begin{equation}
 F =  \frac{r^2}{L^2}  - M 
 \end{equation}
 with $L$ the (anti-deSitter) cosmological length scale and $M$ the (dimensionless) mass. 
 
 The surfaces of constant $t$ have induced space-like metric
\begin{equation}
h_{ij} \dd x^i \dd x^j  = \frac{\dd r^2}{F} + r^2 \dd\phi^2 \label{eq:BTZ_nonrot_h}
\end{equation} 
and vanishing extrinsic curvature:
\begin{equation}
K_{ij}  =0 \; . 
\end{equation}
In the usual way, these space-like 
$\Sigma$ only exist outside
the standard horizon (MOTS) at $r = L \sqrt{M}$. 
Inside (\ref{eq:BTZ_nonrot_h}) becomes time-like. 
Hence in these time slices we can only consider
outside MOTSodesics. However, in those slices, 
the MOTSodesics are Riemannian geodesics.

%As would be expected for these stationary coordinates, several components components of the metric and extrinsic curvature diverge at the event horizon
%\begin{equation}
%r_+ =  \frac{L}{\sqrt{2}} \sqrt{ M+ \sqrt{M^2 - \left(\frac{J}{L} \right)^2 }} \; .
%\end{equation}

The equations are easily developed. 
A unit parameterized curve
\begin{eqnarray}
  r & = P(s) \\
  \phi & = \Phi(s) 
\end{eqnarray}
in a surface of constant $t$ has unit tangent vector
\begin{equation}
\hat{T} = \dot{P} \pdv{r} + \dot{\Phi} \pdv{\phi} 
\end{equation}
where the dot indicates a derivative with respect to arclength $s$. That is 
\begin{equation}
\frac{\dot{P}^2}{F} + P^2 \dot{\Phi}^2 = 1 \label{BTZnrCons}\; . 
\end{equation}
Then the outward oriented for increasing $\Phi$ unit normal is
\begin{equation}
\hat{N} = \frac{P \dot{\Phi}  }{\sqrt{F}} \pdv{r} -\frac{ \dot{P} }{P \sqrt{F}}\pdv{\phi} \, , 
\end{equation}
and the MOTSodesic equations (\ref{MOTS1nr}) take the relatively simple form:
\begin{eqnarray}
\ddot{P} & = \frac{F'}{2F} \dot{P}^2 + P F \dot{\Phi}^2  \label{BTZnr1} \\
\ddot{\Phi} & = - \frac{2}{P} \dot{P} \dot{\Phi}  \label{BTZnr2} \; . 
\end{eqnarray}
Using (\ref{BTZnrCons}) to eliminate $\Phi$ from (\ref{BTZnr1}) we obtain:
\begin{eqnarray}
\ddot{P} = \left(\frac{F'}{2F} - \frac{1}{P} \right) \dot{P}^2 + \frac{F}{P} \; . 
\end{eqnarray}

% \ivan{Here and in other equations we need to make a definite decision about what to do with $r$ vs $P$}

% and this has a simple solution:
% \begin{eqnarray}
% P = L \sqrt{M + (A e^{s/L} - B e^{-s/L} )^2 }  \; ,
% \label{eq:P_BTZ_nonrot}
% \end{eqnarray}
% for some constants $A$ and $B$. Then (\ref{BTZnr2}) can also be solved exactly as:
% \begin{eqnarray}
% \Phi = \phi_o + k\,  \mathrm{arctan} \left(\frac{2  A e^{s/L} (A e^{s/L} - B e^{-s/L} )+M )}{\sqrt{M(4AB - M) }} \right) \label{eq:Phi_BTZ_nonrot} \; . 
% \end{eqnarray}

% Then it is clear that, apart from the 
% standard MOTS at $P=L \sqrt{M}$ ($A=B=0$), 
% there is no closed curve that is also a solution 
% to the MOTSodesic equations. If either $B=0$ and 
% $A\neq 0$ then $P$ is monotonically increasing, 
% exponentially diverging from $r=L \sqrt{M}$ as $s \rightarrow \infty$ and converging as 
% $s \rightarrow - \infty$. Conversely it is 
% monotonically decreasing with the opposite 
% convergence properties if $A=0$ and $B \neq 0$.
% Such curves cannot close. If neither $A$ nor $B$ vanish then 
% $P$ has a minimum at 
% \begin{equation}
% s = 
% \end{equation}
% and is monotonically decreasing before that and increasing afterwards, diverging as $s \rightarrow \pm \infty$. Again such a curve cannot be closed.

% \textcolor{blue}{{\bf [I have looked at this too, and will write my results here in blue. They can be combined, for example.]}}

This equation has a simple exact solution,
\begin{equation}
    P = L \sqrt{M + (M-C_1) \sinh^2 \left[\frac{s-C_2}{L} \right]} \, ,
\end{equation}
where $C_1$ and $C_2$ are constants. Note that to maintain real values of $P$, we must restrict the integration constant $C_1 \le M$. With this solution for $P(s)$, the remaining equations are solved by
\begin{equation}
    \Phi = C_3 \pm \frac{1}{\sqrt{M}} \tanh^{-1} \left[\sqrt{\frac{C_1}{M}} \tanh \left[ \frac{s-C_2}{L}\right] \right] \, ,
\end{equation}
where $C_3$ is a constant. Requiring this solution to be real further restricts the parameter $C_1$ to lie in the range $0 \le C_1 \le M$.

With the exact solution in hand, it is possible to make some precise statements about MOTSodesics in this spacetime. First, we note that the only solution for which $\dot{P} = 0$ for all values of the curve parameter $s$ is the black hole horizon. This corresponds to the choice of constant $C_1 = M$. Thus, this is the only closed MOTS with circle symmetry. If any further closed MOTSs exist in this slicing, they would have to have dependence on both the radial and angular coordinate. However, it is easy to show that this is not the case. If additional closed MOTSs were to exist, then it would be necessary that the function $P(s)$ would have turning points, as a monotonic evolution would imply an open surface. Aside from the obvious case of the event horizon (for which $\dot{P} = 0$ everywhere), the only parameter-dependent turning point occurs for $s = C_2$. Substitution of this value back into the defining equation for $P(s)$ gives $P(C_2) = \sqrt{M}L$, which is again the event horizon and moreover corresponds to a minimum of $P(s)$.

There is further interesting behaviour that is worth commenting on. Since we have the analytic solution, we can study the difference in angle between the $s \to -\infty$ and $s \to +\infty$ limits of the solution. We find that, 
\begin{equation}
\Delta \Phi \equiv \Phi(+\infty) - \Phi(-\infty) =  \frac{1}{\sqrt{M}} \log \frac{1 - \sqrt{\alpha}}{1+\sqrt{\alpha}} \, , \quad \mbox{where} \; \;  \alpha \equiv \frac{C_1}{M} \, .
\end{equation}
Interestingly, as $\alpha$ ranges between $0$ and $1$, this function ranges between $0$ and $\infty$. The implication of this is that as $\alpha \to 1$, the MOTSodesics can wrap the horizon multiple times. The number of times can be computed exactly, and is simply given by the condition $\Delta \Phi = 2 \pi n$, where $n$ is the number of wrappings. The result of this computation is
\begin{equation}
    \alpha_n = M \tanh^2 \left(n \pi \sqrt{M}  \right) \, . \label{eq:alphan}
\end{equation}
We see that the number of wrappings tends toward infinity as $\alpha \to 1$. Strictly, the end point of this is the horizon itself.

\begin{figure}
    \centering
    \includegraphics{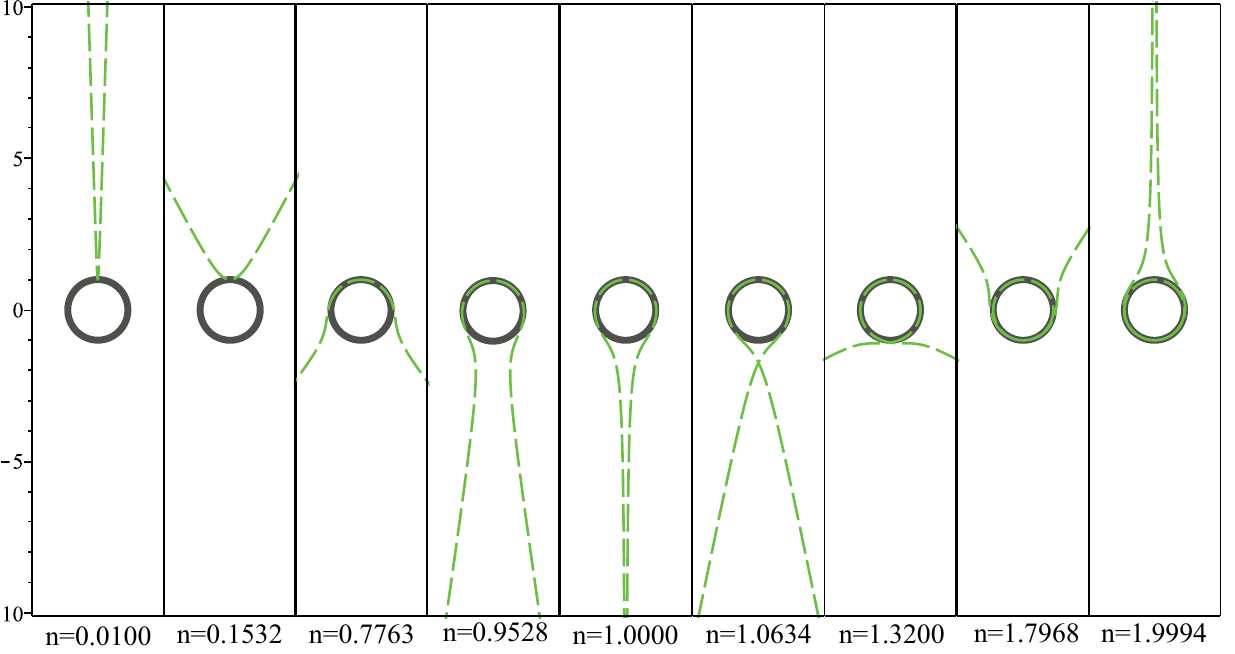}
    \caption{MOTOS wrapping around the horizon for a non-rotating BTZ black hole in static coordinates and with $L=M=1$. The $n$ values should be substituted into  (\ref{eq:alphan})  to find $\alpha$ (and so $C_1$). In these figures $C_2=0$. The grey circle indicates the horizon and 
    the green dashed lines are the MOTSodesics.} 
    \label{fig:BTZ_Wrapping}
\end{figure}

If one imagines adjusting the value of $\alpha$ ``dynamically'', then the creation of each successive wrapping of the horizon occurs as follows: as the value $\alpha_n$ is approached from below, the `endpoints' of the MOTSodesic near infinity gradually approach one another. For $\alpha = \alpha_n$, the endpoints of the MOTSodesic `meet at infinity'. As one further increases $\alpha$, the meeting point moves inward, resulting in a (non-smooth) self-intersection that moves toward the horizon as $\alpha$ increases. Of course, the total number of self-intersections is equal to $n$ --- the number of complete wrappings of the horizon. The development of the first two wrappings is shown in Figure \ref{fig:BTZ_Wrapping}.

It is interesting to compare this behaviour with that observed for Schwarzschild and Schwarzschild-anti de Sitter \cite{Booth:2020qhb,Hubeny:2013gta}. In those cases the near-horizon MOTSodesics did not wrap all the way around the outer horizon, but instead folded back and forth between the north and south pole, as the $\mathcal{K}_N$ term
repelled them from  the $z$ axis. This highlights a qualitative difference between $(2+1)$ and higher dimensions. BTZ solutions do have a rotational 
symmetry, however the MOTSodesics that we have identified are in the symmetric plane. By contrast in $(3+1)$ dimensions, the rotational symmetry (that gave rise to $\mathcal{K}_N$) 
was perpendicular to the $(r, \theta)$ half-plane. Hence with reduction in dimension, the geometry is qualitatively different: there is no extra source of curvature that enables
MOTSodesic turning. However, for higher dimensions we would expect (and will see) behaviour much closer to that observed in $(3+1)$.

Further noteworthy behaviour occurs at the particular value $C_1 = 0$. In this case, the equations determining the angular coordinate along the MOTSodesic are trivially solved by a constant solution, i.e.~$\Phi = C_3$. The radial parameter still exhibits non-trivial $s$-dependence. Therefore, in the BTZ spacetime, the radial rays --- those lines that meet the horizon perpendicularly and extend to infinity --- are all MOTSodesics. Again, this property can be understood as a consequence of searching for MOTSodesics in the symmetric plane.

\subsection{Five-dimensional Schwarzschild black hole}
\label{sec:Schwarz5d}

The Schwarzschild  family of solutions represent the unique static, asymptotically flat vacuum black hole solutions. The simplest $(3+1)$-dimensional case was previously studied using a simpler variant of the above formalism.  We refer the reader to~\cite{Booth:2020qhb,Booth:2021sow}.  In the following we will discuss the five-dimensional case, which is not only 
interesting in its own right but will also be a useful limiting case to 
rotating solutions considered in later sections.

\subsubsection{ $(4+1)$ dimensional Schwarzschild in Painlev\'e-Gullstrand coordinates} 
Consider the well-known five-dimensional Schwarzschild solution, written in 
Painlev\'e-Gullstrand coordinates $(t,r, \theta, \psi, \phi)$,
\begin{eqnarray}
\dd s^2 = - \left( 1- \frac{\mu^2}{r^2} \right) \dd t^2 + \frac{2\mu}{r} \dd r \dd t+  \dd r^2 +  r^2 \dd s^2_{S^3},
\end{eqnarray} where $t \in \mathbb{R}, r > 0$ and the metric on the unit $S^3$ is given by
\begin{equation} 
\dd s^2_{S^3} =  \left(\dd\psi + \frac{\cos\theta}{2} \dd\phi \right)^2  + \frac{1}{4} (\dd\theta^2 + \sin^2\theta \dd\phi^2)
\end{equation} where $\theta \in (0,\pi), \psi \in (0,2\pi),  \phi \in (0, 2\pi)$ are Euler angle coordinates on $S^3$ (such coordinates will be natural when we study MOTSs in the Myers-Perry black holes with equal angular momenta). The parameter $\mu > 0$ is proportional to the ADM mass. This coordinate system is `horizon-penetrating' and hypersurfaces of constant $t$ are intrinsically flat.  This can be seen be rewriting the metric in the ADM form
\begin{eqnarray}\label{sch:Doran}
\dd s^2 &= -\dd t^2 + \left[\dd r + \frac{\sqrt{2\mu }\dd t}{r} \right]^2  + r^2 \dd s^2_{S^3} 
\end{eqnarray} with the event horizon located at $r_+ = \sqrt{2\mu}$.  It is straightforward to read off
\begin{eqnarray}
\uh_{ab} \dd x^a \dd x^b & =\dd r^2  + \frac{r^2\dd \theta^2}{4} \label{uh5DSchwarz}  \\
\uh_{AB}\dd \phi^A \dd \phi^B  &= r^2 \left( \dd\psi + \frac{\cos\theta}{2} \dd \phi \right)^2   + \frac{r^2 \sin^2\theta}{4} \dd \phi^2. \nonumber
\end{eqnarray} 
Note that in contrast to four-dimensional Schwarzschild,  $\tilde{\Sigma}$ is not a Euclidean half-plane. $\uh_{ab}$ is
Ricci flat and $\tilde{\Sigma}$ is certainly a coordinate half-plane, however geometrically it is 
a Euclidean quarter-plane. The easiest way to see this is to 
calculate the circumference of a semi-circle  in $\tilde{\Sigma}$: it will have length $\frac{1}{2} \pi r$ instead of $\pi r$.

\begin{figure}
    \centering
    \includegraphics{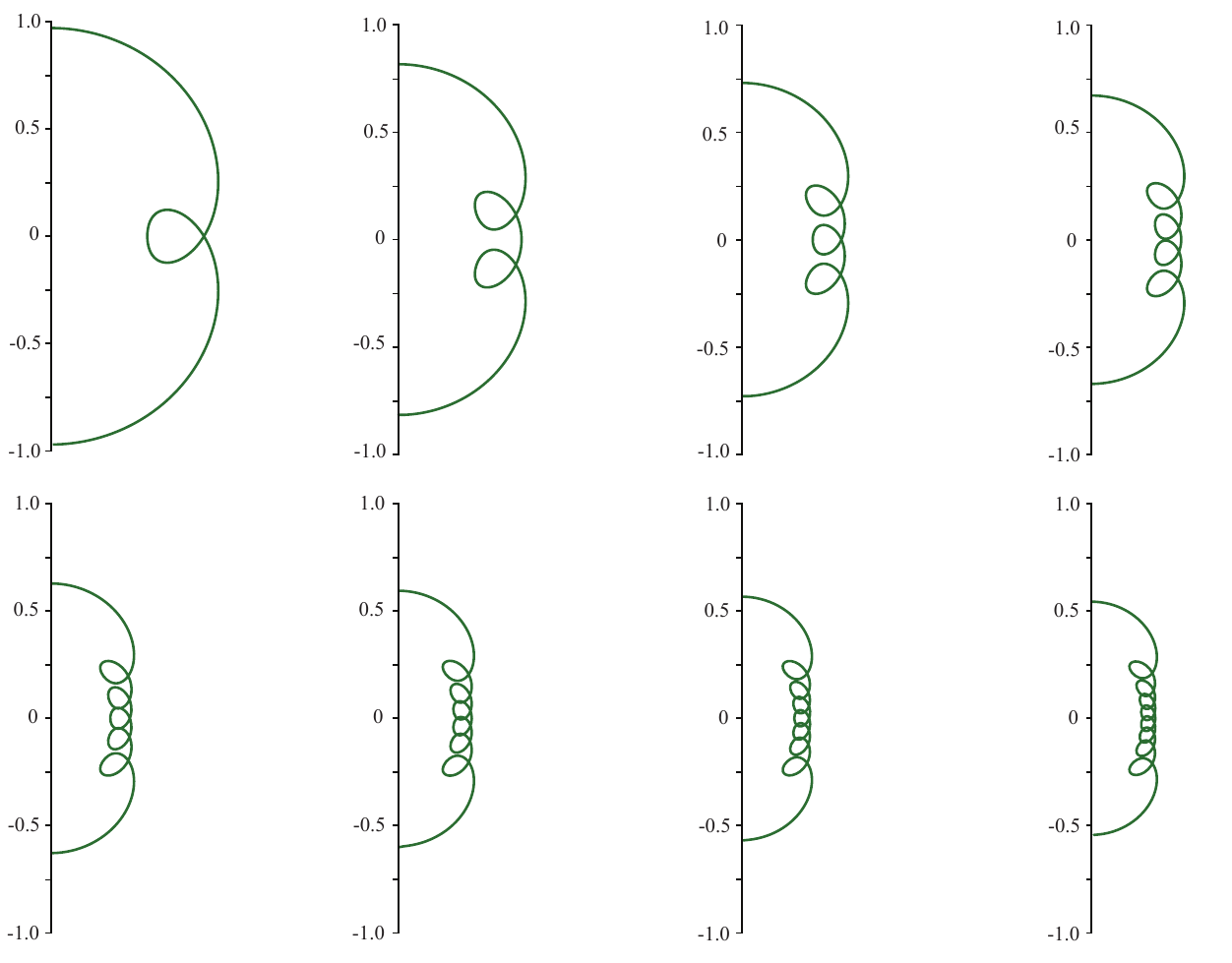}
    \caption{The first eight looping MOTSosdesics in the Painlev\'e-Gullstrand slices of $(4\!+\!
    1)$-dimensional Schwarzschild spacetime with $\mu =1$ so that the event horizon is at $r = \sqrt{2}$. They are plotted via the mapping $x = P \sin \Theta$ and $y=P \cos \Theta$ but keep in mind that as per the discussion following (\ref{uh5DSchwarz}), $\tilde{\Sigma}$ is geometrically a Euclidean quarter-plane (rather than half-plane).
    Hence, one should take care if trying to understand the geometry 
    from the exact shape of the curves. }
    \label{fig:MOTS5DSchwarz}
\end{figure}

The inverse metrics are equally simple, in particular
\begin{eqnarray}
\uh^{AB} \pdv{\phi^A} \pdv{\phi^B} & = \frac{4}{r^2\sin^2\theta} \left( \pdv{\phi} - \frac{\cos\theta}{2} \pdv{\psi}\right)^2 + \frac{1}{r^2} \left(\pdv{\psi}\right)^2.
\end{eqnarray} 
The non-vanishing components of the extrinsic curvature are
\begin{eqnarray}
\uK_{ab} &= \left( {\begin{array}{rr}  \frac{\sqrt{2\mu} }{r^2} & 0 \\ 0 & -\frac{\sqrt{2\mu}}{4} \\ \end{array} } \right) .
\end{eqnarray} 

Now we consider an arclength parameterized curve in $\tilde{\Sigma}$
\begin{eqnarray}
  r &= P(s) \\
  \theta &.= \Theta(s) 
\end{eqnarray}
so that the tangent vector is 
$\hat{T} = \dot{P} \partial_r + \dot\Theta \partial_\theta$, the unit speed condition is 
\begin{equation}
    4\dot{P}^2 + P^2 \dot\Theta^2 = 4
\end{equation} 
and the associated normal is $\hat{N} = (P/2) \dot\Theta \partial_r - 2 \dot{P} \partial_\theta$.  Simple calculations give
\begin{eqnarray}
  \mathcal{K}  &= -\frac{2\sqrt{2\mu}}{P^2}, \\
   \mathcal{K}_{\hat{N}}  &= \dot\Theta - \frac{2 \cot \Theta \dot{P}}{P} , \\
    \mathcal{K}_{\hat{T} \hat{T}}  & = \frac{\sqrt{\mu}}{2\sqrt{2}}\left(-\dot\Theta^2 + \frac{4 \dot{P}^2}{P^2}\right) .
 \end{eqnarray} 
 Then  the MOTSodesic curve equations are
\begin{eqnarray} 
\ddot P &= \frac{P \dot\Theta^2}{4}  + \frac{r \kappa_{\mbox{\tiny{MOTS}}} \dot \Theta}{2} \quad \mbox{and} \\
\ddot \Theta & = -\frac{2 \dot{P} \dot{\Theta}}{P}  - \frac{2 \kappa_{\mbox{\tiny{MOTS}}} \dot P}{P}
\end{eqnarray} where 
\begin{equation}
\kappa_{\mbox{\tiny{MOTS}}} = -\frac{2\sqrt{2\mu}}{P^2} + \dot\Theta - \frac{\sqrt{\mu}}{2\sqrt{2}} \dot \Theta^2 - \frac{2 \cot \Theta \dot P}{P} + \frac{\sqrt{2\mu} \dot P^2}{P^2}. 
\end{equation}  Note that the event horizon arises as the curve $(P(s),\Theta(s)) = \left(\sqrt{2\mu}, \sqrt{\frac{2}{\mu}} s \right)$. 

These equations can easily be solved using, for example \cite{mathematica}. As in previous papers \cite{Booth:2020qhb,Booth:2021sow,Hennigar:2021ogw}, we search for MOTSs by 
solving for MOTSodesics that launch perpendicularly to the symmetry axis: this is a necessary condition for MOTSs that close on that 
axis. The only complication is that the equations are not well-defined at $\theta=0$ (or $\pi$). However this is essentially just a coordinate
problem and no different from that encountered when solving any differential equation in spherical coordinates where the coordinates aren't
well-defined at the poles (and so neither are the equations). As in those earlier papers, we can overcome the problem by finding a series solution off the axis for $P$ and $\Theta$ in $s$ and then using that to fix off-axis boundary conditions for the curves. For more details
see \ref{sec:app}.

Figure \ref{fig:MOTS5DSchwarz} shows the first eight closed and looped MOTSodesics. The equations were solved numerically and, as in \cite{Booth:2021sow}, a 
shooting method was used to identify the closed curves. The main message
to take from this figure is that self-intersecting MOTSs may also be found
in $D=5$ Schwarzschild and they are qualitatively the same as in $D=4$. As noted, the geometry of $\tilde{\Sigma}$ is 
different from in $D\!=\!4$ and so the slightly different
appearance of these curves is not significant one way or the other: a more careful study would need to 
be made of the geometry to draw any definite conclusions.

Note too that these curve equations are covariant in the sense that if one used a different set of angles to parameterize the torus directions, e.g. the familiar Hopf coordinates $(\theta, \phi_1, \phi_2)$ on $S^3$, for which 
\begin{equation}
\dd s^2_{S^3} = \frac{\dd \theta^2}{4} + \sin^2 \frac{\theta}{2} \dd \phi_1^2 + \cos^2\frac{\theta}{2} \dd \phi_2^2
\end{equation} with the identifications $\phi_i \sim \phi + 2\pi$,  one will arrive at the same MOTSodesic equations. This demonstrates the general property that the MOTSodesic equations don't depend on the 
parameterization of $\mathcal{H}$ and so neither do the MOTSs. Calculations explicitly demonstrating these results can be found in \cite{SarahThesis}.

\section{MOTSs in rotating spacetimes of arbitrary dimension}
\label{RotBH}

In this section we consider more general $(\Sigma, h_{ij}, D_i, K_{ij})$ which continue to satisfy the symmetry assumptions of 
Section \ref{Background} but do not satisfy the warped-product assumptions of Section \ref{Warp}.  Namely, the generators of the torus action are not orthogonal to $\tilde\Sigma$. The metric then includes off-diagonal
elements $\uh_{Ab}:= h_{ij} \xi^i_A \xi^j_b = h_{Ab}$:
\begin{eqnarray}
h_{ij} \dd y^i \dd y^j = \underline{h}_{ab} \dd x^a \dd x^b +  \underline{h}_{Aa}( \dd \phi^A \dd x^a \!+\! \dd x^a\dd \phi^A )  + \uh_{AB} \dd \phi^A \dd \phi^B  \label{linel}
\end{eqnarray}
or equivalently
\begin{eqnarray}
h_{ij} = \uh_{ab} \Phi^a_i \Phi^b_j + \uh_{Aa} \left(\Phi^A_i \Phi^a_j + \Phi^A_j \Phi^a_i \right) + \uh_{AB} \Phi^A_i \Phi^B_j \; . 
\end{eqnarray}
And, of course, the inverse metric also includes off-diagonals $\oh^{aB} := h^{ij} \Phi^a_i \Phi^B_j$: 
\begin{eqnarray}
h^{ij} \pdv{y^i}   \pdv{y^j} & = \oh^{ab} \pdv{x^a} \pdv{x^b} +  \oh^{AB} \pdv{\phi^A}  \pdv{\phi^B}  \label{invh}\\
& \phantom{=} + \oh^{aB} \left(\pdv{x^a}   \pdv{\phi^B}  + \pdv{\phi^B}  \pdv{x^a}  \right)  \nonumber 
\end{eqnarray}
or equivalently
\begin{eqnarray}
h^{ij} = \oh^{ab}  \xi_a^i \xi_b^j   + \oh^{aB} \left( \xi_a^i \xi_B^j +  \xi_B^i \xi_a^j  \right) + \oh^{AB} \xi_A^i \xi_B^j \; . 
\end{eqnarray}

The inclusion of these extra components has several implications. These include:
\begin{enumerate}
\item The sets of coordinate basis vectors $\{\xi_a\}$  and $\{\xi_A\}$  are no longer mutually orthogonal:
$$ h_{ij} \xi^i_a \xi^j_B =: \underline{h}_{aB} \neq 0 \, ,    $$
for at least some $a$ and $B$. 
Similarly the sets of coordinate one-forms $\Phi^a$ and $\Phi^A$ are no longer orthogonal to each other:
$$h^{ij} \Phi_i^A \phi_j^B =: \overline{h}^{aB} \neq 0 \, ,  $$ \label{prop_ortho}
for at least some $a$ and $B$. \label{haB}
\item The inverse property (\ref{inverseBD}) no longer holds:
$$\overline{h}^{ab} \neq \left(  \underline{h}_{ab} \right)^{-1} \; \; \mbox{and} \; \; \overline{h}^{AB} \neq \left(  \underline{h}_{AB} \right)^{-1} \; .  $$
That is, the components of the inverse of the pull-backs of the metric to $\tilde{\Sigma}$ and $\mathcal{H}$ are not the same as the corresponding
components of the full inverse metric.  \label{prop_inv}
\item The curve and surface normals are no longer equal:
$$ \hat{N}^i \neq \hat{n}^i \; .  $$
In component terms, the push-forward of $\hat{N}^i = \hat{N}^a \xi_a^i$ still only has $x^a$ components but on raising the index on the surface 
normal $\hat{n}_i$, having $h^{Ab} \neq 0$ means there are also 
$\phi^A$ components:
$$\hat{n}^i = h^{ij} \hat{n}_j = h^{ib} \hat{n}_b = [h^{ab} \hat{n}_b, h^{Ab} \hat{n}_b] \; . $$ \label{prop_normal}
\end{enumerate}
Each of these properties were used in deriving the MOTSodesic formalism of the 
previous section. In particular, the expression for the extrinsic curvature of 
$S$ in $\Sigma$,
\begin{equation}
k_n = q^{ij}  D_i \hat{n}_j = (\hat{T}^i \hat{T}^j + \oh^{ij} ) D_i \hat{N}_j,
\end{equation}
becomes much more complicated if  $\oh^{AB} \neq \uh^{AB}$ and $\hat{n}_i \neq \hat{N}_i$. Most of this section is devoted to working through these issues, 
deriving expressions for $\oh^{ij}$ and $\hat{n}_i$ and then working out the 
associated expansion of the extrinsic curvature.

% \subsection{Quantities characterizing the Killing vectors and 
% $\twoS \times \mathcal{H}$ splitting of $\Sigma$}
 
\subsection{Geometry of the coordinate system}

A careful accounting of the geometry associated with the various coordinate surfaces and 
vector fields will be necessary to deal with these complications. In this subsection we focus on several relevant quantities. 

First consider $\twoS$. Then the metric components of (\ref{linel}) can be 
understood as a set of fields in the various tensor-bundles of $\twoS$. They are
\begin{enumerate}
    \item $\uh_{ab}$: the induced two-metric on $\twoS$,
    \item $\frac{1}{2} (n\!-\!1)(n\!-\!2)$ scalar fields $\uh_{AB}$: these are the 
    dot-products of the Killing vector fields with each other as evaluated over $\twoS$. In particular the $\uh_{AA}$ are the squares of the lengths of these vectors and \label{hAB}
    \item $(n\!-\!2)$ one-form fields $\uh_{Ab}$: these record the dot-products of the 
    Killing vector fields with $T \twoS$.
\end{enumerate}
For these fields, the capital latin indices should be thought of a labels while the lower case indices are 
tensor indices. The number of $h_{AB}$ takes symmetries into account. 
The full covariant derivative $D_i$ induces the $\uh_{ab}$-compatible covariant derivative $\uD_a$ on $\twoS$. Component-wise this this is defined by the Christoffel symbols  (\ref{C2}). 

The extrinsic geometry of the surfaces of constant $\phi^A$ is also important. These
have (non-unit) normal vectors $\Phi_a$ and so their extrinsic geometry is 
characterized by the derivatives:
\begin{equation}
\chi^A_{ij} := D_i \Phi^A_j %=-  \Gamma^k_{ij} \Phi_k^A 
=  -  \Gamma^A_{ij}\label{Gamma1}  \; . 
\end{equation}
Given that these are Christoffel symbols of the full spacetime, they are very easy to calculate.
For a given $A$, if one chooses the non-$A$ components for $i,j$, 
then $\chi^A_{ij}$ is the extrinsic curvature of those surfaces in $\Sigma$. As for the 
metric, on $\twoS$ we can break these quantities up into a two-index tensor,  $(n\!-\!2)$ one-form fields and $\frac{1}{2} (n-2)(n-3)$ scalar fields:
\begin{eqnarray}
\underline{\chi}^C_{ab} : =  \chi^C_{ij} \xi_a^i \xi_b^j \; , \; \;  \underline{\chi}^C_{Ab} : = \chi^C_{ij} \xi_A^i \xi_b^j \; \mbox{and} \; \;   \underline{\chi}^C_{AB} : = \chi^C_{ij} \xi_A^i \xi_B^j \; . \label{eq:chis}
\end{eqnarray}

% we can decompose these into an intrinsic two-index tensor to $\tilde{\Sigma}$, $(n\!-\!2)$ one-form fields and either
% $\frac{1}{2} (n-2)(n-3)$ (for $\underline{\chi}^C_{AB}$) or $\frac{1}{2}(n-3)(n-4)$ (for $\underline{\omega}^C_{AB}$) scalar fields. We have

We will need one more quantity over $\twoS$. For each $A$ we define
a scalar field
\begin{equation}
    \Omega_A := \epsilon^{ab} \xi_a^i \xi_b^j D_{i} \xi_{Aj} \label{eq:Omega}
\end{equation}
where 
\begin{equation}
    \epsilon^{ab} = \frac{1}{\sqrt{\det (\uh_{cd})}} 
    \left[
      \begin{array}{rr}
          0 & 1 \\ -1 & 0 
      \end{array}
    \right] 
    \label{eq:LCup}
\end{equation}
is the indexed-raised version of the Levi-Civita tensor
defined in (\ref{eq:LCdn}). 

For the special case $n=2$, all of (\ref{eq:chis}) and (\ref{eq:Omega}) vanish: 
there are no symmetry directions and hence no $\xi_A^i$ or $\Phi^A_i$! However
for higher dimensions they will generally be non-zero. 

There are other components of the geometry besides these quantities. However these are the ones that we will need 
in future sections.

\subsection{Hybrid coordinate bases}

For computational purposes, we can partially recover some of the convenience of properties
(\ref{prop_ortho})-(\ref{prop_normal}) by changing the basis that we use to expand our tensors. 
In particular, in this section we mix the coordinate vectors and one-forms to construct a (non-coordinate) 
basis with which tensors can be broken up into components parallel and perpendicular to either $\twoS$ or the $\mathcal{H}_x$. This will allow us to recover block-diagonal forms of the metric. 

In standard coordinate-based calculations the $\{ \xi_a, \xi_A \}$ are used as a basis for vectors and other `upstairs' tensors, while the  $\{ \Phi^a, \Phi^A \}$ are used as a basis for one-forms and other `downstairs' 
tensors (including the metric). Then calculating components after `raising' and `lowering' indices really
corresponds to a transformation between these bases. 
Each are complete bases in their own right. 
For example,  we don't have to use the $\{\xi_a, \xi_A\}$ basis
for ``upstairs'' tensors. We can perfectly well write
\begin{eqnarray}
h^{ij} & = \uh_{ab} \Phi^{ai} \Phi^{bj} + \uh_{Aa} \left(\Phi^{Ai} \Phi^{aj} + \Phi^{Aj} \Phi^{ai} \right) + \uh_{AB} \Phi^{Ai} \Phi^{Bj} \;. 
\end{eqnarray}
With respect to this basis, the inverse metric has the same components as the usual metric written in 
respect to the coordinate one-forms; this is 
possible as the coordinate one-forms have now been raised into vectors\footnote{While this is essentially a 
trivial statement, it is not something that we usually think about. Most introductions to Riemannian geometry 
employ coordinate bases in the standard way and so that use in deeply ingrained. 
Hence using them in this non-standard way may initially be confusing.}. 

% The components will, of course, be different from (\ref{invh}) but this is still a valid expansion of the metric. %Similarly in general
%\begin{eqnarray}
%\iuh^{ab} \neq \oh^{ab}  \; \; \mbox{and} \; \; \iuh^{AB} \neq \oh^{ab} \; . 
%\end{eqnarray}
%The components of the inverse induced metrics are usually not the same as the components of inverse of the full metric (with respect to a 
%different basis!). 

We can also use hybrid coordinate bases  $\{ \xi_a^i, \Phi^A_i\}$ or $\{ \Phi^a_i, \xi_A^i\}$. These are useful as they are well adapted to the geometry that we are studying. 
For example, when focused on $\twoS$, 
the $\xi_a^i \in T \twoS$ while the $\Phi^A_i$ are normal to those surfaces. 
Conversely, if we are focused on the $\mathcal{H}$ then the $\xi_A^i \in T\mathcal{H}$ and the  $\Phi^a_i$ are perpendicular to those tangent vectors.  

First consider expanding a vector with respect to a mixed bases
\begin{eqnarray}
z^i = Z^a \xi_a^i + Z_A \Phi^{Ai} 
\end{eqnarray}
for some $Z^a$ and $Z_A$.  Then contracting with $\xi_a^i$ and $\Phi_j^B$ we can easily solve for
\begin{eqnarray}
Z^a & = \underline{h}^{ab} (z^j \xi_{bj} ) = \uh^{ab} \underline{z}_b \; \; \mbox{and} \\
Z_A  & = \ioh_{AB} (z^j \Phi_j^B) = \ioh_{AB} \overline{z}^B\; .  \label{dv1}
\end{eqnarray}
Alternatively, relative to $\{ \Phi^{ai}, \xi_A^i \}$,
\begin{eqnarray}
z^i = \tilde{Z}_a \Phi^{ai} + \tilde{Z}^A \xi_A^i,
\end{eqnarray}
where
\begin{eqnarray}
\tilde{Z}_a &=  \ioh_{ab} (z^j \Phi^{b}_j ) =\ioh_{ab} \overline{z}^b\; \; \mbox{and} \\
 \tilde{Z}^A  &= \uh^{AB} (z^j \xi_{jB}) = \uh^{AB} \underline{z}_B \; . 
\end{eqnarray}

Note that we have used different notations for the inverses of $\uh_{ab}$ and $\uh_{AB}$ versus
those of $\oh^{ab}$ and $\oh^{AB}$. $\uh_{ab}$ and $\uh_{AB}$ are metrics written in the standard way 
and we continue with standard notation and indicate their inverses simply by raising indices:
$ \uh^{ab} := (\uh^{-1})^{ab} $ and $\uh^{AB} := (\uh^{-1})^{AB} $.
These are the metrics used to raise and lower indices. However for $\oh^{ab}$ and $\oh^{AB}$ we continue
to write the inverses as  $(\oh^{-1})_{ab}$ and $(\oh^{-1})_{AB}$. This avoids a notational
ambiguity in which, for example, 
$\oh_{AB}$ could be interpreted as either the inverse of 
$\ioh_{AB}$ or $\oh^{CD} \uh_{AC} \uh_{BD}$. It turns out that it is only the
inverse that shows up in our calculations, but
to avoid any confusion we will always  
explicitly write $\ioh_{AB}$. 

The metric can also be expanded with respect to the hybrid bases. Then
\begin{eqnarray}
h_{ij} = X_{ab} \Phi^a_i \Phi^b_j + Y_a^B (\Phi^a_i \xi_{Bj} + \Phi^a_j \xi_{Bi}) + Z^{AB} \xi_{Ai} \xi_{Bj} 
\end{eqnarray}
for some $X_{ab}$, $Y_a^B$ and $Z^{AB}$. We
contract with $\Phi^{ci} \Phi^{dj}$, $\Phi^{ci} \xi_D^j$ and $\xi_C^i \xi_D^j$ to solve for the components and find
\begin{equation}
 h_{ij}= \ioh_{ab} \Phi^a_i \Phi^b_j + \uh^{AB} \xi_{Ai} \xi_{Bj}  \; .   \label{BlkDg}
\end{equation}
Alternatively if we expand
\begin{eqnarray}
h_{ij} = \tilde{X}^{ab} \xi_{ai} \xi_{bj} + \tilde{Y}^a_B (\xi_{ai} \Phi^B_j + \xi_{aj} \Phi^Bi) + \tilde{Z}_{AB} \Phi^A_i \Phi^B_j
\end{eqnarray}
and contract with $\xi_C^i \xi_D^j$,  $\xi_C^i \Phi^{Dj}$ and $\Phi^{Ci} \Phi^{Dj}$ we find
\begin{equation}
 h_{ij} =\uh^{ab} \xi_{ai} \xi_{bj} + \ioh_{AB} \Phi^A_i \Phi^B_j \;   \label{BlkDg2}
\end{equation}
with $\ioh_{AB} = \mbox{matrix inverse of} \;  \oh^{AB}$. 
Hence we recover a foliation-adapted, block-diagonal-type form for the metric. Using
this will simplify future calculations.

\subsection{Metrics and inverses relative to the curve dyad}

Let us now consider how the curve dyad $\{ \hat{T}^a, \hat{N}^a \} \in T \twoS$ interacts with this more 
complicated geometry. The initial set-up is unchanged  from Section \ref{curves1} and in the usual 
way we can expand
\begin{equation}
    \uh_{ab} = \hat{T}_a \hat{T}_b +  \hat{N}_a \hat{N}_b.
\end{equation}
However we now also have other quantities to consider. Ultimately we want equations for $\gamma$ with explicit $\hat{T}$ dependencies and 
so expand $T\twoS$-quantities in terms of $\hat{T}$ and $\hat{N}$. 

First, the $(n\!-\!2)$ metric cross-term one-forms $\uh_{Aa}$ can be expanded in terms of the orthonormal curve dyad as
\begin{equation}
\underline{h}_{Ab} = \ubeta_A \hat{T}_b + \uV_A \hat{N}_b \label{eq:hAb}
\end{equation}
for
\begin{eqnarray}
\ubeta_A := \uh_{Aa} \hat{T}^a = \xi^i_A h_{ij} \hat{T}^j \; \; \mbox{and} \; \uV_A := \uh_{Aa} \hat{N}^a =  \xi_A^i h_{ij} \hat{N}^j \; .  \label{betaV}
\end{eqnarray}
As noted earlier, $h_{Ab}$ describes how $\twoS$ intersects the $\mathcal{H}_x$. This expansion is 
then a convenient way to study these intersections along $\gamma$ (and so ultimately on $S$). 

In Section \ref{ArbDim} with $\uh_{Ab}= 0$, $\xi_A^i \parallel \Phi^{Ai}$. 
This is no longer the case and we can show that $\ubeta_A$ and $\uV_A$ parameterize the failure of 
these vectors to be parallel. Expanding $\xi_A^i$ in terms of the $\{ \xi_a^i, \Phi_A^i \}$ basis and 
applying
(\ref{dv1}), (\ref{betaV}) and $\xi^i_A \Phi_i^C = \delta^C_A$ it is straightforward to show that
\begin{eqnarray}
\xi_A^i 
%& = \ubeta_A \hat{T}^i + \uV_A \hat{N}^i + \ioh_{AB} \Phi^{Bi}   \; . %\label{xi1}
%\nonumber \\
= \left( \ubeta_A \hat{T}^a + \uV_A \hat{N}^a \right) \xi^i_a+ \ioh_{AB} \Phi^{Bi}   \; . \label{xi2}
\end{eqnarray}
Clearly if $\ubeta_A = \uV_A =0$ then $\xi_A^i$ and $\Phi^{Ai}$ return to being parallel. 

For non-zero $\uh_{Ab}$, we also have  $\uh_{AB} \neq \ioh_{AB}$ and  $\oh^{AB} \neq \uh^{AB}$. The 
$\ubeta_A$ and $\uV_{A}$ enter here as well. First, from (\ref{BlkDg}) we write the full metric along 
$\gamma$  as 
\begin{eqnarray}
h_{ij}  = & (\hat{T}^a \hat{T}^b + \hat{N}^a \hat{N}^b) \xi_{ai} \xi_{bj} + \ioh_{AB} \Phi^A_i \Phi^B_j \; . 
%+ \uh_{AB} \Phi^A_i \Phi^B_j \\
%& + \ubeta_A (\hat{T}_i \Phi^A_j + \hat{T}_j \Phi^A_i) + \uV_A (\hat{N}_i \Phi^A_j + \hat{N}_j \Phi^A_i) \, ,    \nonumber
\end{eqnarray}
Then contracting both sides with $\xi^i_A$ and $\xi^i_B$ and applying (\ref{betaV}), it follows that
\begin{eqnarray}
\underline{h}_{AB} = h_{ij} \xi_A^i \xi_B^j = \ubeta_A \ubeta_B + \uV_A \uV_B + \ioh_{AB} 
\end{eqnarray}
so that
\begin{eqnarray}
 \ioh_{AB} = \underline{h}_{AB} -  \ubeta_A \ubeta_B - \uV_A \uV_B \; . \label{ioh}
\end{eqnarray}
We can then invert this matrix to obtain $\oh^{AB}$. Making the ansatz
\begin{equation}
\oh^{AB} = \uh^{AB}  + X \ubeta^A \ubeta^B + Y \uV^A \uV^B + Z (\ubeta^A \uV^B + \uV^A \ubeta^B)
\end{equation}
for some functions $X$, $Y$ and $Z$, we can solve $ \ioh_{AB} \oh^{BC} = \delta_A^C$
for $X$, $Y$ and $Z$ to find that:
\begin{eqnarray}
&\oh^{AB} \label{oh}  \\ =&  \uh^{AB} 
 + \frac{(1-\uV^{\, 2}) \ubeta^A \ubeta^B + (1-\ubeta^2) \uV^A \uV^B + \ubeta \cdot \uV (\ubeta^A \uV^B + \uV^A \ubeta^B)}{(1-\ubeta^2)(1-\uV^{\, 2})- (\ubeta \cdot \uV)^2} \, , \nonumber
\end{eqnarray}
where
\begin{equation} \ubeta^2 := \uh^{AB} \ubeta_A \ubeta_B \; , \; \; \uV^{\, 2} = \uh^{AB} \uV_A \uV_B \; \mbox{and} \; \; \ubeta \cdot \uV = \uh^{AB} \ubeta_A \uV_B \; . 
\end{equation}
Though explicit, this is a fairly complicated expression and so we 
make a couple of consistency checks. First if $h_{Aa} = 0$ then $V_A = \beta_A = 0$ and 
so we recover $\uh^{AB} = \oh^{AB}$ as in (\ref{inverseBD}). 
Second, under rotations of the $\hat{T}$ and $\hat{N}$ (and  the associated redefinition of $\beta_A$ and $V_A$),
the relations  (\ref{xi2}), (\ref{ioh}) and (\ref{oh}) are each invariant in form. This is as it should be: they relate quantities that are independent of $\gamma$
and so they should be independent of the particular choice of the dyad. 

\begin{figure}
\centering
\includegraphics[scale=1]{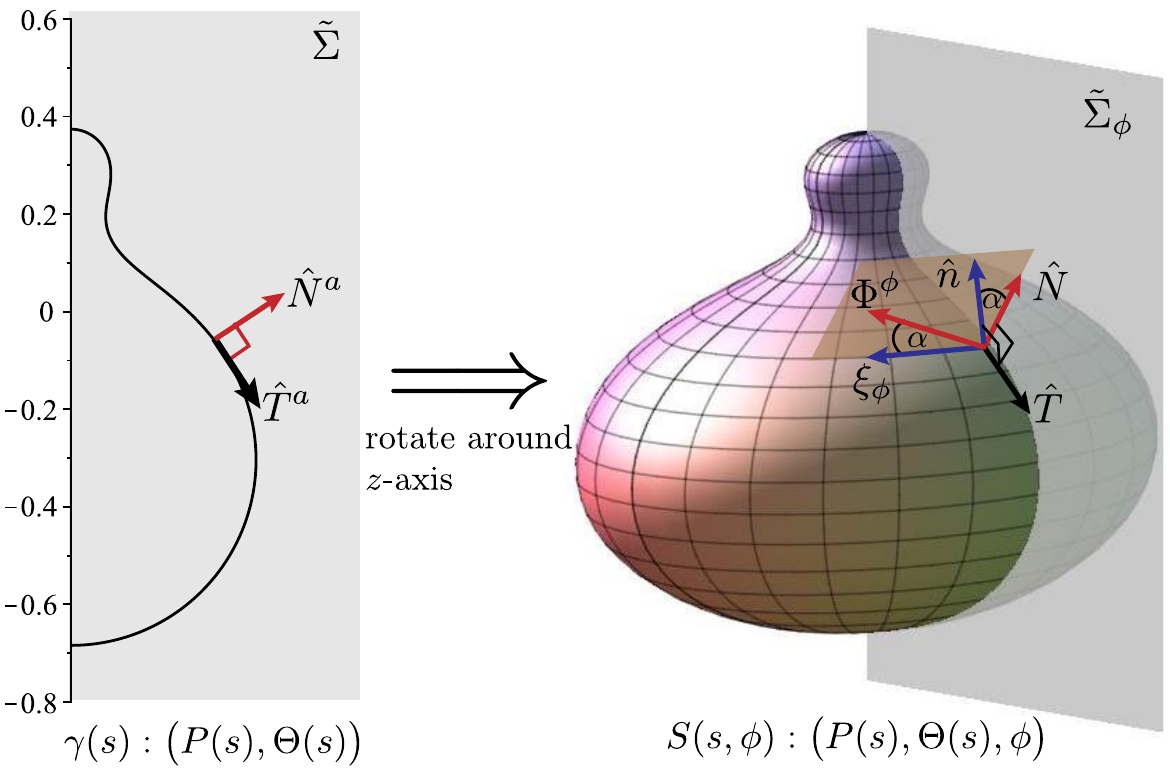}
\caption{The rotation of a curve $\gamma$ into a surface $S$ and the resulting two surface-associated triads. $\{ \hat{T}, \hat{N}, \Phi^\phi \}$
are tied to the curve with $\hat{T}, \hat{N} \in T \tilde{\Sigma}_{(\phi)}$ and $\Phi^a$ orthogonal to that surface while 
$\{ \hat{T}, \hat{n}, \xi_\phi \}$ are tied to the surface with $\hat{T}, \xi_\phi \in TS$ and $\hat{n}$ normal to the surface.  }.
\label{fig:nonorthocoords}
\end{figure}

\subsection{Metric and normal to $S$}

Next we return $S$ and consider how non-zero $\uh_{Ab}$ affects its unit normal $\hat{n}_i$ and 
inverse induced metric $q^{ij}$. This is in preparation for calculating
\begin{equation}
    k_n = q^{ij} D_i \hat{n}_j \; . 
\end{equation}
For the MOTSodesic calculation, we need all of the pieces expressed in terms of quantities defined only 
as tensors over $S$ (and then $\twoS$). 

First, $S$ is foliated 
by leaves of constant $s$ (the arclength parameter along the curve $\gamma$) and so the leaves have 
unit normal
\begin{equation}
\hat{\zeta} = \frac{\bd s}{|| \bd s ||} \; .  
\end{equation}
As in Section \ref{sec:2.4}, the black-board font is used as a reminder that this one-form lives in $T^\star S$,
not $T^\star \Sigma$. $\{ \hat{\zeta}^\alpha, \xi_A^\alpha \}$ span $TS$ with $\hat{\zeta} \cdot \xi_A = 0$
and so we can expand
\begin{equation}
\left( \pdv{s} \right)^{\alpha}= \hat{T}^\alpha = \nu \hat{\zeta}^\alpha + \ubeta^A \xi_A^\alpha \label{TX}
\end{equation}
where 
\begin{equation}
\ubeta^A = \uh^{AB} \ubeta_B =  (\uh^{AB} \uh_{Bb}) \hat{T}^b \label{betaup} 
\end{equation}
%is the projection of $\hat{T}^\alpha$ into $T \twoS$
for $\ubeta_B$ as defined in (\ref{betaV}) and the scaling constant $\nu$ defined by
\begin{equation}
\nu^2   := 1 -   \ubeta^2 \; .  \label{nudef}
\end{equation}

The $\nu$ and $\ubeta^A$ are
the analogues of lapse and shift for the $(s, \phi^A)$ coordinate system on $S$. Thus we can write the induced metric $q_{\alpha \beta}$ in lapse-shift form as
\begin{eqnarray}
& q_{\alpha \beta} \bd \vartheta^\alpha \bd \vartheta^\beta \nonumber \\
 = & (\hat{T} \cdot \hat{T}) \bd s^2 + (\hat{T} \cdot \xi_A) (\bd s \, \bd \phi^A + \bd \phi^A \, \bd s) + (\xi_A \cdot \xi_B) \bd \phi^A \, \bd \phi^B  \nonumber \\
 = &  \bd s^2 + \ubeta_A (\bd s \, \bd \phi^A + \bd \phi^A \, \bd s) + \uh_{AB} \bd \phi^A \, \bd \phi^B 
%
%= \tilde{\Phi}^s_\alpha \tilde{\Phi}^s_\beta + \beta_A \left(\tilde{\Phi}^s_\alpha \tilde{\Phi}^A_\beta + \tilde{\Phi}^A_\alpha \tilde{\Phi}^s_\beta  \right) + \uh^{(x)}_{AB} \tilde{\Phi}^A_\alpha \tilde{\Phi}^B_\alpha
\end{eqnarray}
and the inverse metric is
\begin{eqnarray}
q^{\alpha \beta} &= \hat{\zeta}^\alpha \hat{\zeta}^\beta + \uh^{AB} \xi_A^\alpha \xi_B^\beta \nonumber \\
& = \frac{1}{\nu^2} \Big(T^\alpha - \ubeta^A \xi_A^\alpha \Big) \Big( T^\beta - \ubeta^B \xi_B^\beta \Big) +  \uh^{AB} \xi_A^\alpha \xi_B^\beta \nonumber \\
& =  \frac{1}{\nu^2} \hat{T}^\alpha \hat{T}^\beta - \frac{\ubeta^A}{\nu^2} \left(\xi^\alpha_A \hat{T}^\beta + \hat{T}^\alpha \xi^\beta_A \right) +   \ueta^{AB} \xi^\alpha_A \xi^\beta_B \;  \label{qbasic}
\end{eqnarray}
where
\begin{equation}
\ueta^{AB} =  \uh^{AB} + \frac{1}{\nu^2}  \ubeta^A \ubeta^B \label{uq} \; . 
\end{equation}
%and one should continue to keep in mind that $\uh^{AB} \neq h^{AB}$. 
This pushes-forward from $S$ into $\Sigma$ to give $q^{ij}$ in the desired form:
\begin{equation}
q^{ij} = \frac{1}{\nu^2} \hat{T}^i \hat{T}^j - \frac{\ubeta^A}{\nu^2} \left(\xi^i_A \hat{T}^j + \hat{T}^i \xi^j_A \right) + \ueta^{AB} \xi^i_A \xi^j_B \; . \label{qij}
\end{equation}

%Making use of (\ref{thorn}) we could also have directly written
%\begin{equation}
%q^{ij} = \frac{1}{\nu^2} \tthorn^i_a \tthorn^j_b \hat{T}^a \hat{T}^b + \uh^{AB} \xi_A^i \xi_B^j \; ,  \label{qup}
%\end{equation}
%since 
%\begin{equation}
%\hat{\zeta}^i = \frac{1}{\nu} \tthorn^i_a \hat{T}^a \; . 
%\end{equation}

Next, we construct the unit normal to $S$ from the relations:
\begin{equation}
T^i \hat n_i = \xi^i_A \hat n_i = 0 \; . \label{xin}
\end{equation} 
%Hence $q^{ij} \hat{n}_j = 0$.
% Given that the $\xi_A$ are symmetries of both $S$ and $\{ \Sigma, h_{ij}, K_{ij} \}$ they are also symmetries of this normal. That is
%\begin{equation}
%\mathcal{L}_{\xi_A} \hat{n}_i = 0 \;  \Longrightarrow \;  \xi_A^j D_j \hat{n}_i + \hat{n}_j D_i \xi^j_A = 0 \; . \label{Lien}
%\end{equation}
%From a coordinate perspective, this is simply saying that $\hat{n}_i$ does not depend on the symmetry coordinates $\phi^A$. 
From these relations we can find
\begin{equation}
\hat{n}_i = \cos \! \alpha  \left(\hat{N}_i - V_i \right) \, ,  \label{hatn}
\end{equation}
where
\begin{equation}
\cos \alpha =h_{ij}  \hat{N}^i \hat{n}^j \; \mbox{and} \; \; V_i  = \uV_A \Phi^A_i %\; \; \mbox{with} \; \: V_A =- \hat{N}^i h_{ij} \xi^j_A  = - (\vec{h}_{Aa} \overline{\epsilon}^a_{\phantom{a} b})\hat{T}^b\label{Vup}
\end{equation}
for $\uV_A$ defined in (\ref{betaV})
%Considering $\hat{N}_i = \hat{N}^a \xi_{ai}$, this is a $\{ \xi_{ai}, \Phi^A_i\}$ expansion. 
and $\alpha$ the angle formed between $\hat{n}^i$ (the normal to $S$) and $\hat{N}^i$ (the normal to $\gamma$ 
in $\twoS$ pushed-forward into $\Sigma$) as shown in Figure \ref{fig:nonorthocoords}.
Since $\hat{N}$ and $\hat{n}$ are both unit length, we can dot $\hat{n}$ with itself and so find:
\begin{eqnarray}
 \cos^2 \! \alpha & = \frac{1}{1 + \oh^{AB} \uV_A \uV_B}   \label{VcA}  
  = \frac{(1 - \ubeta^2)(1-\uV^{\, 2}) - (\ubeta \cdot \uV\,)^2 }{1 - \ubeta^2} 
\end{eqnarray}
where for the second equality we have applied (\ref{oh}). Note that if $h_{Ab} = 0$, then $V_A = \beta_A = 0$ 
and so $\alpha = 0$ and the two normals coincide.

%Note that there is a potential confusion involving $\beta^A$ and $V_A$ as, despite the common index label, they naturally define objects in
% different spaces. Specifically:
%\begin{equation}
%\beta^i = \beta^A \xi_A^i  \in T\; \; \mbox{and} \; V_i = V_A \Phi^A_i \, , 
%\end{equation}
%
%*****************

\subsection{Extrinsic curvature of $S$ in $\Sigma$} 
We are now ready to calculate
\begin{eqnarray}
k_n  = q^{ij} D_i \hat{n}_j  \; . 
\end{eqnarray}
By  (\ref{qij}) and (\ref{hatn}), this can be expanded as
\begin{eqnarray}\label{kn}
k_n  = \cos \alpha \left(\frac{1}{\nu^2} \hat{T}^i \hat{T}^j - \frac{2\ubeta^A}{\nu^2} \hat{T}^i \xi_A^j + \ueta^{AB} \xi_A^i \xi_B^j \right) D_i (\hat{N}_j - V_j) 
\end{eqnarray}
where we have used  the fact that $q^{ij} n_j = 0$ to both extract the $\cos \alpha$ from inside the derivative and combine 
the $\hat{T}^j \xi_A^i$ and $\hat{T}^i \xi_A^j$ terms. 

We analyze this in pieces, starting with the $\hat{N}_j$ term in (\ref{kn}):
\begin{eqnarray}
& \left(\frac{1}{\nu^2} \hat{T}^i \hat{T}^j - \frac{2\ubeta^A}{\nu^2} \hat{T}^i \xi_A^j + \ueta^{AB} \xi_A^i \xi_B^j \right) D_i \hat{N}_j \label{big1}\\
= &  - \frac{\kappa}{\nu^2}  - \frac{2\ubeta^A}{\nu^2} \hat{T}^i D_i \uV_{\!A}  + \frac{2\ubeta^A}{\nu^2} \hat{T}^i \hat{N}^j D_i \xi_{Aj}
+\ueta^{AB} \xi_A^i D_i \ubeta_B  \nonumber \\ 
& \phantom{- \frac{\kappa}{\nu^2}  - \frac{2\ubeta^A}{\nu^2} \hat{T}^i D_i \uV_{\!A}  + \frac{2\ubeta^A}{\nu^2} \hat{T}^i \hat{N}^j D_i \xi_{Aj}} - \ueta^{AB} \xi_A^i \hat{N}^j D_i \xi_{Bj}   \nonumber \\
= &   - \frac{\kappa}{\nu^2}  - \frac{2\ubeta^A}{\nu^2} \hat{T}^i D_i \uV_{\!A} - \frac{\ubeta^A}{\nu^2} \epsilon^{ij}  D_i \xi_{Aj}     + \ueta^{AB} \xi_A^i \hat{N}^j D_j \xi_{Bi}   \nonumber \\
= &   - \frac{\kappa}{\nu^2}  - \frac{2\ubeta^A}{\nu^2} \hat{T}^i D_i \uV_{\!A} -   \frac{\ubeta^A }{\nu^2} \underline{\Omega}_A   
+ \frac{1}{2} \ueta^{AB} \hat{N}^i D_i \uh_{AB}   \nonumber
\end{eqnarray}
where $\kappa$ retains its definition (\ref{kappa}).
%\begin{equation}
%\kappa = \hat{N}_j \hat{T}^i D_i \hat{T}^j = \hat{N}_b \hat{T}^a \uD_a \hat{T}^b \label{kappa}
%\end{equation}
In the second line we have used $\hat{T}^i \hat{N}_i = 0$, $\xi_A^i \hat{T}_i = \ubeta_A$, $\xi^i_A N_i = \uV_{\! A}$ and the Leibniz rule to extract
$\hat{N}_j$ from inside the derivative. In the third line we have used that the $\xi^i_A$ are Killing vectors to anti-commute $D_i \xi_{Aj} = - D_j \xi_{Ai}$ and eliminate $\xi_A^i D_i \ubeta_B$. 
In the last line we again used the Leibniz rule, along with the definition of $\uh_{AB}$.
$\epsilon^{ij}$ and $\underline{\Omega}_A$ were defined in (\ref{eq:LCdn}) and  (\ref{eq:Omega}).
In the last line we  again used the Leibniz rule, along with the definition of $\uh_{AB}$. 
Note that $\underline{\Omega}_A$ is independent of $\gamma$ and $\hat{T}^a$. It depends only on the geometry of the space and its Killing vectors.

Next, examining the $V_j$ term in (\ref{kn}) and expanding $V_j = \uV_C \Phi^C_j$
\begin{eqnarray}
&  -\left(\frac{1}{\nu^2} \hat{T}^i \hat{T}^j - \frac{2\ubeta^A}{\nu^2} \hat{T}^i \xi_A^j + \ueta^{AB} \xi_A^i \xi_B^j \right) D_i (\uV_C \Phi^C_j) \label{big2}  \\
= &  -\left(\frac{1}{\nu^2} \hat{T}^i \hat{T}^j - \frac{2\ubeta^A}{\nu^2} \hat{T}^i \xi_A^j + \ueta^{AB} \xi_A^i \xi_B^j \right) ( \uV_C D_i  \Phi^C_j + \Phi^C_j D_i \uV_C) \nonumber  \\
= &  - \uV_{\!C} q^{ij} D_i \Phi^C_j + \frac{2\ubeta^C}{\nu^2} \hat{T}^i  D_i \uV_{\!C} - \ueta^{AC}  \xi_A^i D_i \uV_{\!C}    \nonumber  \\
= &   - V_C \underline{\chi}^C  + \frac{2\ubeta^C}{\nu^2} \hat{T}^i  D_i \uV_{\!C}   \nonumber  
\end{eqnarray}
where in the second line we used Leibniz to expand out the $\uV_{\!C} \Phi^C_i$. In the third we recombined the terms into $q^{ij}$ for the 
first term and applied $\hat{T}^i \Phi_i^A = 0$ and $\xi^i_A \Phi^C_i = \delta^A_C$. In the fourth line we used
the fact that the $\xi_A^i$ are Killing vectors to eliminate $\xi^i_{B} D_i \uV_{\!C}$. We have also defined a traced extrinsic
curvature quantity:
\begin{equation}
 \slashed{\chi}^C= q^{ij} \chi^C_{ij} = q^{ij} D_i \Phi^C_j \; . 
\end{equation}
This quantity does depend on $\gamma$ (through the $\hat{T}$s that help to determine $q^{ij}$). In Section \ref{MOTSodesics} we will examine the dependence more carefully. 

Combining (\ref{big1}) and (\ref{big2}) we  obtain
\begin{eqnarray}
\sec\!\alpha \, k_n =  - \frac{\kappa}{\nu^2} -   \frac{\ubeta^A }{\nu^2} \underline{\Omega}_A   
+ \frac{1}{2} \ueta^{AB} \hat{N}^i D_i \uh_{AB}  - \uV_{\! A} \slashed{\chi}^A  \, , 
\end{eqnarray}
which can be solved for $\kappa$ as:
\begin{equation}
\kappa =  -   \ubeta^A  \underline{\Omega}_A  + \nu^2 \left(   \frac{1}{2} \ueta^{AB} \hat{N}^i D_i \uh_{AB}  -   \sec \!\alpha \, k_n   -  \uV_{\! A}  \slashed{\chi}^A \right)  \; .  \label{kappaXX}
\end{equation}

%Previous papers \cite{} have studied non-rotating spacetime in which $\ubeta^A = \uV_{\!A} = 0$ and so $\nu = 1$ and $\alpha = 0$. In those cases it 
%then also follows that  $\ueta^{AB} = \uh^{AB}$ and so 
%\begin{equation}
%\kappa^{\mbox{\tiny{NR}}} = - k_n + \frac{1}{2} \uh^{AB} \hat{N}^i D_i \uh_{AB} = -k_n +  N^a \uD_a \ln (\sqrt{ \uh } )\, , 
%\label{kappaXX0} 
%\end{equation}
%where the non-indexed $\uh = \det \left( \uh_{AB} \right) $ and we have applied the Jacobi formula for the derivative of a determinant:
%$\partial_a \ln \uh =  \, \uh^{AB} \partial_a \uh_{AB}$. 

\subsection{MOTSs and MOTSodesics}
\label{MOTSodesics}
We now have all the pieces necessary to derive the MOTSodesic 
equations. 

% As noted earlier, 
% $S$ is a marginally outer trapped surface if 
% \begin{equation}
% \theta_{\ell} = 0  \; \; \Longrightarrow \; \;  k_{u} + k_{n} = 0 \label{Meq}
% \end{equation}
% where 
% \begin{equation}
% k_{u} = q^{ij} K_{ij}  \; . 
% \end{equation}
% In this expression we assume that $\hat{n}$ is pointing in the direction of vanishing expansion.  

As noted in (\ref{eq:Meq}), $k_n = -k_u$ on a MOTS, and so for
a MOTSodesic we can replace the $k_n$ in (\ref{kappaXX}) with $-k_u$:
%\begin{equation}
%\kappa_{\mbox{\tiny{MOTS}}} =  \nu^2  \sec \!\alpha \, k_u  -   2\ubeta^A  \underline{\Omega}_A  + \frac{1}{2}\nu^2 \ueta^{AB} \hat{N}^i D_i \uh_{AB}  + \nu^2 \uV_{\! A} \underline{\chi}^A \; .
%\end{equation}
\begin{equation}
\kappa_{\mbox{\tiny{MOTS}}} =-   \ubeta^A  \underline{\Omega}_A +  \nu^2\left(\frac{1}{2} \ueta^{AB} \hat{N}^a \uD_a \uh_{AB}  +   \sec \!\alpha \, k_u    -   \uV_{\! A} \slashed{\chi}^A \right)  \; . \label{kappaMOTS}
\end{equation}

For some calculations this may be the most convenient form for $\kappa_{\mbox{\tiny{MOTS}}}$,  
however in other circumstances it is useful to explicitly display the dependence of each of the terms on  $\gamma$. 
This manifests as a dependence on $\hat{T}^a$ and 
$\hat{N}^a$ (which is also really just $\hat{T}^a$ dependence
since 
$\hat{N}^a = \underline{\epsilon}^a_{\phantom{a} b} \hat{T}^b$
from (\ref{hN})).

For the first term of (\ref{kappaMOTS}) this is straightforward. Applying (\ref{betaup}) we obtain
\begin{equation}
-  \ubeta^A \underline{\Omega}_A = (-\underline{\Omega}_A \uh^{AB} \underline{h}_{Ba} ) \hat{T}^a \; . 
\end{equation}

The second term requires more work. Applying (\ref{betaup}), (\ref{nudef})
 and  (\ref{uq}) we have
 \begin{eqnarray}
 \nu^2 \ueta^{AB} & = \uh^{AB} - (\uh^{CD} \ubeta_C \ubeta_D) \uh^{AB}  + \ubeta^A \ubeta^B \label{uqup} \\
 & = \uh^{AB} + \Xi^{AB}_{ab} \hat{T}^a  \hat{T}^b \nonumber 
% & =   \uh^{AB} - \underline{\epsilon}^{AD} \underline{\epsilon}^{BC} \underline{h}_{Ca} \underline{h}_{Db} \hat{T}^a  \hat{T}^b \nonumber \; . 
 \end{eqnarray}
 where 
 \begin{equation}
\Xi^{AB}_{ab} := (  \uh^{AC} \uh^{BD}  - \uh^{AB} \uh^{CD} )  \underline{h}_{Ca} \underline{h}_{Db}  \; . 
\end{equation}
Hence
%T version
%\begin{eqnarray}
%& \frac{1}{2} \nu^2 \underline{q}^{AB} \hat{N}^i D_i \uh_{AB} \\
% =  & \hat{N}^a \uD_a \ln \sqrt{\uh} + \frac{1}{2} \left( \ubeta^A \ubeta^B - (\uh^{CD} \ubeta_C \ubeta_D ) \uh^{AB} \right) 
%\hat{N}^a \uD_a \uh_{AB} \nonumber \\
% = &  \left(\overline{\epsilon}^b_{\phantom{b} a} \uD_b \ln \sqrt{\uh} \right) \hat{T}^a  \\
% &  + \frac{1}{2}  \left( (  \uh^{AC} \uh^{BD}  - \uh^{AB} \uh^{CD} )  \underline{h}_{Ca} \underline{h}_{Db} 
% \left(\overline{\epsilon}^d_{\phantom{b} c} \uD_d \uh_{AB}  \right)  \right)  \hat{T}^a \hat{T}^b \hat{T}^c 
%\end{eqnarray}
\begin{eqnarray}
& \frac{1}{2} \nu^2 \ueta^{AB} \hat{N}^i D_i \uh_{AB} \\
 =  & ( \uD_a \ln \sqrt{\uh} ) \hat{N}^a+ \frac{1}{2} \Xi^{AB}_{ab}
 \uD_c \uh_{AB}     \hat{T}^a \hat{T}^b \hat{N}^c \nonumber 
\end{eqnarray}
where for the first part we have reused (\ref{LNh}).

Next applying (\ref{qij}) along with (\ref{uqup}), the third term of (\ref{kappaMOTS}) becomes
%\begin{eqnarray}
%\nu^2 k_u & = \nu^2 q^{ij} K_{ij}  \\
%& = \tilde{K}_{ab} \hat{T}^a \hat{T}^b + \nu^2 \underline{h}^{AB} \underline{K}_{AB} \nonumber \\
%& = \left(\tilde{K}_{ab} - tr\right) \hat{T}^a \hat{T}^b+ \underline{\tr}\underline{K} \, ,  \nonumber 
%\end{eqnarray}
%
\begin{eqnarray}
\nu^2 k_u & = \nu^2 q^{ij} K_{ij}  \\
& = \left( \hat{T}^i \hat{T}^j - \ubeta^A \left(\xi^i_A \hat{T}^j + \hat{T}^i \xi^j_A \right) + \nu^2 \ueta^{AB} \xi^i_A \xi^j_B \right) K_{ij} \nonumber \\
%& = {K}_{ij} (\hat{T}^i - \beta^i) (\hat{T}^j - \beta^j) + (1- \uh_{kl} \beta^k\beta^l)  \underline{\tr}\underline{K} \nonumber \\
%& = \left(\tilde{K}_{ab} -  \underline{\tr}\underline{K} \,  \underline{h}^{AB} \underline{h}_{Aa} \underline{h}_{Bb}  \right) \hat{T}^a \hat{T}^b+ \underline{\tr}\underline{K} \, ,  \nonumber \\
& = \left( \underline{K}_{ab} - 2 \underline{h}^{AB} \underline{h}_{Aa} \underline{K}_{Bb}  + \Xi^{AB}_{ab} \underline{K}_{AB} \right) 
\hat{T}^a \hat{T}^b + \underline{\tr}\underline{K} \nonumber \;,
\end{eqnarray}
where
$\underline{\tr}\underline{K} =  \uh^{AB}  \underline{K}_{AB}$. % and between the third and fourth lines we have expanded 
%out $\tilde{K}_{ab} = \tthorn_a^i \tthorn_b^j K_{ij}$ with (\ref{thorn}). 

Similarly for the fourth term in (\ref{kappaMOTS})
% T version
%\begin{eqnarray}
%\nu^2 \uV_C \underline{\chi}^C = &  - \left(\underline{h}_{Cc} \overline{\epsilon}^c_{\phantom{c} d} \hat{T}^d \right) \\
%& \times \bigg( \Big(\tilde{\chi}_{ab}^C + (\underline{\chi}^{CAB} - \utrchi^C \, \uh^{AB}) \underline{h}_{Aa} \underline{h}_{Bb}  \Big) \hat{T}^a \hat{T}^b+ \utrchi^C \bigg) \nonumber
%\end{eqnarray}
\begin{eqnarray}
 -\nu^2  \slashed{\chi}^C 
%= &  - \underline{h}_{Cc}  \hat{N}^c 
% \bigg( \Big(\tilde{\chi}_{ab}^C + \Xi^{AB}_{ab} \chi^C_{AB} \Big) \hat{T}^a \hat{T}^b+ \utrchi^C \bigg) \nonumber \\
  =  - \left(\underline{\chi}^C_{ab} - 2 \underline{h}^{AB} \underline{h}_{Aa} \underline{\chi}^C_{Bb}  + \Xi^{AB}_{ab} \underline{\chi}^C_{AB}  \right) \hat{T}^a \hat{T}^b- \utrchi^C 
  \end{eqnarray}
where
\begin{equation}
%\tilde{\chi}_{ab}^C = \tthorn_a^i \tthorn_b^j \chi^C_{ij} \;, \; \;  \underline{\chi}_{AB}^C = \xi_A^i \xi_B^j \chi^C_{ij} \; \mbox{and} \; \; 
\utrchi^C =  \uh^{AB}  \underline{\chi}^C_{AB} \label{cT} \; . 
\end{equation} 

Finally we can put all of this together to obtain
 \begin{eqnarray}
 \kappa_{\mbox{\tiny{MOTS}}} = \sec \! \alpha\, \left( \mathcal{K} +  \mathcal{K}_{\hat{T} \hat{T} } \right) +  \mathcal{K}_{\hat{T}} + \mathcal{K}_{\hat{N}}   + \mathcal{K}_{\hat{T} \hat{T} \hat{N}} \label{eq:kappaMOTS}
 \, , 
 \end{eqnarray}
 for 
 \begin{eqnarray}
 \mathcal{K}  &=    \underline{\tr} \underline{K}\\
 \mathcal{K}_{\hat{T}} & = \left( -  \underline{\Omega}_A \uh^{AB} \underline{h}_{Ba} \right) \hat{T}^a\\
  \mathcal{K}_{\hat{N}} & =\left( \uD_a  (\ln \! \sqrt{\uh})   - \utrchi^C \underline{h}_{Ca}  \right)  \hat{N}^a\\
 \mathcal{K}_{\hat{T} \hat{T}}  &=  \left( \underline{K}_{ab} - 2 \underline{h}^{AB} \underline{h}_{Aa} \underline{K}_{Bb}  + \Xi^{AB}_{ab} \underline{K}_{AB}  \right) \hat{T}^a \hat{T}^b
  \\
 \mathcal{K}_{\hat{T} \hat{T} \hat{N}}  &=  \left( \frac{1}{2} \Xi_{ab}^{AB} \uD_c \uh_{AB}     
   - \left(\underline{\chi}^C_{ab} - 2 \underline{h}^{AB} \underline{h}_{Aa} \underline{\chi}^C_{Bb}  + \Xi^{AB}_{ab} \underline{\chi}^C_{AB}  \right) \underline{h}_{Cc} \right) \nonumber\\
  & \phantom{==}  \times \hat{T}^a \hat{T}^b \hat{N}^c  
 \end{eqnarray}
 and
 \begin{equation}
\sec \! \alpha = \sqrt{1+ \oh^{AB}  (\underline{h}_{Aa}  \underline{h}_{Bb})  \hat{N}^a\hat{N}^b} \; . 
\end{equation}
%which introduces some extra $\hat{T}^a$ dependence. 
With these expressions we have achieved our goal: writing
$\kappa_{\mbox{\tiny{MOTS}}}$ explicitly in terms of 
geometric quantities defined over $\twoS$ along with 
$\hat{T}$ and $\hat{N}$. As a consistency check note that if $\underline{h}_{Aa} = 0$, then this reduces to $\kappa_{\mbox{\tiny{MOTS}}}^\perp$ from (\ref{kM0}).

Then, as for the orthogonal case, the MOTSodesic equations are
\begin{equation}
\hat{T}^a \uD_a \hat{T}^b = \kappa_{\mbox{\tiny{MOTS}}} \hat{N}^b  \;  
\end{equation} 
which may be explicitly expanded into a pair of second order ordinary differential equations as:
\begin{equation}
\frac{\mathrm{d}^2 X^a}{\mathrm{d}s^2}  =-  \underline{\Gamma}_{bc}^{a} \frac{\mathrm{d} X^b}{\mathrm{d} s} \frac{\mathrm{d}X^c}{\mathrm{d}s} +  {\kappa}_{\mbox{\tiny{MOTS}}}  \underline{\epsilon}^a_{\phantom{a} b} \frac{\mathrm{d} X^b}{\mathrm{d} s} \, , 
\end{equation}
where it should be kept in mind that the $\underline{\Gamma}_{bc}^{a}$ are the two-dimensional 
Christoffel symbols associated with $\uh_{ab}$ over $\twoS$.

We can now apply this formalism to examples. 

\section{MOTSodesic equations in $(2+1)$-dimensional spacetimes}
\label{(2+1)}

\subsection{General considerations}

For a $(2+1)$-dimensional spacetime the hypersurfaces $\Sigma$ are two-dimensional. Then 
there are no $\phi^A$ coordinates or $\Phi^A$ normals. Hence neither are there any symmetry assumptions:
for any two-dimensional $\Sigma$, all MOTSs may be found by MOTSodesic methods. 

Thus, the  MOTSodesic equations are  very simple. All terms involving capital Latin indices vanish and the normals are equal ($\hat{n} = \hat{N} \Leftrightarrow  \alpha =0 $ ). Then
\begin{equation}
{\kappa}_{\mbox{\tiny{MOTS}}}^{\mbox{\tiny{2D}}}  =  {K}_{ab} \hat{T}^a \hat{T}^b 
\end{equation} 
and so the MOTSodesic equations are simply
\begin{equation}
\hat{T}^a \uD_a \hat{T}^b = ({K}_{cd} \hat{T}^c \hat{T}^d ) \hat{N}^b  \label{MOTS1} \; . 
\end{equation} 

In the special case where $\Sigma$ is extrinsically flat ($K_{ij} = 0$) the MOTSodesic equations are the geodesic equations. So in $(2+1)$ dimensions
with $K_{ij} = 0$, space-like geodesics are identical with MOTOS and 
closed space-like geodesics are identical with MOTSs. This was the situation 
for the MOTOS in non-rotating BTZ in static coordinates as found in Section \ref{sec:BTZnonrot}.

\subsection{ Rotating BTZ black hole spacetime: stationary coordinates}

We return to $(2+1)$ dimensions and BTZ black holes but this time allow for rotation \cite{Banados:1992gq,Banados:1992wn}. These take the form
\begin{equation}
 \dd s^2 = -F \dd t^2 + \frac{\dd r^2}{F} + r^2 \left(\dd\phi - \frac{J}{2r^2} \dd t \right)^2
 \end{equation} where
 \begin{equation}
 F =  \frac{r^2}{L^2}  - M + \frac{J^2}{4 r^2} 
 \end{equation}
 with $L$ the (anti-deSitter) cosmological length scale, $M$ the mass and $J$ the angular momentum. 
% \textcolor{blue}{{\bf [There is a typo in the above metric for rotating BTZ --- it should be $J^2/(4r^2)$ in the metric function. If it was just $J$: (1) it would be dimensionally inconsistent, (2) you could generate horizons just by rotating in the opposite direction. Hopefully just a typo in the document. ]}}
 
% 
% 
% 
%***********
%
%
%
%
%
%A well known solution of gravity with a negative cosmological constant is the BTZ black hole solution which satisfies the Einstein equations 
%\begin{equation}
%R_{ab} = -\frac{2}{\ell^2} g_{ab}
%\end{equation}  where $\ell$ is the AdS$_3$ radius. The BTZ metric is given in 2+1 form as 
%\begin{equation}
%g = -N^2 \dd t^2 + \rho^2(\dd \phi + N^\phi \dd t)^2 + \frac{r^2}{N^2 \rho^2} \dd r^2
%\end{equation} where $\phi$ has period $2\pi$, $t \in \mathbb{R}$, and
%\begin{equation}
%N^2 = \frac{r^2(r^2 - r_+^2)}{\ell^2 \rho^2}, \qquad N^\phi = - \frac{4 J } {\rho^2}, \qquad \rho^2 = r^2 + 4 M \ell^2 - \frac{r_+^2}{2}
%\end{equation}  and there is a nondegenerate horizon at $r = r_+ > 0$. If $M = |J|/\ell$, $r_+ =0$, and there is an extreme horizon at $ r =0$. 
%\begin{equation}
%r_+^2 = 8\ell \sqrt{ M^2 \ell^2 - J^2}
%\end{equation}  
The surfaces of constant $t$ have induced space-like metric
\begin{equation}
h_{ij} \dd x^i \dd x^j  = \frac{\dd r^2}{F} + r^2 \dd\phi^2
\end{equation} 
and associated extrinsic curvature 
\begin{equation}
K_{ij} \dd x^i \dd x^j = - \frac{J}{r \sqrt{F}}  \dd r \dd \phi \; . 
\end{equation}
For $J \neq 0$ this extrinsic curvature is non-zero and so, unlike the 
non-rotating case we considered earlier, MOTSodesics are not geodesics. 

As would be expected for these stationary coordinates and slices which end at the event horizon
\begin{equation}
r_+ =  \frac{L}{\sqrt{2}} \sqrt{ M+ \sqrt{M^2 - \left(\frac{J}{L} \right)^2 }} \, ,
\end{equation}
several components of the metric and extrinsic curvature diverge 
at that value of $r$. 

A unit parameterized curve $(r=P(s),\phi=\Phi(s))$ in a surface of constant $t$ has unit tangent vector
\begin{equation}
\hat{T} = \dot{P} \pdv{r} + \dot{\Phi} \pdv{\phi} 
\end{equation}
where the dot indicates a derivative with respect to arclength $s$. That is 
\begin{equation}
\frac{\dot{P}^2}{F} + P^2 \dot{\Phi}^2 = 1 \; .
\label{eq:Norm_BTZ_stationary}
\end{equation}
Then the outward oriented for increasing $\Phi$ unit normal is
\begin{equation}
\hat{N} = \frac{P\dot{\Phi}  }{\sqrt{F}} \pdv{r} -\frac{ \dot{P} }{P \sqrt{F}}\pdv{\phi} \, , 
\end{equation}
and the MOTSodesic equations (\ref{MOTS1}) take the relatively simple form:
\begin{eqnarray}
\ddot{P} & = \frac{F'}{2F} \dot{P}^2 + P F \dot{\Phi}^2 - J \dot{P} \dot{\Phi}^2 \label{eq:MOTSodesic_BTZ_stationary}\\
\ddot{\Phi} & = - \frac{2}{P} \dot{P} \dot{\Phi} + \frac{J}{F P^2} \dot{P}^2 \dot{\Phi}  \nonumber \; . 
\end{eqnarray}

\begin{figure}
    \centering
    \includegraphics[scale=1.0]{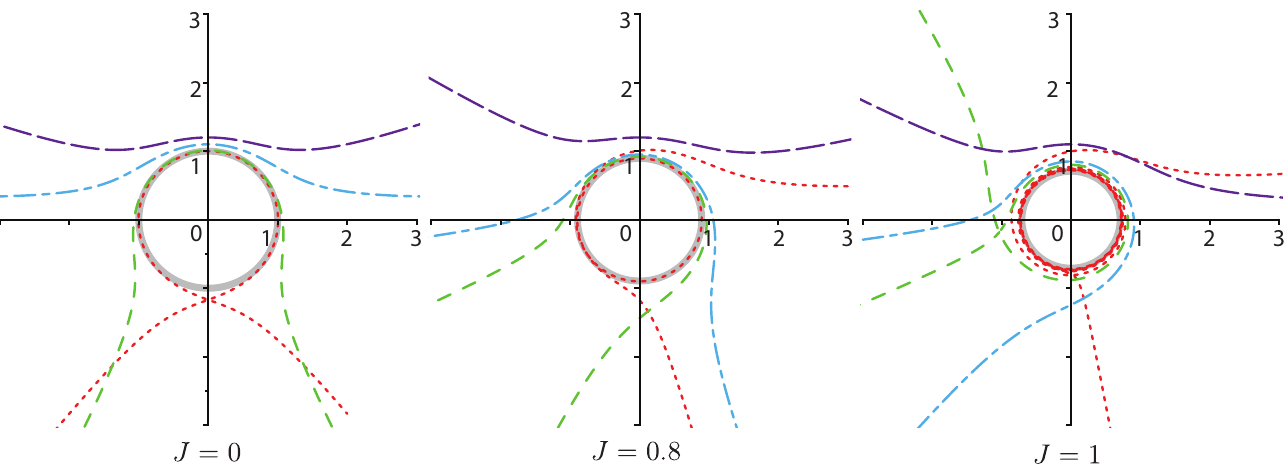}
    \caption{Representative (open) MOTSodesics for rotating BTZ and stationary time slices. For all of these figures $\ell = M = 1 $. In the expected way, there are no local maxima of $P$ and so there are no closed MOTSs in these slices (apart from the standard outer black hole horizon). }
    \label{fig:BTZ_standard}
\end{figure}
Unlike the non-rotating case of Section \ref{sec:BTZnonrot} we do not have an exact solution for these equations.
However we can still see that there are no closed MOTSosdesics. Any closed curve would be compact
and so necessarily have a maximum and minimum
value of $P$ at which $\dot{P}=0$. At any
maximum  $\ddot{P} < 0$ while at any minimum 
$\ddot{P}>0$. However from
(\ref{eq:Norm_BTZ_stationary}) and
(\ref{eq:MOTSodesic_BTZ_stationary})
and 
\begin{equation}
  \left.   \ddot{P} \right|_{\dot{P}=0} = P F \dot{\Phi}^2 = \frac{F}{P} \; . 
\end{equation}
For these time slices, $F>0$ and so no maximum is
possible. Hence there can be no closed curves and there are no
exotic MOTSs in these slices. 

A few representative numerical solutions of the equations are shown in Figure \ref{fig:BTZ_standard}. 

\subsection{BTZ black hole: Doran-like coordinates}

Of course, by experience we expect any 
exotic closed MOTS to be contained inside the 
BTZ black hole and to properly probe this 
region we will
need horizon-crossing coordinates with 
space-like time slices. 
We find them in the usual way: making a transformation from ingoing Eddington-Finkelstein-type coordinates.
For
BTZ these are
 \begin{equation}
 \dd s^2  = -F dv^2 + 2 dv dr + r^2 \left( d\phi   - \frac{J}{2 r^2} dv \right)^2
 \end{equation} 
with $F$ as in the previous section.

We can then consider transformations of the form
 \begin{equation}
 v = \tilde{t} - g(r), \qquad \psi = \phi - q(r) \; . 
 \end{equation} 
Choosing 
 \begin{equation}
 q'(r) = -\frac{g'(r) J}{2r^2} \label{qp}
 \end{equation} 
 (with the prime indicating a derivative with respect to $r$)
 to eliminate the $g_{r\psi}$ term this becomes
 \begin{equation}
 \dd s^2 = -F  \dd \tilde{t}^2 + p(r) \dd r^2 + 2\sqrt{1 - F p(r)} \dd \tilde{t} \dd r + r^2 \left(\dd \psi - \frac{J}{2r^2} \dd \tilde{t} \right)^2  \end{equation} 
 where 
 \begin{equation}
 p(r) = g'(2 - f^2 g') \; . \label{gp}
 \end{equation}
 More conveniently we can choose a $p(r)$ and view (\ref{qp}) and (\ref{gp}) as differential equations defining the 
 corresponding transformation. 
It is easy to check that any $p(r)$ works
 and so we can forget about $g(r)$ and $q(r)$ and just work with $p$. 
 Note that such transformations are closely analogous to the corresponding generalized Painlev\'e-Gullstrand coordinates  introduced in \cite{Hennigar:2021ogw}). 

Taking the unit normal to the surfaces of constant $\tilde{t}$:
\begin{equation}
u_\alpha = - \frac{1}{\sqrt{p}} \dd \tilde{t} \; .
\end{equation}
$p$ completely characterizes the acceleration of that set of observers:
\begin{equation}
a^\alpha = u^{\beta} \nabla_\beta u^\alpha = - \frac{p'}{2p^2} \pdv{r} \; \; \Longrightarrow  \; \; \| a \|^2  = \frac{(p')^2}{4 p^3} \; . 
\end{equation}
As in the non-rotating case, for constant $p$ the $u^\alpha$ are the unit tangent vectors for time-like geodesics.

Given a $p$, the induced metric on the surfaces of constant $\tilde{t}$ is
\begin{equation}
h_{ij} \dd x^i \dd x^j = p \dd r^2 + r^2 \dd \psi^2 
\end{equation}
and
\begin{equation}
K_{ij} \dd x^i \dd x^j = \frac{p^2 F' + p'}{2 \sqrt{p(1-pF)}}  -  \frac{J \sqrt{p}}{r} \dd r \dd \psi - r \sqrt{\frac{1-pF}{p}} \dd \psi^2 \; . 
\end{equation}
Then for an arclength parameterized curve $(r=P(s),\psi=\Psi(s))$ the unit tangent vector and (left-handed) 
unit normal vector are
\begin{equation}
\hat{T} = \dot{P} \pdv{r} + \dot{\Psi} \pdv{\psi}  \; \; \mbox{and} \; \;
\hat{N} = \frac{P\dot{\Psi}}{\sqrt{p}}  \pdv{r} - \frac{\sqrt{p}\dot{P}}{P}  \pdv{\psi}
\end{equation}
with the arclength condition is
\begin{equation}
    p \dot{P}^2 + P^2 \dot{\Psi}^2 = 1 \label{eq:BTZ_Doran_arc} \; . 
\end{equation}

Then the MOTSodesic equations (\ref{MOTS1}) are:
\begin{eqnarray}
\ddot{P} & = - \frac{p'}{2p} \dot{P}^2 + \frac{P}{p} \dot{\Psi}^2 + \frac{P K_{TT} }{\sqrt{p}} \dot{\Psi} \quad \mbox{and}\\
\ddot{\Psi} & = - \frac{2}{P} \dot{P} \dot{\Psi} - \frac{\sqrt{p} K_{TT} }{P} \dot{P}
\end{eqnarray}
where
\begin{equation}
K_{TT} = \frac{p^2 F' + p'}{2 \sqrt{p(1-pF)}} \dot{P}^2 - \frac{J \sqrt{p}}{P} \dot{P} \dot{\Psi} - P \sqrt{ \frac{1-pF}{p}  }\dot{\Psi}^2 \; . 
\end{equation}
 
Let us now consider a concrete choice for $p(r)$. To keep all terms well-defined for $r>0$ we want both $p>0$ and 
$1-pF>0$. That is
\begin{equation}
\left\{ \begin{array}{ll}  
p > 0 & \mathrm{if } \; F < 0  \\
0 < p < \frac{1}{F}  & \mathrm{if } \; F > 0
\end{array}\right\} \; . 
\end{equation} 
 If we define $p(r)$ by an auxiliary function $\nu(r)$ via
 \begin{equation}
 p(r) = \frac{4 L^2 r^2}{(4r^4 + J^2 L^2) + 4 L^2 \nu r^2} \;,
 \end{equation}
 then 
 \begin{equation}
1-pF = p( \nu + M ) \; . 
 \end{equation}
 For $M>0$, an obvious choice is $\nu = 0$ for which
\begin{equation}
 p = \frac{4 L^2 r^2}{ J^2 L^2 + 4 r^4 } \; . 
 \end{equation} 
One advantage of this choice is that is it symmetric under 
$J \rightarrow - J$. Then the extrinsic curvature term
\begin{equation}
   \kappa =  K_{TT} = - \frac{2JL  \dot{P} \dot{\Psi}}{\sqrt{4P^4+J^2L^2}} - P \sqrt{M}\dot{\Psi}^2 \; .
\end{equation}
  Other choices are also possible\footnote{Another possibility
 would be $\nu = \frac{J}{L}$ for which
 \begin{equation}
 p = \frac{4 L^2 r^2}{\left(2r^2 + J L \right)^2} \label{eq:p}\; . 
 \end{equation} 
 The choice turns $p$ into a perfect square which 
 simplifies some of our expressions (and is analogous to choices made in \cite{Hennigar:2021ogw}). However it has the downside
 that for $J<0$, this $p$ (and so the time-slicing) diverges at $r^2 = |JL/2|$. } 
and correspond to other slicings of spacetime, however we will use $\nu = 0$ here.

% The one-form $\dd T$ is manifestly everywhere timelike with  $g^{TT}  = -g_{rr} = -p(r) <0$. In ADM form the metric is
%  \begin{equation}
%  \dd s^2 = -\frac{1}{p_1}\dd T^2 + p_1 \left[ \dd r^2 + \sqrt{\frac{M}{p_1}} \dd T \right]^2 + r^2 \left(\dd \psi - \frac{J \dd T}{2 r^2}\right )^2
%  \end{equation}  Note that $p_1(r_+) = 1/M$.  In this form the lapse and shift  are
%  \begin{equation}
%  N = p_1^{-1/2}, \qquad  N^r = \sqrt{\frac{M}{p_1}}, \qquad N^\psi  = - \frac{J}{2r^2}.
%  \end{equation} 
 
For (\ref{eq:p}), the MOTSodesic equations are
 \begin{eqnarray}
 \ddot P  &= -\frac{p'}{2p} \dot P^2 + \frac{P}{p} \dot \Psi^2  -\sqrt{\frac{M}{p}} P^2 \dot \Psi^3 - J \dot P \dot \Psi^2 \;
 \mbox{and} \label{eq:BTZ_Doran_P}\\
 \ddot \Psi & = - \frac{2 \dot P \dot \Psi}{P} + \sqrt{M p} \dot P \dot \Psi^2 + \frac{p J \dot P^2 \dot \Psi}{P^2} \, , 
  \label{eq:BTZ_Doran_psi}
 \end{eqnarray} 
 where one should keep in mind that here $p=p(P(s))$ and $p'= p'(P(s))$.
 
 As a simple test we can check that the standard MOTS at $P(s)=r_+$
 is a solution. Then $\dot{P} = 0$ and by the arclength condition (\ref{eq:BTZ_Doran_arc}), 
 $\dot{\Psi} = \frac{1}{r_+}$. Then (\ref{eq:BTZ_Doran_psi}) 
 holds trivially and a short calculation (using 
 $p (r_+)= \frac{1}{M}$) shows that (\ref{eq:BTZ_Doran_P})
 holds as well. 
 %\robie{What is $p_1$? Should this just be %$p$?}

As usual we are mainly interested in closed MOTSodesics. However for these more complicated equations we can neither solve them 
exactly as we did for (\ref{eq:MOTSodesic_BTZ_stationary})
nor simply prove the absence of closed curves as we did for 
(\ref{eq:BTZ_Doran_P}). Instead we will plot a few 
representative curves which suggest the absence of closed 
curves and leave a definitive proof for future works.

\begin{figure}
\centering
\includegraphics[width=\hsize]{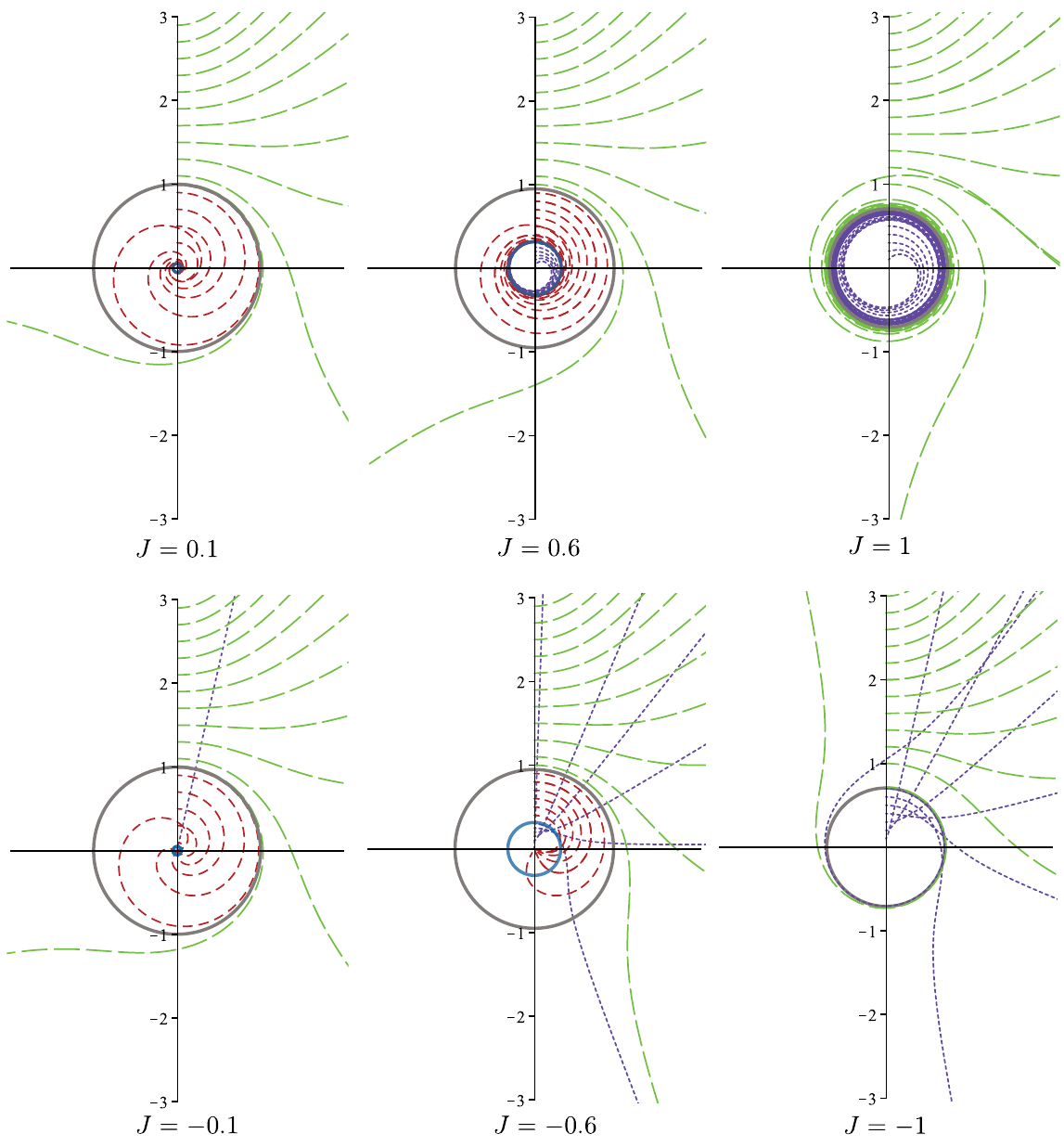}
\caption{Representative (open) MOTSodesics for rotating BTZ in 
Doran-like time slices. For all of these figures $L=M=1$ and 
$p(r)=p_1(r)$. Again there are no closed MOTSodesics apart from
the standard horizons. All curves either converge to one of the horizons, diverge to infinity or converge to the origin. Green curves start outside the outer horizon, red ones start between the inner and outer
horizons and purple ones start inside the inner horizon. For $J=0.1$ the inner horizon is barely visible
(and we plot only one curve that starts inside) while the $J=1$ the horizon is extremal and so there
is no intermediate region. 
For visual clarity, MOTSodesics starting on $y$-axis outside the black hole horizon are long-dashed and green, those starting between 
the inner and outer horizons are short-dashed and red and those 
starting inside the inner horizon are dotted and purple. The outer and inner horizons are respectively gray and blue circles.
This colour and line-type labelling is also used in later figures in this paper. }
\label{fig:BTZ_all}
\end{figure}

As for our proof in static coordinates, we note that a compact MOTSodesic
would necessarily have a maximum and minimum $P$ and hence 
two points where $\dot{P}=0$. Without loss of generality we can assume that at least one of those points lies on the positive $y$-axis 
and so it is sufficient to consider MOTSodesics launched 
perpendicularly that axis. If no such curve closes, then there are no closed MOTSodesics. 

A few sample curves are shown in \ref{fig:BTZ_all}. Note that in this case it is necessary to consider
$J>0$ and $J<0$ separately. In higher dimensions, the direction of rotation doesn't matter. Geometrically
this is because the direction of rotation is perpendicular to the symmetry axis. However in $(2+1)$
this isn't the case: the MOTSodesic can be oriented with $s$ increasing either in the direction of 
the rotation or against it and this breaks the symmetry. 

In these cases (and all other cases that we considered)
the curves either diverge to infinity, converge onto one of the
horizons or plunge into the singularity. In no case did we 
find a closed MOTS (apart from the horizons). This is consistent with the corresponding non-rotating BTZ examples
of Section \ref{sec:BTZnonrot} and again highlights the key role played by the catenoid-like turning from the axis
enabled by a non-zero $\mathcal{K}_{\hat{N}}$ term. MOT(O)S in $(2 \! + 1 \!)$ are curves, not surfaces and so cannot
make use of opposing principal curvatures to manipulate to modify the size (and sign) of $k_n$.

% Hence there are no exotic MOTS in the standard black hole solutions of $(2+1)$ gravity. While at first this might be a 
% surprise, a little further thought shows that it should have been expected. The absence of symmetry axes in $(2+1)$-gravity
% means that we cannot obtain the turnarounds from the $z$-axis that are so familiar from higher dimensional examples. 
% Those turnarounds play the same role as the necks of catenoids in standard minimal surface theory...

% \ivan{Maybe one could choose a foliation for which $k_u$ becomes very negative close to the $z$ axis and then recover
% an exotic MOTS. Thinking about this...}

\section{MOTSodesic equations in $(3+1)$-dimensional spacetimes}
\label{(3+1)}

\subsection{General considerations}
\label{sec:3p1GC}

For a four-dimensional spacetime the hypersurfaces $\Sigma$ are three-dimensional and there is a single $\phi$-coordinate. Hence the normal
space to $\twoS$ is one-dimensional and there is a single normal one-form
\begin{equation}
\Phi^\phi_i = [d \phi]_i 
\end{equation}
and single rotational Killing vector field
\begin{equation}
\xi_\phi^i = \left[\pdv{\phi} \right]^i  \, .  
\end{equation}
If $0 \leq \phi  < 2\pi$ and $\xi_\phi^i$ generates a closed orbit (the symmetry and coordinate labelling that would usually be considered) then the 
circumference of the orbit is $C = 2 \pi \sqrt{h_{\phi \phi}}$ and so it is geometrically insightful to define the circumferential radius $R := \sqrt{h_{\phi \phi}}$. Then the single
component of  $\uh_{AB}$ is
\begin{equation}
\uh_{\phi \phi} = R^2  \; \; \Longrightarrow \uh^{\phi \phi} = \frac{1}{R^2} \, ,
\end{equation}
%Then $\ueta^{AB}$ also has a single component 
%\begin{equation}
%\ueta^{\phi \phi} = \frac{1}{\nu^2 R^2}   \; \; \mbox{for} \; \;  \nu^2 = 1 - R^2 (\ubeta^\phi)^2
%\end{equation}
and since all capital latin indices are $\phi$, it is convenient to just drop some of those indices:
\begin{equation}
 \chi_{ij} := \chi^\phi_{ij}  \andd \Omega := \Omega_{\phi}  \, , 
\end{equation}
and it also follows that $\Xi^{AB}_{ab} = 0$. 
Then for the usual 
\begin{eqnarray}
 \kappa_{\mbox{\tiny{MOTS}}}^{\mbox{\tiny{3D}}} = \sec \! \alpha\, \left( \mathcal{K} +  \mathcal{K}_{\hat{T} \hat{T} } \right) +  \mathcal{K}_{\hat{T}} + \mathcal{K}_{\hat{N}} + \mathcal{K}_{\hat{T} \hat{T} \hat{N}}    \, , 
 \end{eqnarray}
 the components are
 \begin{eqnarray}
 \mathcal{K}  &=    \underline{\tr} \underline{K}\\
 \mathcal{K}_{\hat{T}} & =  - \left( \frac{\underline{\Omega} }{R^2}   \underline{h}_{\phi a} \right)  \hat{T}^a \\
  \mathcal{K}_{\hat{N}} & =\left( \uD_a \ln R   - \utrchi \underline{h}_{\phi a}  \right)  \hat{N}^a\\
  \mathcal{K}_{\hat{T} \hat{T}}  &=  \left( \underline{K}_{ab} - \frac{2}{R^2} \underline{h}_{\phi a} \underline{K}_{\phi b}  \right) \hat{T}^a \hat{T}^b
  \\
 \mathcal{K}_{\hat{T} \hat{T} \hat{N}}  &=      
   - \left(\underline{\chi}_{ab} - \frac{2}{R^2} \underline{h}_{\phi a} \underline{\chi}_{\phi b}  \right) \underline{h}_{\phi c}   \hat{T}^a \hat{T}^b \hat{N}^c  
 \end{eqnarray}
 with
 \begin{equation}
\sec \! \alpha = \sqrt{1+  \left(\overline{h}^{\phi \phi} \underline{h}_c  \hat{N}^c \right)^2}\; . 
\end{equation}

%
%and so 
%\begin{equation}
%\kappa_{\mbox{\tiny{MOTS}}} =  \nu^2  \sec \!\alpha \, k_u  -   2\ubeta^\phi  \underline{\Omega}_\phi  + N^a D_a (\ln R)  + \nu^2 \uV_{\! \phi} \underline{\chi}^\phi \; .
%\end{equation}

$\mathcal{K}_{\hat{N}}$ can be understood more easily by considering special cases. In the three-dimensional non-rotating cases discussed in previous papers\cite{Booth:2020qhb,Booth:2021sow,Hennigar:2021ogw}, $\underline{h}_{\phi a} = 0$ and so 
\begin{equation}
\kappa_{\mbox{\tiny{MOTS}}}^\perp =   \underline{\tr} \underline{K} +  N^a D_a (\ln R) +  {K}_{ab}   \hat{T}^a \hat{T}^b \; . \label{kappa3perp}
\end{equation}
Further specializing to $K_{ij} = 0$:
\begin{equation}
\kappa^{\perp_0}_{\mbox{\tiny{MOTS}}}  =  \hat{N}^a \!  D_a (\ln R)  \; . \label{kappa30}
\end{equation}
We can then refer back to Figure \ref{fig:catenoid} to interpet this term. 
In this simplest case, MOTSs are simply two-dimensional minimal surfaces in a three-dimensional Riemannian 
background. The axisymmetry means that the principal directions are $\hat{T}$ and $\xi_\phi$. 
The curves of constant $s$ are circles of circumferential radius $R$ and the generating curve $\gamma$ has curvature 
$\kappa^{\mbox{\tiny{NR}}_o}_{\mbox{\tiny{MOTS}}}$ in the $\hat{N}$ direction in order to balance off that curvature. 
These principal curvatures are in opposite directions. 
Switching to the MOTSodesic perspective, the $\mathcal{K}_{\hat{N}}$ term generates a repulsion from that 
axis (assuming $R$ increases moving away from the $\phi=0$ axis). 

In more general spacetimes, of course, things become more complicated. However $\mathcal{K}_{\hat{N}}$ continues to play a key role in 
turning curves back from the $z$-axis. 

% \ivan{I'll pick it up here tomorrow...}

% \robie{I'm guessing that someone is planning to finish this section, but I wanted to leave a comment just in case}

\subsection{Kerr spacetime: Boyer-Lindquist coordinates}

As our first example, consider Kerr in the standard Boyer-Lindquist coordinate system. Then on a surface of constant $t$:
\begin{eqnarray}
h_{ij} \dd y^i \dd y^j = &  \frac{\varrho_+^2}{\Delta} \dd r^2 + \varrho_+^2 \dd \theta^2 + \left(\frac{\rho^4- a^2 \Delta \sin^2 \! \theta}{\varrho_+^2} \right) \sin^2 \! \theta \dd \phi^2 \label{BLmetric}
\end{eqnarray}
for $\Delta = r^2 -2 m r +a^2$ and $\varrho_+^2  = r^2+ a^2 \cos^2 \!\theta$, and
\begin{eqnarray}
&  K_{ij} \dd y^i \dd y^j \\
 = &  \frac{1}{\varrho_+^3  \sqrt{\rho^4-a^2 \Delta \sin^2 \! \theta}} \nonumber \\
& \times \Bigg( \frac{am \sin^2 \! \theta \left(a^2 \cos^2 \! \theta (a^2-r^2) - r^2(3r^2+a^2) \right)}{\sqrt{\Delta}}  (\dd \phi \dd r + \dd r \dd \phi)   \nonumber \\
& \phantom{\times \Bigg( xx}  + (2a^3 m r  \sqrt{\Delta} \sin^3 \! \theta  \cos \theta) \,  (\dd \phi \dd \theta + \dd \theta \dd \phi)  \Bigg) \; .  \nonumber
\end{eqnarray}

For this metric, lower case Latin indices run over $(r, \theta)$ while capital indices are all $\phi$. For this diagonal metric things are quite 
simple and we have
%\begin{eqnarray}
%\underline{h}_{ab} \dd x^a \dd x^b  = & \frac{\varrho_+^2}{F} \dd r^2 + \varrho_+^2 \dd \theta^2  \, , \\
% \uh_{\phi a} \dd x^a  = & 0 \andd \nonumber\\
% \uh_{AB} \dd \phi^A \dd \phi^B  = & \left(\frac{(r^2+a^2)^2- a^2 F \sin^2 \! \theta}{\varrho_+^2} \right) \sin^2 \! \theta \dd \phi^2 \nonumber \; . 
%\end{eqnarray}
\begin{eqnarray}
\underline{h}_{ab}  = & \left[ \begin{array}{cc} \frac{\varrho_+^2}{\Delta} & 0 \\ 0  &  \varrho_+^2 \dd \theta^2  \end{array} \right] \, , \\
 \uh_{\phi a}  = & 0 \andd \nonumber\\
 \uh_{\phi \phi}   = & \left(\frac{\rho^4- a^2 \Delta \sin^2 \! \theta}{\varrho_+^2} \right) \sin^2 \! \theta \nonumber \; . 
\end{eqnarray}
Further, again because the metric is diagonal, we have
\begin{eqnarray}
\oh^{\phi \phi} = \frac{1}{\uh_{\phi \phi}}  \; . 
\end{eqnarray}
Hence in this case all the care that we have taken to distinguish between these quantities is unnecessary!

 With $\uh_{\phi a} = 0$,  $\kappa_{\mbox{\tiny{MOTS}}}$ is determined by (\ref{kappa3perp}). However things 
become even simpler as we also have
\begin{equation}
 \underline{\tr K} = 0  \andd K_{ab} = 0 \; . 
\end{equation}
Hence we obtain the same equation 
\begin{equation}
\kappa^{\mbox{\tiny{Kerr BL}}}_{\mbox{\tiny{MOTS}}} = \hat{N}^a D_a \ln R \label{kappaBL}
\end{equation}
  as if $K_{ij} = 0$, with
\begin{equation}
R = \sqrt{h_{\phi \phi}} = \sin \! \theta \sqrt{\frac{\rho^4- a^2 \Delta \sin^2 \! \theta}{\varrho_+^2} } \; . 
\end{equation}

Then for $\gamma$ parameterized as $(r=P(s), \theta=\Theta(s))$
\begin{equation}
\hat{T} = \dot{P} \pdv{r} + \dot{\Theta} \pdv{\theta} \andd \hat{N} = \sqrt{\Delta} \dot{\Theta} \pdv{r} - \frac{\dot{P}}{\sqrt{\Delta}} \pdv{\theta} \, ,  \label{TNBL}
\end{equation}
with the constraint that
\begin{equation}
\varrho_+^2 \left(\frac{\dot{P}^2}{\Delta} + \dot{\Theta}^2 \right) = 1 \,, 
\end{equation}
we have equations of motion
\begin{eqnarray}
 \ddot{P}  +  \frac{1}{\varrho_+^2} \Bigg(& \frac{a^2(m-P)\cos^2 \! \Theta +a^2m - mP^2}{\Delta} \dot{P}^2 \label{Eq_BL}  \\
&  -2 a^2 \sin \! \Theta \cos \! \Theta \dot{P} \dot{\Theta}   -P\Delta \dot{\Theta}^2 \Bigg) 
 =  \kappa \sqrt{\Delta} \dot{\Theta} \nonumber \\
 \ddot{\Theta}  + \frac{1}{\varrho_+^2} \Bigg(& \frac{a^2 \sin \! \Theta \cos \! \Theta}{\Delta} \dot{P}^2  
+ 2 P \dot{P} \dot{\Theta} - a^2 \sin \! \Theta \cos \! \Theta \dot{\Theta}^2 \Bigg) 
= - \frac{\kappa  \dot{P}}{\sqrt{\Delta}} \nonumber
\end{eqnarray}
with 
\begin{eqnarray}
\kappa = &  \left(- \frac{P \Delta}{\varrho_+^2} + \frac{\sqrt{\Delta} \left( 2 a^2 (P-m) \cos^2 \! \Theta + 2a^2(m+P) + 4P^3\right) }{\rho^4- a^2 \Delta \sin^2 \! \Theta } \right) \dot{\Theta} \nonumber \\
& + \left(- \frac{\cot \! \Theta}{\sqrt{\Delta}} - \frac{a^2 \sin \! \Theta \cos \! \theta}{\sqrt{\Delta} \varrho_+^2} + \frac{2a^2 \sqrt{\Delta} \cot \! \Theta}{\rho^4- a^2 \Delta  } \right) \dot{P} 
\end{eqnarray}
as defined by (\ref{kappaBL})-(\ref{TNBL}). To save space we have dropped the extra labels on $\kappa$ and it should be kept in mind that all 
$r$ and $\theta$ appearing in $\rho$,  $\varrho_+$ and $\Delta$ should be replaced with $P(s)$ and $\Theta(s)$ respectively. 

Though fairly complicated in appearance, these equations can easily be solved numerically using standard commands from, for example, 
Maple\cite{maple}. As was discussed in Section \ref{sec:Schwarz5d} the curves are launched perpendicularly from the $z$-axis at $\theta=0$ and 
a shooting method is used to search for MOTSodesics that similarly re-intersect the $z$-axis at a right angle. Further, we again have to 
deal with coordinate problems at the axis and so again use a series solution (\ref{sec:app}) to fix off-axis boundary conditions.

% The only complication is that they are not well-defined on the $z$-axis where $\theta =0$ where the $\cot \theta$ is not defined. 
% However for an axis-symmetric MOTS that intersects the $z$-axis at a right angle, we would also have $\dot{P} = 0$ there and it is not 
% hard to see that the $\theta = 0$ limit is well-defined for such curves. We can iteratively solve for
% \begin{eqnarray}
% P(s) & = P_0 +P_1 s + P_2 s^2 + P_3 s^3 + P_4 s^4 + P_5 s^5 \dots \\
% \Theta(s) & = \Theta_0 + \Theta_1 s + \Theta_2 s^2 + \Theta_3 s^3 + \Theta_4 s^4 + \Theta_5 s^5
% \end{eqnarray}

\begin{figure}
\centering
\includegraphics{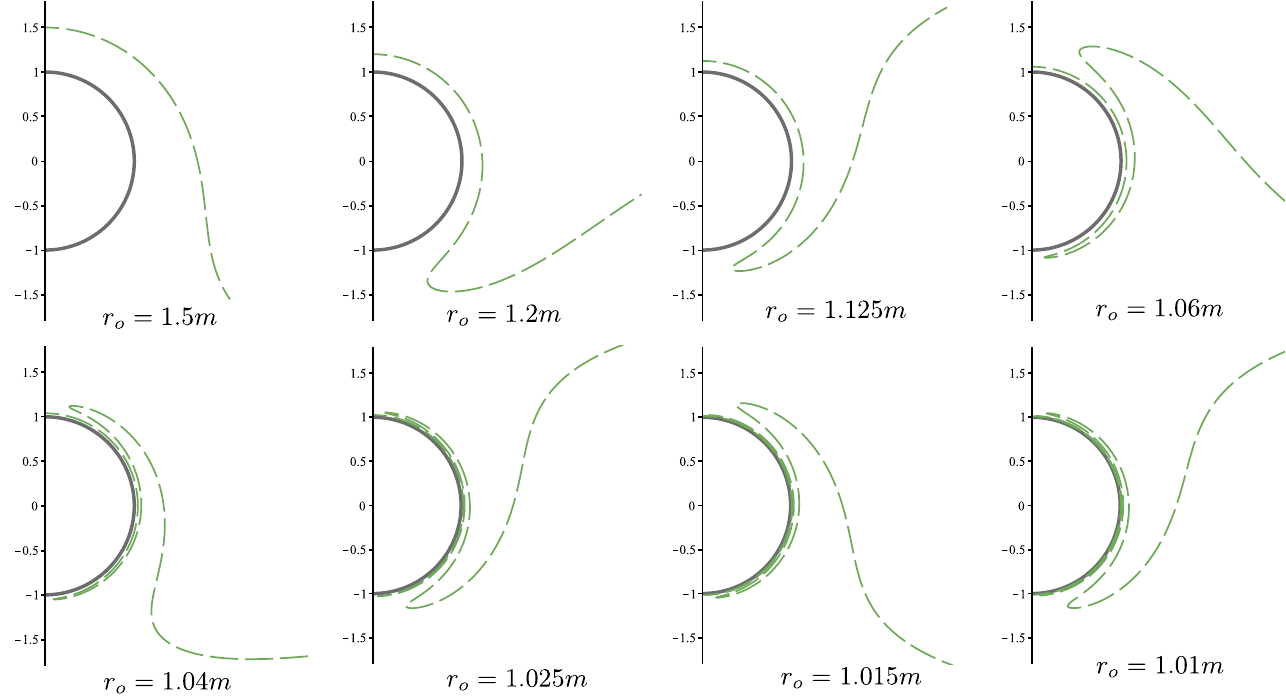}
\caption{Kerr Boyer-Lindquist wrapping for an extremally rotating $(a=m)$ black hole $a=m$. As for all Kerr diagrams in this paper, 
keep in mind that the induced metric (\ref{BLmetric}) in the $(r,\theta)$-plane is not flat. Hence the cross-section of the horizon is only 
a coordinate circle (not a geometric one). 
}
\label{BL_ex}
\end{figure}

Some sample MOTOS for the extremally rotating case are shown in Figure \ref{BL_ex}. 
The structure of MOTOS in the Boyer-Lindquist slicing is qualitatively similar to previous results for static black holes in slicings that are not horizon penetrating, at least insofar as we have explored here (and also in the non-extremal cases that we have investigated but not included here). In particular, we see the the ``folding'' of MOTOS in the near horizon region that has been previously observed for Schwarzschild~\cite{Booth:2020qhb}.

\subsection{Kerr spacetime: Doran coordinates}

More interesting MOTSs can be found if we adopt a horizon-crossing coordinate system. Doran coordinates are one such  system for 
Kerr in which the surfaces of constant time are both  space-like and horizon-crossing \cite{Doran:1999gb}. A surface $\Sigma$ of constant time 
in this coordinate system has induced metric
\begin{eqnarray}
h_{ij} \dd x^i \dd x^j  = & \frac{\varrho_+^2}{\rho^2} \dd r^2   \label{Doranhij}
%& - 2 \frac{\sqrt{2mr}a \sin^2 \theta}{\sqrt{\rho^2}} \dd r \dd \phi\\
+ \varrho_+^2 \dd \theta^2   +\left(\frac{\cGp}{\varrho_+^2}  \right) \sin^2 \! \theta \dd \phi^2   \\
  &  -  \frac{ a \sin^2 \! \theta \sqrt{2mr}}{\rho}  (\dd r \dd \phi + \dd \phi \dd r) \nonumber
\end{eqnarray}
%which has inverse
%\begin{eqnarray}
%  h^{ij} \pdv{x^i} \pdv{x^j}  
%& = & \frac{r^4 + a^2(r^2 + 2mr) + a^2 \! \cos^2 \! \theta \,  (\rho^2 - 2mr)}{\varrho_+^4} \left[ \pdv{r} \right] ^2 \\
%& &  + \frac{a \sqrt{2mr} }{\rho \varrho_+^2} \left[ \pdv{r} \right]  \left[ \pdv{\phi} \right] 
%+ \frac{1}{\rho^2 \sin^2 \! \theta } \left[ \pdv{\phi} \right] ^2 
%  + \frac{1}{\varrho_+^2} \left[ \pdv{\theta} \right] ^2 \nonumber
%\end{eqnarray}
%along with the extrinsic curvature
%\begin{equation}
%& K_{ij} \dd x^i \dd x^j \nonumber \\
%= & \sqrt{\frac{m}{2r}} \left(\frac{\varrho_-^2 }{\rho \varrho_+^2} \right) \dd r^2  
%- 2 \sqrt{2mr} \left( \frac{a^2 \sin \theta \cos \theta}{\rho \varrho_+^2} \right) \dd r \dd \theta \nonumber \\
%& - 2 a m \sin^2 \theta \left(\frac{\varrho_-^2 }{\varrho_+^2} \right) \dd r \dd \phi \nonumber - 2 \rho \frac{\sqrt{2mr^3}}{\varrho_+^2} \dd \phi^2  \nonumber \\
%&  +2 \frac{2mra^3 \sin^3 \theta \cos \theta}{\rho^4} \dd \theta \dd \phi \nonumber \\
%& -  \frac{\rho}{\varrho_+^2}  \sqrt{\frac{2m}{r}} \sin^2 \theta \nonumber \\
%& \quad \times \left(r^2 -   \frac{ a^2 mr \sin^2 \theta \varrho_-^2}{\varrho_+^4} \right)  \dd \phi^2 \nonumber \, , 
%\end{equation}
%
and extrinsic curvature
\begin{eqnarray}
& K_{ij} \dd x^i \dd x^j  \label{DoranKij} \\
 = &   \frac{1}{\rho \varrho_+^2}\sqrt{\frac{m}{2r}}  \Big( \varrho_-^2  \dd r^2  - 2 a^2 r \sin\! \theta \cos \!\theta  (\dd r \dd \theta + \dd \theta \dd r)  - 2 r^2 \rho^2 \dd \theta^2 
 \Big) \nonumber \\
& 
 + \frac{ a m \sin^2 \! \theta}{\varrho_+^4} \Big(-\varrho_-^2  (\dd r \dd \phi + \dd \phi \dd r)    + 2a^2 r \sin\! \theta \cos \! \theta (\dd \theta \dd \phi + \dd \phi \dd \theta)\Big) \nonumber \\
 &  -   \rho  \sqrt{2mr}  \frac{\mathcal{H}}{\varrho_+^6}  \sin^2 \! \theta   \dd \phi^2 \nonumber 
\end{eqnarray}
%
%
%
%\begin{eqnarray}
%& K_{ij} \dd x^i \dd x^j  \\
% = &   \sqrt{\frac{m}{2r}} \left(\frac{\varrho_-^2 }{\rho \varrho_+^2} \right) \dd r^2 -  \sqrt{2mr} \left( \frac{a^2 \sin\! \theta \cos \!\theta}{\rho \varrho_+^2} \right) (\dd r \dd \theta + \dd \theta \dd r)  -  \rho \frac{\sqrt{2mr^3}}{\varrho_+^2} \dd \theta^2 \nonumber \\
%& 
% -  a m \sin^2 \! \theta \left(\frac{\varrho_-^2 }{\varrho_+^4} \right) (\dd r \dd \phi + \dd \phi \dd r)    + \frac{2mra^3 \sin^3\! \theta \cos \! \theta}{\varrho_+^4} (\dd \theta \dd \phi + \dd \phi \dd \theta) \nonumber \\
% &  -  \rho  \sqrt{\frac{2m}{r}} \sin^2 \! \theta \! \left( \!  \frac{r^2 \varrho_+^4- a^2 mr  \varrho_-^2 \sin^2 \! \theta}{\varrho_+^6} \! \right)  \dd \phi^2 \nonumber 
%\end{eqnarray}
where $\rho_{+}^2 := r^2+ a^2 \cos^2 \!\theta$ as in the previous section, $\varrho_-^2 :=  r^2 -  a^2 \cos^2\! \theta$ and $\rho^2 := {r^2+a^2}$.
The notation $\varrho_-^2$ is chosen for ease of checking dimensions, so that all the $\rho^2$ and $\varrho^2$ terms will have the same dimensions of length squared. For sufficiently small $r$ it can be negative (however its square root never appears). Further,
\begin{equation}
\cGp := \rho^2 \varrho_+^2 + 2 a^2 m r \sin^2 \! \theta \andd \cH : = r \varrho_+^4- a^2 m  \varrho_-^2 \sin^2 \! \theta
\end{equation}
which respectively have dimensions length to the fourth and length to the fifth. 

Note that due to the $\sqrt{r}$-terms, both the metric and extrinsic curvature become ill-defined if one tries to pass through the $\theta = \pi/2$ ring singularity 
and enter the $r<0$ region. Further, geometrically it is straightforward to check that the trace of the extrinsic curvature
\begin{equation}
\mbox{tr} K = h^{ij}K_{ij} = - \sqrt{ \frac{m}{2r} } \left( \frac{3r^2 + a^2}{\rho \varrho_+^2} \right) \label{trK}
\end{equation}
diverges as $r \rightarrow 0$: the extrinsic geometry becomes singular everywhere on the disk that ``fills-in'' the ring singularity. By contrast
the intrinsic geometry of these surfaces diverges only on the ring singularity. We will find that MOTSodesics may terminate on this coordinate 
singularity.

Note that such a (coordinate) disk singularity does not exist in the
Eddington-Finkelstein-type form of the metric which can be continued to negative $r$. One way to understand its appearance is to note that in 
the construction of Doran from Eddington-Finkelstein
the coordinate transformations include $\sqrt{r}$ terms and that is the origin of this coordinate singularity \cite{Doran:1999gb,Baines:2020egv}\footnote{In the first reference it is claimed that the transformation is well-defined for all $r$. It appears that the author is implicitly assuming $r>0$ 
since the transformation becomes complex for $r<0$. The second reference correctly discusses the $r<0$ case. }. It is possible that there
exist space-like slicings of Kerr that do not exhibit this singular behaviour but searching for them is outside the scope of this paper. 

To find the MOTSs for this non-diagonal metric, we require more of the structure developed in the previous sections than was necessary in the our first examples. We have
%\begin{eqnarray}
%\underline{h}_{ab} \dd x^a \dd x^b  = &  \frac{\varrho_+^2}{\rho^2} \left( \dd r^2 + \rho^2 \dd \phi^2  \right) \, , \\
% \uh_{\phi a} \dd x^a  = & - \frac{ a \sin^2 \! \theta \sqrt{2mr}}{\rho}  \dd \phi \andd \nonumber\\
% \uh_{AB} \dd \phi^A \dd \phi^B  = &\left(\rho^2 +   \frac{2mra^2 \sin^2 \! \theta}{\varrho_+^2} \right) \sin^2 \! \theta \dd \phi^2 \nonumber \; . 
%\end{eqnarray}
\begin{eqnarray}
&\underline{h}_{ab}   = & \left[ \begin{array}{cc}  \frac{\varrho_+^2}{\rho^2} & 0 \\ 0 &  \varrho_+^2 \end{array}  \right] \, , \\
& \uh_{\phi a}   =  & - \left( \frac{ a \sin^2 \! \theta \sqrt{2mr}}{\rho} \right)  \left[ \; 1 \; \; \;   0 \;  \right] \andd \nonumber\\
& \uh_{\phi \phi}  =& R^2  = \left(\frac{\cGp}{\varrho_+^2} \right) \sin^2 \! \theta   \nonumber \; . 
\end{eqnarray}
In this case
\begin{equation}
\oh^{\phi \phi} =\frac{1}{ (r^2+a^2) \sin^2 \! \theta} \, , 
\end{equation}
and so clearly  $\oh^{AB} \neq \uh^{AB}$. 

Note that  $\tilde{\Sigma}$ with metric $\underline{h}_{ab}$ is flat:  a Euclidean half-plane. 
Intrinsically it 
is identical to a plane of constant $t$ and $\phi$ in Boyer-Lindquist Kerr with $m=0$. That is, flat space in oblate spheroidal coordinates.   The equivalence can be  seen explicitly from the coordinate transformation
\begin{eqnarray}
\left\{
\begin{array}{l}
  x = \rho \sin \theta \\
  z = r \cos \theta 
  \end{array} \right\} \; \; \mbox{by which} \; \; 
      \frac{\varrho_+^2 }{\rho^2} \dd r^2 + \varrho_+^2 \dd \theta^2  =  \dd x^2 + \dd z^2 \, , \label{148}
\end{eqnarray} 
for $0 \leq x \leq \infty$ and $- \infty \leq z \leq \infty$.  In the usual way this transformation breaks the coordinate 
degeneracy at $r=0$ so that that apparent point becomes an  open segment $0 \leq  x < a$ with a curvature singularity at 
$x = a$. Under a $\pdv{\phi}$ rotation this becomes the well-known Kerr ring singularity. For a nice discussion of these coordinates see, for example, \cite{Griffiths:2009dfa}.  However,  with Doran not well defined for $r < 0$, it is also true that crossings from positive to negative $z$ (or 
vice versa) are not possible: for $0 \leq x < a$, $\tilde{\Sigma}$ inherits the coordinate singularity at $z=0$. Intrinsically $\tilde{\Sigma}$ is 
a Euclidean half-plane, however it is embedded in the larger $\Sigma$ which isn't well-defined beyond $z=0$ in that region.

%It is well known that in Eddington-Finkelstein-type coordinates, crossing that $z=0$ segment means entering the $r < 0$ region of spacetime. However in changing from those coordinates to Doran, the coordinate transformation functions all include a  $\sqrt{r}$ \cite{Doran,MattKerr}. This introduces a coordinate singularity for $r=0, \theta \neq \pi/2$ (equivalently 
%$0 \leq x < a$ and $z=0$). This is in the metric (\ref{Doranhij}) and extrinsic curvature (\ref{DoranKij}). 
%

The relevant extrinsic curvature quantities are
\begin{eqnarray}
 \underline{\tr} \underline{K} = \uh^{AB} \uK_{AB} = \frac{\uK_{\phi \phi}}{\uh_{\phi \phi}} = 
  - \frac{   \rho}{\varrho_+^4}  \sqrt{2mr}   \left( \frac{ \cH }{\cGp}  \right)\nonumber 
\end{eqnarray}
(note that this is different from (\ref{trK}) as it contracts with $\uh^{ab}$ rather than $h^{ij}$) and 
\begin{equation}
\underline{K}_{ab} =  \frac{1}{\rho \varrho_+^2} \sqrt{\frac{m}{2r}}\left[ \begin{array}{cc} 
\varrho_-^2  & - 2r a^2 \sin\! \theta \cos \!\theta \\
-  2ra^2 \sin\! \theta \cos \!\theta &  - 2r^2  \rho^2 
\end{array} \right]
\end{equation}
along with
\begin{equation}
\uK_{\phi a} =\frac{am\sin^2 \! \theta}{\varrho_+^4}  \left[ \begin{array}{cc} -   \varrho_-^2 &   2a^2 r \sin \! \theta \cos \! \theta \\ \end{array} \right]  \; . 
\end{equation}
All of these are ill-defined at the curvature singularity $\varrho_+ = 0$ but note that $\underline{\tr} \underline{K}$ and $\underline{K}_{ab}$ 
are also undefined at $r=0$. 

The rotational Killing vector field is
\begin{equation}
\xi^i = \left[\pdv{\phi}\right]^i = \left[ \begin{array}{c} 0 \\ 0 \\ 1 \\\end{array} \right] \; 
\end{equation}
and the induced area form on $T \twoS$ is 
\begin{equation}
\underline{\epsilon} = \frac{\varrho_+^2}{\rho} \dd r \wedge \dd \theta \; . 
\end{equation}
Then 
\begin{equation}
\omega_{ab} = \frac{a \sqrt{2mr} \sin \theta \cos \theta}{\rho} \left[
\begin{array}{cc}
0 & 1 \\
-1 & 0 \\
\end{array} \right] 
\end{equation}
and so 
\begin{equation}
\underline{\Omega} = \frac{2a  \sqrt{2mr} \sin \! \theta \cos \! \theta}{\varrho_+^2}.
\end{equation}

The  increasing-in-$\phi$ normal to these surfaces is
\begin{equation}
\Phi_i = [\dd \phi]_i = \left[ \begin{array}{ccc} 0 & 0 &  1 \\ \end{array} \right];
\end{equation}
then
\begin{equation}
\utrchi = \frac{\underline{\chi}_{\phi \phi}}{\uh_{\phi \phi}}= \underline{h}^{AB} \underline{\chi}_{AB}  =
  \frac{a }{\rho \varrho_+^4}\sqrt{2mr} \left( \frac{\cH}{\cGp } \right)
\end{equation}
and
\begin{equation}
\underline{\chi}_{ab} = -\frac{a}{ \rho^3 \varrho_+^2} \sqrt{\frac{m}{2r}} \left[
\begin{array}{cc}
\varrho_-^2 & -2r \rho^2 \cot \theta \\
-2r  \rho^2 \cot \theta &- 2 \rho^2 r^2\\
\end{array} \right]\; ,
\end{equation}
while
\begin{equation} 
\underline{\chi}_{\phi a} =  \left[ \begin{array}{cc} \frac{1}{\rho^2} \left( -r + \frac{a^2 m \varrho_-^2 \sin^2 \! \theta  }{\varrho_+^4}  \right)
&   -\cot \! \theta - \frac{2 a^2 m r \sin \! \theta \cos \! \theta}{\varrho_+^4} \\ \end{array} \right] \; . 
\end{equation}

Then for $\gamma$ arclength parameterized as $(r=P(s), \theta=\Theta(s))$, the unit tangent and normal vectors are
\begin{equation}
\hat{T} = \dot{P} \pdv{r} + \dot{\Theta} \pdv{\theta} \andd \hat{N} = \rho \dot{\Theta} \pdv{r} - \frac{\dot{P}}{\rho} \pdv{\theta} 
\end{equation}
with the constraint that
\begin{equation}
\frac{\varrho_+^2}{\rho^2} \left(\dot{P}^2 + \rho^2 \dot{\Theta}^2 \right) = 1  \;  . 
\end{equation}
Hence 
\begin{eqnarray}
\ubeta_\phi &= \hat{T}^i h_{ij} \xi^j = -  \frac{a\sqrt{2mP} \sin^2 \! \Theta \dot{P}}{\rho}  \andd \\
\uV_\phi &= \hat{N}^i h_{ij} \xi^j = -a \sqrt{2mP} \sin^2 \! \Theta \, \dot{\Theta}
\end{eqnarray}
while 
\begin{equation}
\sec \! \alpha = \sqrt{ 1 + \frac{2mPa^2 \sin^2 \! \Theta \, \dot{\Theta}^2 }{\rho^2} }\; . 
\end{equation}

Combining all of the above, we have equations of motion
\begin{eqnarray}
\ddot{P} +   \frac{1}{\varrho_+^2} \left (\frac{a^2 P \sin^2 \! \Theta}{\rho^2} \dot{P}^2 - 2 a^2 \sin \! \Theta \cos \! \Theta \dot{P} \dot{\Theta} - P \rho^2  \dot{\Theta}^2 \right)  = \rho \kappa \dot{\Theta} \label{Eq_Doran}\\
\ddot{\Theta} + \frac{1}{\varrho_+^2} \left(\frac{a^2 \sin \! \Theta \cos \! \Theta}{\rho^2} \dot{P}^2 + 2 P \dot{P} \dot{\Theta} - a^2 \sin \! \Theta \cos \! \Theta \dot{\Theta}^2 \right) = -  \frac{\kappa}{\rho} \dot{P} \nonumber
\end{eqnarray}
where 
\begin{eqnarray}
 \kappa = \sec \! \alpha\, \left( \mathcal{K} +  \mathcal{K}_{\hat{T} \hat{T} } \right) +  \mathcal{K}_{\hat{T}} + \mathcal{K}_{\hat{N}} + \mathcal{K}_{\hat{T} \hat{T} \hat{N}}    \, , 
 \end{eqnarray}
and  the components are
 \begin{eqnarray}
 \mathcal{K}  &=    -  \frac{   \rho}{\varrho_+^4} \sqrt{2mP}    \frac{\cH}{\cGp}  \\
 \mathcal{K}_{\hat{T}} & = \left(  \frac{4a^2mP \sin \! \theta \cos \! \Theta }{\rho \mathcal{G}_+} \right) \dot{P} \\
  \mathcal{K}_{\hat{N}} 
  & = \left(- \frac{\rho \cot \! \Theta}{\varrho_+^2} + \frac{a^2 \Delta \sin \! \Theta \cos \! \Theta}{\rho \mathcal{G}_+} \right) \dot{P} \\
  & \phantom{=} + \left(P - \frac{a^2 m \varrho_-^2 \sin^2 \! \Theta}{ \varrho_+^4} \right) \frac{\dot{\Theta}}{\rho } \nonumber \\
  \mathcal{K}_{\hat{T} \hat{T}}  
  & = \sqrt{\frac{m}{2P}}\left(\frac{ \varrho_-^2}{ \rho \varrho_+^2} \right) \left( \frac{ \mathcal{G_-} }{ \mathcal{G}_+ } \right)  \dot{P}^2   \\
  & \phantom{=} - \left( \frac{2 a^2 \sqrt{2mP} \rho \sin \! \Theta \cos \! \Theta}{\mathcal{G}_+} \right) \dot{P} \dot{\Theta}  - \left( \frac{\sqrt{2m} \rho P^{3/2} }{\varrho_+^2} \right) \dot{\Theta}^2 \nonumber
\\
 \mathcal{K}_{\hat{T} \hat{T} \hat{N}}  &=  - \frac{a^2m \sin^2 \! \Theta}{ \rho^3 \varrho_+^2} \left(\varrho_-^2 + \frac{4r\cH}{\cGp} \right) \dot{P}^2 \dot{\Theta} \\
 & \phantom{=} + \left(\frac{4a^4mP\sin^3 \! \Theta \cos \! \Theta}{\rho \cGp} \right) \dot{P} \dot{\Theta}^2 + \left(\frac{2a^2mP^2\sin^2 \! \Theta}{\rho \varrho_+^2} \right) \dot{\Theta}^3 \nonumber
 \end{eqnarray}
with
\begin{equation}
\cGm = \rho^2 \varrho_+^2  - 2a^2 mP \sin^2 \!\Theta
\end{equation}
 defined analogously to $\cGp$ and
\begin{equation}
\sec \! \alpha = \sqrt{1 + \frac{2a^2mP\sin^2 \! \Theta \, \dot{\Theta}^2}{\rho^2}}  \; . 
\end{equation}

\begin{figure}
\centering
\includegraphics{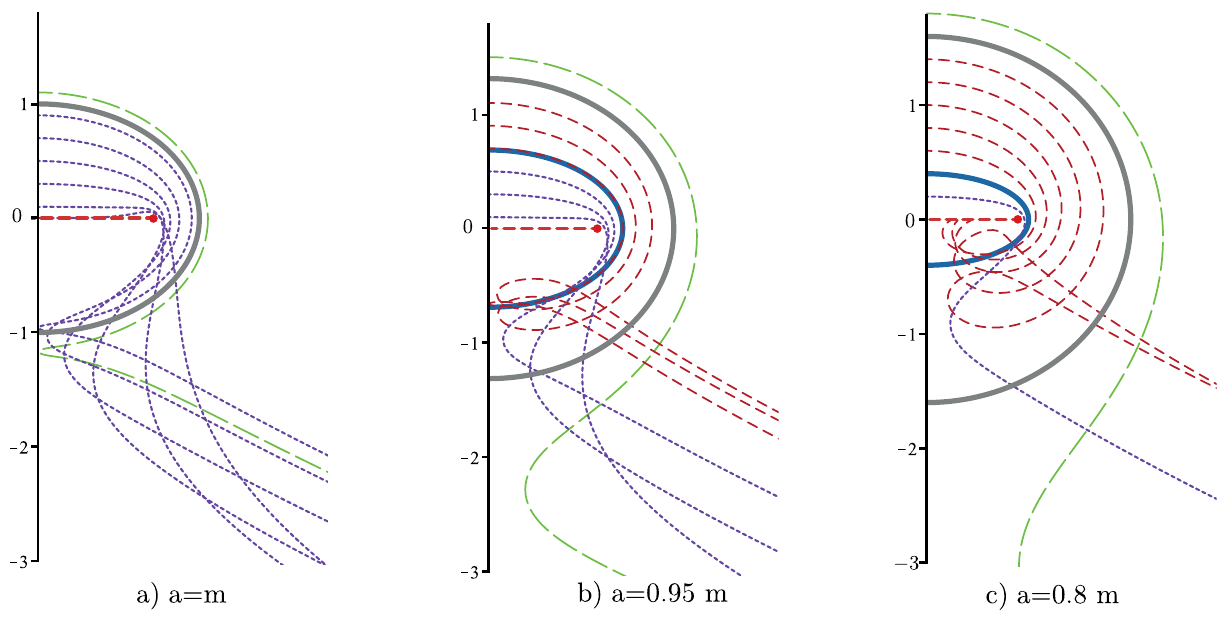}
\caption{Examples of high rotation Kerr spacetimes for which there are no exotic MOTSs in the Doran slicing: all MOTSodesics either 
diverge to infinity or run into the disk singularity. 
The colour and linestyle scheme from Figure \ref{fig:BTZ_all} is repeated here: outer horizon is gray, inner horizon is light blue and 
MOTSodesics are presented different colours and linestyles depending on the point from which they leave the $z$-axis. In this case 
we also have time slices disk singularity (red dashed line) which fills the ring singularity (big red dot). 
}
\label{highrot}
\end{figure}

\begin{figure}
\centering
\includegraphics{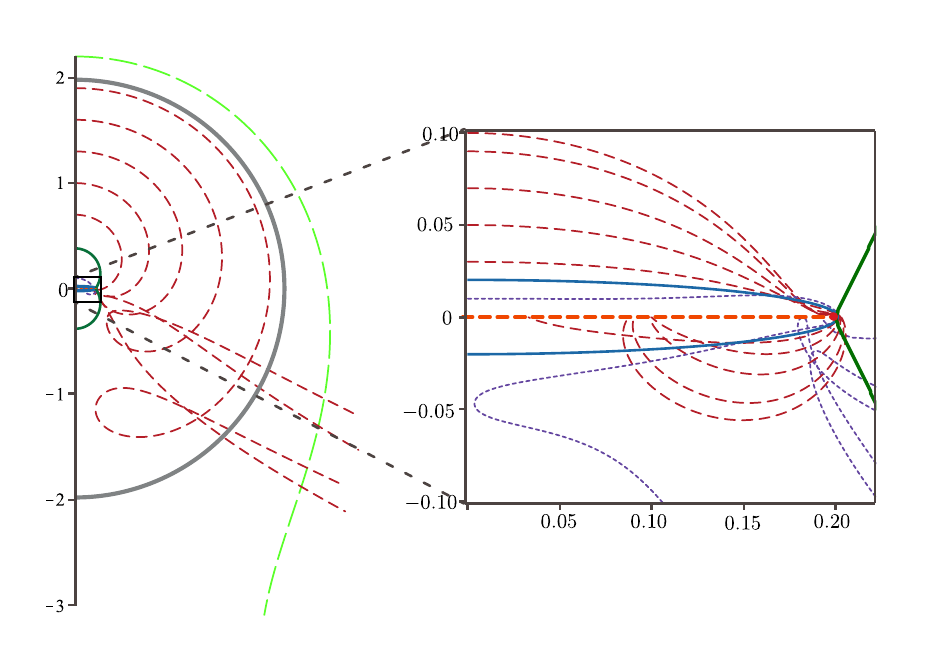}
\caption{Non-horizon exotic MOTSs in Kerr-Doran are only found for $a \lesssim 0.2m$. Shown here is the case where $a=0.2m$ (with $m=1$). This is the 
analogue of the one-loop MOTS from the Schwarzschild cases but instead of a loop we have a repulsion from the ring-singularity. Apart from 
this MOTS and the two horizons, all other corresponding MOTSodesics either diverge to infinity or run into the disk singularity. }
\label{firstclosed}
\end{figure}

\begin{figure}
\centering
\includegraphics{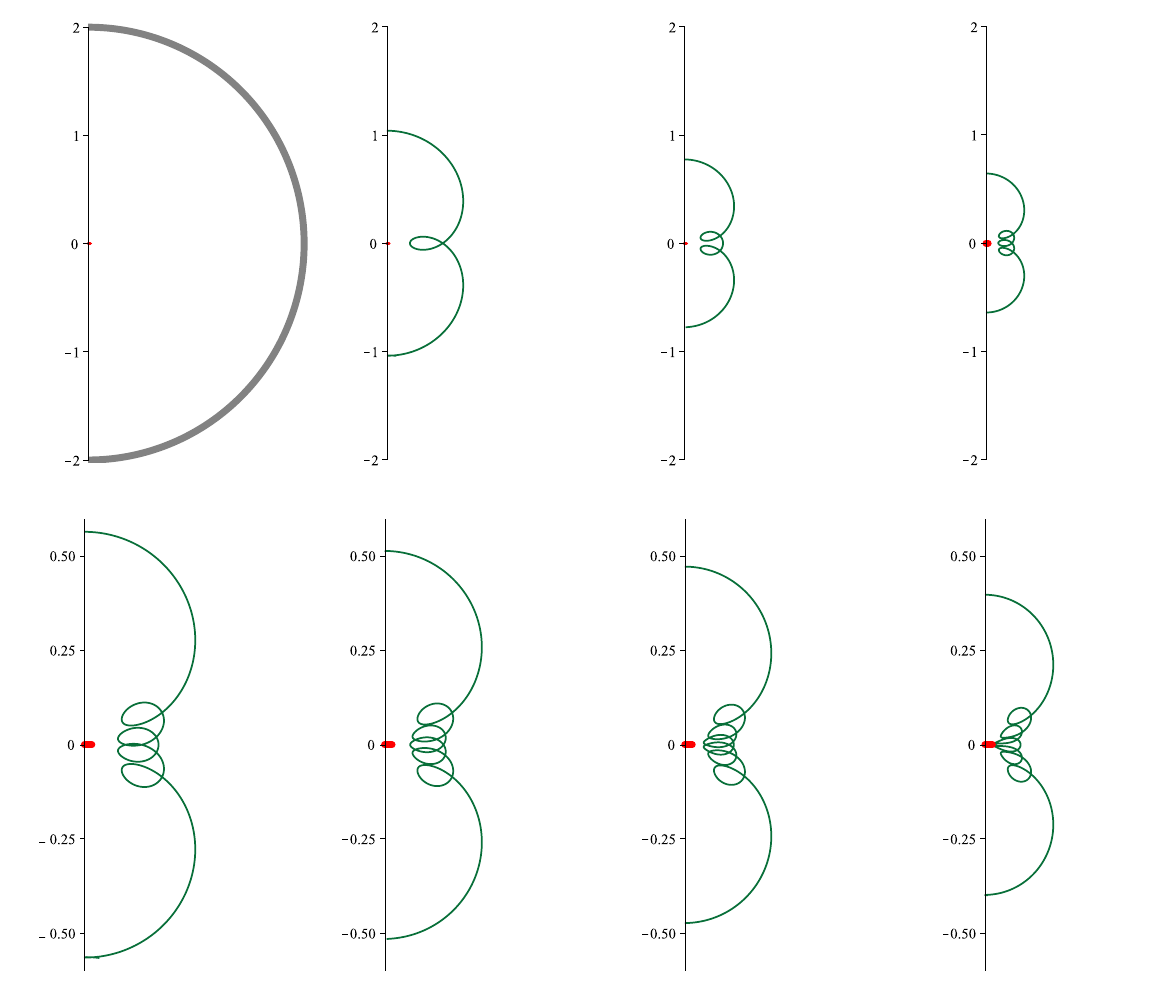}
\caption{For low rotation Kerr-Doran the familiar picture from Schwarzschild  partially reasserts itself: many exotic self-intersecting MOTS 
inside the outer horizon. Shown here are the eight MOTSs for $a=0.02m$ (plotted with $m=1$). In this case these are the only eight. Beyond this
point, the MOTSodesics begin to interact with the inner horizon and singularity and these interactions prevent the formation of higher order loops. 
}
\label{lowrot}
\end{figure}

Again these equations are complicated in appearance but can be easily integrated numerically (using the now familiar 
series expansion off the $z$-axis to establish boundary conditions). Some sample MOTSs are shown in 
Figures \ref{highrot}-\ref{lowrot}. These are not intended to exhaustively explore the space of all possible Kerr MOTSs (even in the current
coordinate system) but rather just give some idea of possible behaviours and how they compare to MOTSs seen in non-rotating 
black holes.

Many of the features of MOT(O)S seen in the Doran slicing are qualitatively similar to those seen in other examples of two-horizon black hole spacetimes in horizon-penetrating coordinates, especially Reissner-Nordstrom~\cite{Hennigar:2021ogw}. In particular, as seen in Figure~\ref{lowrot}, the number of exotic MOTSs present in the interior is highly dependent on the rotational parameter $a$. We find that for any finite value of the rotation parameter there are a finite number (including zero) of closed, self-intersecting MOTSs in the interior. The number of such surfaces is larger for smaller values of the rotation parameter, with at least a single, closed surface existing provided that  $a \lessapprox 0.2 M$. 
Qualitatively, they exhibit the same type of looping behaviours as those seen in the archetypal case of the Schwarzschild interior~\cite{Booth:2020qhb}.

However, there are also notable differences --- we have illustrated an example in Figure~\ref{highrot}. Here we show the effect of the time slice
disk singularity, which appears in the figures as the dashed, horizontal red curve. We find that it is possible for MOTOS to impact this singularity. For time slices of spacetime that extended through the ring-singularity into the asymptotic region (with negative mass) it is almost certain that the analogous MOTOSs would also penetrate that region.   However, because the Doran coordinates do not cover this region, we are unable to follow such curves any further. 
Our numerical investigations suggest that it is only MOTOSs that begin within the event horizon that have the possibility of exiting through the ring singularity. Moreover, the fraction of curves that possess this behaviour increases as the rotation is decreased from its extremal value (see the progression, left-center-right of Figure~\ref{highrot}).

\section{MOTS in the rotating Myers-Perry black hole spacetime}
\label{(4+1)} 
\subsection{The equal angular momenta Myers-Perry black hole} The Myers-Perry black hole is a solution of the vacuum Einstein equations in five spacetime dimensions. It is a natural extension of the Kerr solution and represents a spinning, asymptotically flat stationary black hole with two axial symmetries.  The solution is characterized by three physical parameters: its mass $m$ and two independent angular momenta $J_1, J_2$ in each orthogonal plane of rotation.  We will focus on the subfamily with equal angular momenta $J_1 = J_2$. In this case, the isometry group is enhanced from $\mathbb{R} \times U(1) \times U(1) \to \mathbb{R} \times SU(2) \times U(1)$. This simplification only occurs in odd dimensions and has no analogue in four dimensions, so we will focus on this case as it is qualitatively different to Kerr. 

We find it convenient to use Doran-type coordinates for the Myers-Perry metric rather than the more familiar Boyer-Lindquist coordinates (see \cite{Finch:2013gha}).  As in the case of Doran coordinates for the Kerr solution,  these coordinates are horizon penetrating. The spacetime metric is
\begin{eqnarray}\label{mp:Doran}
\dd s^2 &= -\dd t^2 + \frac{r^2}{r^2 + a^2}\left[\dd r + \frac{\sqrt{2\mu}}{r}\left(\dd t - a (\dd\psi + \frac{\cos\theta}{2} \dd\phi) \right)\right]^2 \\
& + (r^2+a^2)(\dd\psi + \frac{\cos\theta}{2} \dd\phi)^2 + \frac{r^2+a^2}{4} (\dd\theta^2 + \sin^2\theta \dd\phi^2).
\end{eqnarray}  Here $t \in \mathbb{R}$,  $\phi \sim \phi + 2\pi, \psi \sim \psi + 2\pi$ are angular coordinates associated to the commuting vector fields $\partial_\phi, \partial_\psi$  and $\theta \in (0, \pi)$ is a polar coordinate on $S^3$.  The solution is parameterized by two parameters $(\mu, a)$ which are proportional to the ADM mass and angular momentum respectively.  We assume $\mu >0, a \geq 0$ without loss of generality.  In the exterior region, the  radial coordinate $r \in (r_+, \infty)$ with $r \to \infty$ and $r = r_+$ corresponding to an asymptotically flat region and to the event horizon, respectively.  This can be seen by computing 
\begin{equation}
\dd r \cdot \dd r = g^{-1}(\dd r, \dd r) =  \frac{(r^2+a^2)^2 - 2 r^2\mu}{r^2(r^2+a^2)}
\end{equation} so that $\dd r$ becomes null when 
\begin{equation}
r^2_\pm = \mu - a^2 \pm \sqrt{\mu(\mu - 2a^2)}\; ,
\end{equation} where the $+$ sign gives the location of the event horizon and the $-$ sign to an inner horizon.  Note that we must take $\mu \geq 2 a^2$, and $\mu = 2a^2$ is the extreme case.  The null generator of the event horizon is the Killing vector field
\begin{equation}
V = \partial_t + \Omega_H \partial_\psi, \qquad \Omega_H = \frac{a}{2(r_+^2 + a^2)}.
\end{equation} Note that the coordinate system breaks down at $r =0$, however this is merely a coordinate singularity which can be removed by switching to a 
new radial coordinate $\tilde{r} = r^2 + a^2$. Then the $r=0$ singularity disappears but a genuine curvature singularity remains at $\tilde{r}=0$. We will return to this alternate coordinate later in this section but for now proceed with the regular one.

In the above coordinate chart, the constant $t$ surfaces have unit lapse function. The constant $t$ hypersurfaces have the induced Riemannian metric expressed in the coordinates $y^i = (r, \theta, \psi,\phi)$: 
\begin{eqnarray}
h_{ij} \dd y^i \dd y^j &= \frac{r^2}{r^2 + a^2}\left[\dd r - \frac{\sqrt{2\mu} a }{r}\left( \dd\psi + \frac{\cos\theta}{2} \dd\phi \right)\right]^2 \\
& + (r^2+a^2)\left(\dd\psi + \frac{\cos\theta}{2} \dd\phi\right)^2 + \frac{r^2+a^2}{4} (\dd\theta^2 + \sin^2\theta \dd\phi^2) \nonumber.
\end{eqnarray} We are interested in finding $U(1) \times U(1)$-invariant MOTSs within this initial data set. We split the coordinates as $x^a = (r,\theta)$ with $a = 1,2$ and $\phi^A = (\psi, \phi)$ with $A =3, 4$.  This gives the identifications 
\begin{equation}
\xi_1 =  \pdv{r}, \quad \xi_2 =  \pdv{\theta}, \quad \xi_3 =  \pdv{\psi}, \quad \xi_4 =  \pdv{\phi}.
\end{equation} The metric components split up as
\begin{eqnarray}
\uh_{ab} \dd x^a \dd x^b & =\frac{r^2 \dd r^2}{r^2 +a^2}  + \frac{(r^2+a^2)\dd \theta^2}{4} , \\
\uh_{Ab} \dd \phi^A \dd x^b & =   - \frac{\sqrt{2\mu } a r }{(r^2+a^2)} \left( \dd\psi + \frac{\cos\theta}{2} \dd\phi \right) \dd r\\
\uh_{AB}\dd \phi^A \dd \phi^B & = \left[ r^2 + a^2 + \frac{2 a^2 \mu}{r^2 + a^2} \right] \left( \dd\psi + \frac{\cos\theta}{2} \dd\phi \right)^2 \\ & + \frac{(r^2 + a^2) \sin^2\theta}{4} \dd \phi^2, \nonumber
\end{eqnarray} and the inverse metric splits as 
\begin{eqnarray}
\oh^{ab} \pdv{x^a} \pdv{x^b} &= \left(\frac{ (r^2 + a^2)^2 + 2 a^2 \mu}{r^2(r^2+a^2)}\right) \left(\pdv{r}\right)^2 + \frac{4}{r^2+a^2} \left(\pdv{\theta}\right)^2\\
\oh^{Ab} \pdv{\phi^A} \pdv{x^b} & = \frac{\sqrt{2\mu} a}{r (r^2+a^2)} \pdv{\psi}\pdv{r} \\
\oh^{AB} \pdv{\phi^A} \pdv{\phi^B} & = \frac{4}{\sin^2\theta(r^2+a^2)} \left( \pdv{\phi} - \frac{\cos\theta}{2} \pdv{\psi}\right)^2 \\ &+ \frac{1}{r^2+a^2} \left(\pdv{\psi}\right)^2.
\end{eqnarray} Note that 
\begin{equation}
\uh^{ab} = \left( {\begin{array}{cc}  \frac{r^2+a^2}{r^2} & 0 \\ 0 & \frac{4}{r^2+a^2} \\ \end{array} } \right) \neq \oh^{ab}
\end{equation} 
and $\uh^{\psi\phi} = \oh^{\psi\phi}, \uh^{\phi\phi} = \oh^{\phi\phi}$, but 
\begin{equation}
\uh^{\psi\psi} = \oh^{\psi\psi} - \frac{2 a^2\mu}{(r^2+a^2)((r^2+a^2)^2 + 2 a^2\mu)}.
\end{equation} Within this basis, the extrinsic curvature has components
\begin{eqnarray}
\uK_{ab} &= \left( {\begin{array}{cc}  \frac{\sqrt{2\mu} r^2}{(r^2+a^2)^2} & 0 \\ 0 & -\frac{\sqrt{2\mu}}{4} \\ \end{array} } \right) , \\\ \uK_{AB} &= -\frac{\sqrt{2\mu}}{2r} \partial_r \uh_{AB}
\end{eqnarray}
and 
\begin{equation}
\uK_{r\psi} =-\frac{2 a \mu r}{(r^2+a^2)^2}, \quad \uK_{r\phi} = \frac{\cos\theta}{2} \uK_{r\psi},
\end{equation} or equivalently
\begin{equation}
\uK_{Aa} \dd \phi^A \dd x^a = -\frac{2 a \mu r}{(r^2+a^2)^2} \cdot \dd r \left( \dd\psi + \frac{\cos\theta}{2} \dd \phi \right).
\end{equation}  It can be verified directly that at the intersection $S$ of the spatial slice and the event horizon, $\theta^+ = H + q^{ij} K_{ij} =0$ where $H$ is the mean curvature and $q^{ij}$ is the induced metric  on $S$. 

% The connection coefficients $\chi^A_{ij}$ and $\omega_{Aij}$ are readily computed. \robie{Are these to be added? Or is this just a remark? If it is just a remark, does it deserve an independent paragraph?}  

Now consider a smooth unit speed curve $(r=P(s),\theta=\Theta(s))$ in $\tilde{\Sigma}$.  The tangent vector will be written
\begin{equation}
\hat{T} = \dot{P} \pdv{r} + \dot{\Theta} \pdv{\theta}
\end{equation} where 
\begin{equation}\label{MPunitspeed}
\frac{P^2}{P^2+a^2} \dot{P}^2 + \frac{P^2+a^2}{4} \dot\Theta^2 = 1.
\end{equation} 
%By a slight abuse of notation we will not explicitly pull-back functions evaluated on the curve, e.g. we will write $r$ rather than $r = P(s)$.
We then have for the unit normal
\begin{equation}
\hat{N} = \frac{P}{2} \left(\frac{(P^2+a^2)\dot\Theta}{P^2} \pdv{r} - \frac{4 \dot{P}}{P^2+a^2} \pdv{\theta} \right).
\end{equation} For simplicity, we will just give the results of the computations described in \ref{MOTSodesics}. We find
\begin{equation}
\ubeta_\psi = -\frac{\sqrt{2\mu} a P \dot{P}}{P^2+a^2} ,\qquad \ubeta_\phi = \frac{\cos\Theta}{2} \ubeta_\psi
\end{equation} and 
\begin{equation}
\uV_\psi = -\sqrt{\frac{\mu}{2}} a \dot\Theta, \qquad \uV_\phi = \frac{\cos\Theta}{2} \uV_\psi.
\end{equation} Note $\ubeta$ and $\uV$ are both multiples of the $SU(2)$-left invariant form $(\dd\psi + \cos\theta \dd \phi/2)$.  We compute the scalars
\begin{eqnarray}
\ubeta^2 &= \frac{2 \mu a^2 P^2 \dot P^2}{(P^2+a^2)((P^2+a^2)^2 + 2a^2\mu)}, \\ \uV^2& = \frac{\mu a^2 (P^2+a^2) \dot\Theta^2}{2((P^2+a^2)^2 + 2a^2\mu)}\end{eqnarray} 
and 
\begin{equation} 
\ubeta \cdot\uV =  \frac{\mu a^2 P \dot P \dot \Theta}{(P^2+a^2)^2 + 2a^2\mu}.
\end{equation}
We then compute
\begin{equation}
\cos^2\! \alpha = \frac{2(P^2+a^2)}{(2(P^2+a^2) + a^2\mu \dot{\Theta}^2)}\;,
\end{equation} 
where we used the unit speed condition (\ref{MPunitspeed}).  
If $a=0$ we get $\alpha =0$, as expected for a constant time slice of Schwarzschild in Painlevé-Gullstrand coordinates. 

A computation gives 
\begin{equation}
\Omega_\psi =0,\qquad \Omega_\phi = -\frac{\sqrt{2\mu} a \sin\theta}{r^2+a^2}.
\end{equation} We have
 \begin{eqnarray}
 \kappa_{\mbox{\tiny{MOTS}}} = \sec \! \alpha\, \left( \mathcal{K} +  \mathcal{K}_{\hat{T} \hat{T} } \right) +  \mathcal{K}_{\hat{T}} + \mathcal{K}_{\hat{N}}   + \mathcal{K}_{\hat{T} \hat{T} \hat{N}}
 \, , 
 \end{eqnarray}
\begin{eqnarray}
  \mathcal{K}  &= \underline{\tr} \underline{K} = -\frac{2\sqrt{2\mu} (P^2+a^2)}{(P^2+a^2)^2 + 2a^2 \mu}, \\
 \mathcal{K}_{\hat{T}} & = 0, \\
   \mathcal{K}_{\hat{N}} & = \dot\Theta -\frac{2P \cot \Theta \dot{P}}{P^2+a^2},  \\
    \mathcal{K}_{\hat{T} \hat{T}}  & = \frac{\sqrt{\mu}}{2\sqrt{2}}\left(-\dot\Theta^2 + \frac{4P^2 \dot{P}^2}{(P^2+a^2)^2 + 2 a^2 \mu}\right), 
  \\
   \mathcal{K}_{\hat{T} \hat{T} \hat{N}}  &=  \frac{\mu a^2 \dot \Theta^3}{4(P^2+a^2)} + \frac{4a^2 \mu P^3  \cot\Theta \dot{P}^3}{(P^2+a^2)^2((P^2 + a^2)^2  + 2a^2\mu)} \\ & - \frac{4\mu a^2 P^2 \dot P^2 \dot\Theta}{(P^2+a^2)((P^2+a^2)^2 + 2 a^2\mu)}. \nonumber
 \end{eqnarray} Here we used that since $\uh_{A\theta} =0$, the quantities $\Xi^{AB}_{ab}$ all vanish unless $(a,b) =(rr)$. 
 
 We also have the Christoffel symbols $ \underline{\Gamma}_{bc}^{a}$ associated to the 2d metric $\uh_{ab}$: 
 \begin{equation}
\underline\Gamma^{ \theta}_{~r\theta} = \frac{r}{r^2+a^2} \qquad \underline\Gamma^{ r}_{~rr} = \frac{a^2}{r(r^2+a^2)}, \qquad \underline\Gamma^{r}_{~\theta\theta} = -\frac{r^2+a^2}{4 r}.
\end{equation}  The curve equations are then
\begin{equation}
\frac{\mathrm{d}^2 X^a}{\mathrm{d}s^2}  =-  \underline{\Gamma}_{bc}^{a} \frac{\mathrm{d} X^b}{\mathrm{d} s} \frac{\mathrm{d}X^c}{\mathrm{d}s} +  {\kappa}_{\mbox{\tiny{MOTS}}}  \underline{\epsilon}^a_{\phantom{a} b} \frac{\mathrm{d} X^b}{\mathrm{d} s}
\end{equation} with $X^a(s) = (P(s), \Theta(s))$.  Explicitly, for Myers-Perry, since
\begin{equation}
\epsilon^r_b = \frac{r^2+a^2}{2r} \delta^\theta_{~b}, \qquad \epsilon^\theta_b = -\frac{2r}{r^2+a^2} \delta^r_{~b},
\end{equation} the MOTSodesic equations are
\begin{eqnarray}
\ddot P &= -\frac{a^2 \dot P^2}{P(P^2+a^2)} + \frac{(P^2+a^2) \dot\Theta^2}{4P} + {\kappa}_{\mbox{\tiny{MOTS}}}  \frac{P^2+a^2}{2P} \dot\Theta \\
\ddot \Theta & = - \frac{2 P \dot P \dot \Theta}{P^2+a^2} - {\kappa}_{\mbox{\tiny{MOTS}}} \frac{2 P}{P^2+a^2} \dot P.
\end{eqnarray}

\begin{figure}
    \centering
    \includegraphics{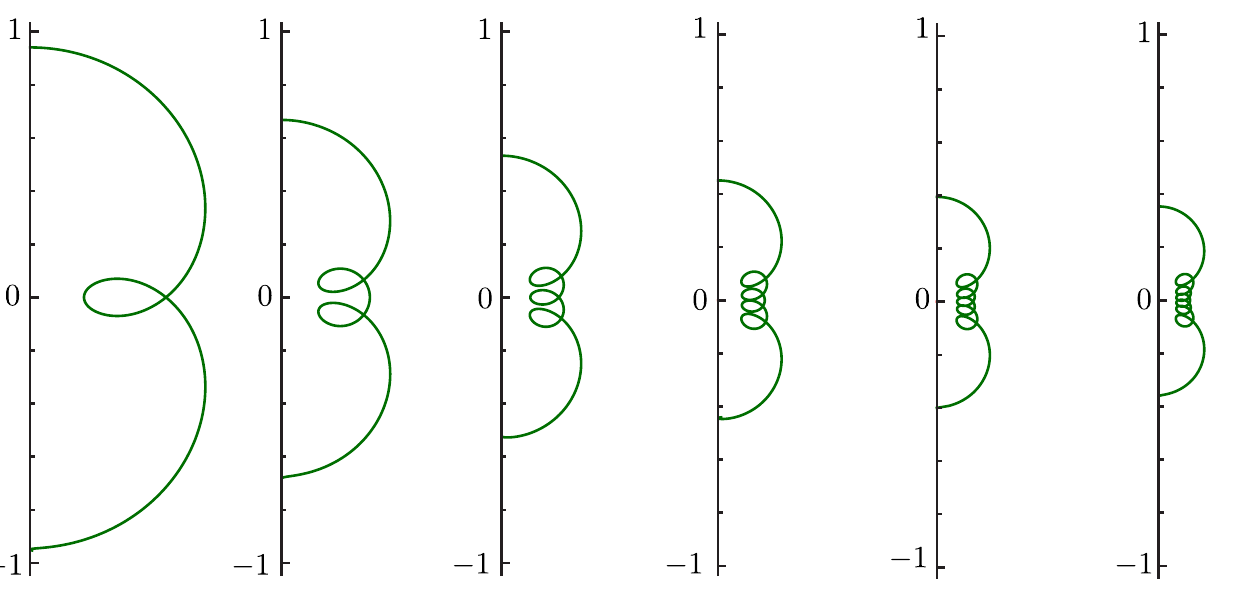}
    \caption{The first six exotic looping MOTSs for the Myers-Perry black hole in a Doran-type slicing for $a=0.001\sqrt{\mu}$. These are essentially the same as for five-dimensional Schwarzschild and there are many more of these with more loops. The outer horizon is at $\mathcal[P]_+ \approx 1.999999\mu$ and the inner horizon at
    $\mathcal{P}_- \approx 1 \times 10^{-6}\mu$ is too small to be visible in this figure. }
    \label{fig:ManyLoops}
\end{figure}

\begin{figure}
    \centering
    \includegraphics{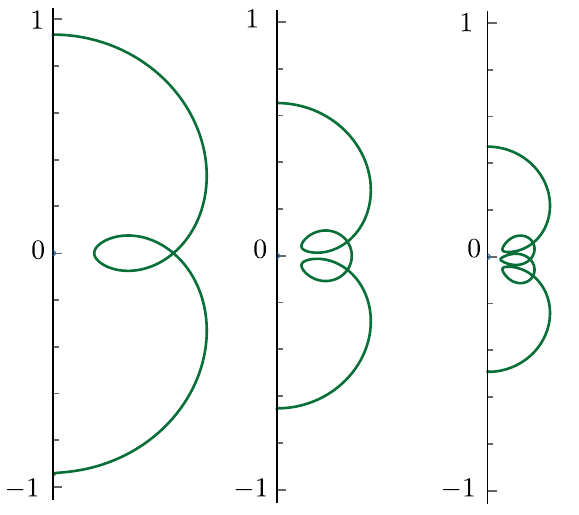}
    \caption{The only three symmetric looping exotic MOTSs for the Myers-Perry black hole in a Doran-type slicing for $a=0.1\sqrt{\mu}$. Higher loop candidates intersect with the singularity before they can close. The outer horizon is at $\mathcal{P}_+ \approx 1.99\mu$ and the inner horizon at
    $\mathcal{P}_- \approx  0.01\mu$ (which is barely visible).  }
    \label{fig:MPa0p1}
\end{figure}

\begin{figure}
    \centering
    \includegraphics{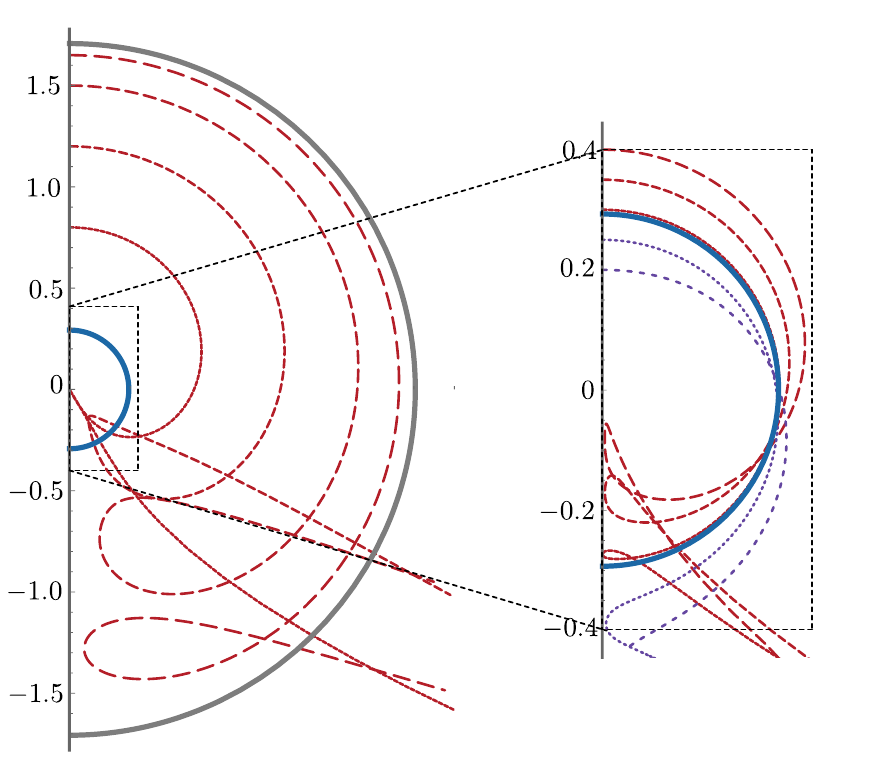}
    \caption{A search for looping exotic MOTSs for Myers-Perry with $a = 0.5 \sqrt{\mu}$. 
    In this case we find 
    no looping MOTSodesics. As for large rotation Kerr solutions or high charge Reissner-Nordstr\"om as studied in \cite{Hennigar:2021ogw}, in the approach to extremality the looping MOTSs like those seen in Schwarzschild disappear. }
    \label{fig:MPa0p5}
\end{figure}

\begin{figure}
    \centering
    \includegraphics{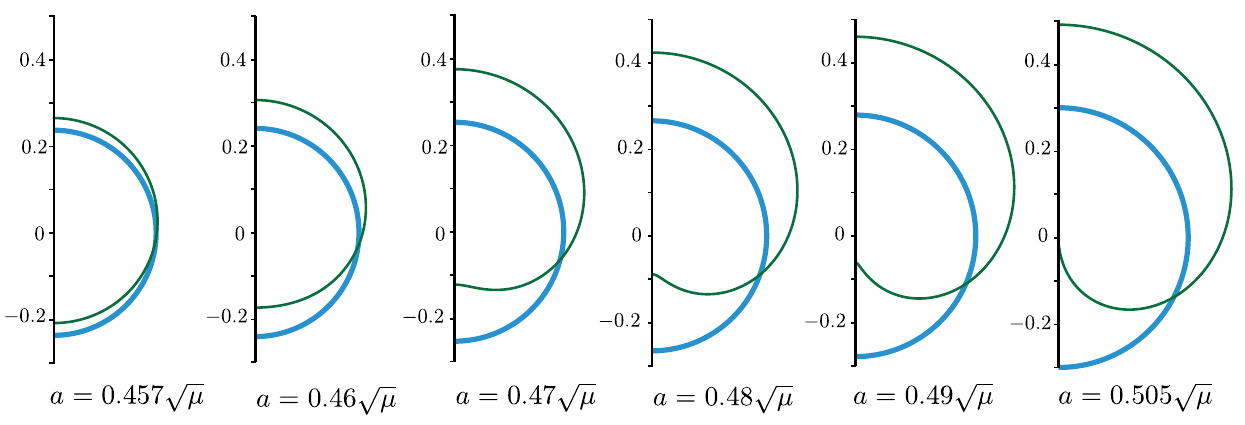}
    \caption{While the Schwarzschild-like exotic MOTSs disappear for larger rotation Myers-Perry, other non-standard MOTSs appear in the vicinity of the inner horizon. Here we see MOTSodesics that intersect the inner horizon a single time. }
    \label{fig:weird1}
\end{figure}

\begin{figure}
    \centering
    \includegraphics{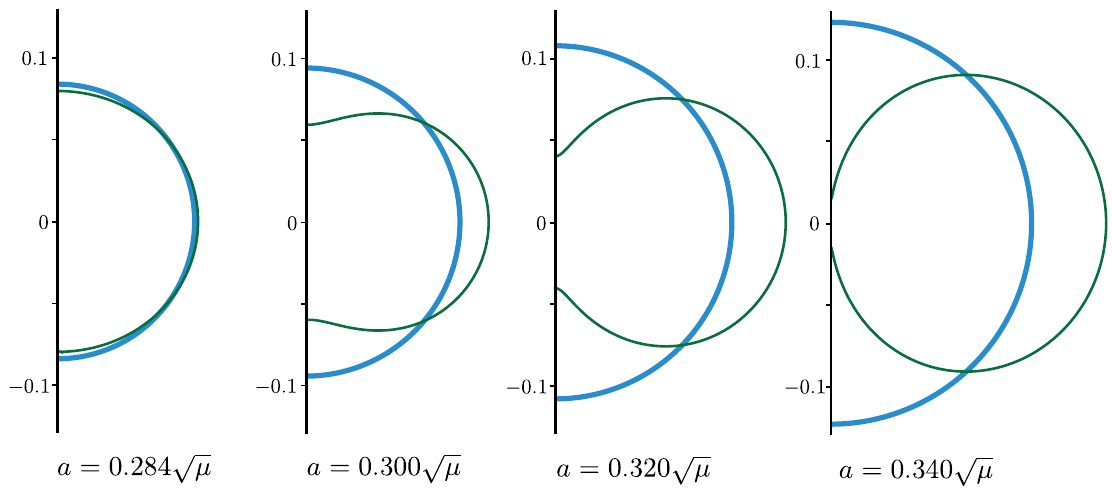}
    \caption{Myers-Perry with an intermediate value of the rotation parameter $a$. Here we see MOTSodesics that intersect the inner horizon twice. }
    \label{fig:weird2}
\end{figure}

\begin{figure}
    \centering
    \includegraphics{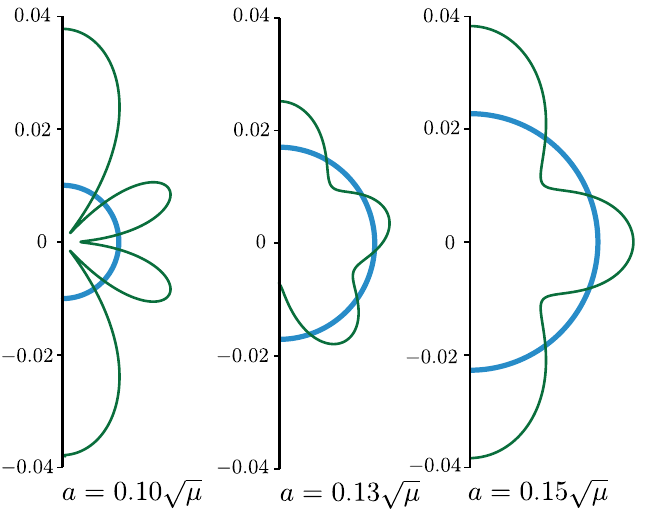}
    \caption{Myers-Perry with low values of the rotation parameter $a$. Here there are multiple 
    intersections with the inner horizon. See \cite{Hennigar:2021ogw} to compare with the 
    corresponding Reissner-Nordstr\"om case. Note that when comparing $a=0.10 \sqrt{\mu}$ to Figure 
    \ref{fig:MPa0p1}, the four-lobed figure would be just barely visible due to the difference in scales.}
    \label{fig:InnerUnstable}
\end{figure}

In the above coordinate system, the spacetime and spatial metrics have a coordinate singularity at $r = 0$ and a curvature singularity at $r^2 = -a^2$. At the latter singularity, the vector fields $\partial_\phi, \partial_\theta$ degenerate whereas the circle generated by $\partial_\psi$  grows infinitely large.  To avoid the problems caused by the coordinate singularity, we switch to a new radial coordinate 
$ \tilde{r} = r^2 + a^2$. At the level of the curve equations, this means changing from $P(s)$ to a new variable $\mathcal{P}(s):= P(s)^2 + a^2$. This reduces to the substitutions
\begin{equation}
P \dot P = \frac{\dot{\mathcal{P}}}{2}, \qquad P \ddot P = \frac{\ddot{\mathcal{P}}}{2}  - \frac{\dot{\mathcal{P}}^2}{4 (\mathcal{P} - a^2)},
\end{equation} transforming the MOTSodesic equations into 
\begin{eqnarray}
\ddot{\mathcal{P}} &= \frac{\dot{\mathcal{P}}^2}{2 \mathcal{P}} + \frac{\mathcal{P} \dot \Theta^2}{2}  + {\kappa}_{\mbox{\tiny{MOTS}}}  \mathcal{P} \dot\Theta 
\\
\ddot \Theta & = - \frac{\dot{\mathcal{P}} \dot \Theta}{\mathcal{P}}  - {\kappa}_{\mbox{\tiny{MOTS}}} \frac{\dot{\mathcal{P}}}{\mathcal{P}}
\end{eqnarray} 
with unit speed condition $\dot \mathcal{P}^2 + \mathcal{P}^2 \dot\Theta^2 = 4\mathcal{P}$. In terms of $\mathcal{P}$, the components 
of $\kappa$ are
\begin{eqnarray}
\cos^2\alpha &= \frac{2\mathcal{P}}{2\mathcal{P} +  a^2 \mu \dot\Theta^2}  \\
\mathcal{K}  &= -\frac{2\sqrt{2\mu} \mathcal{P} }{\mathcal{P}^2 + 2a^2 \mu} \\
\mathcal{K}_{\hat{T}}  & = 0 \\
   \mathcal{K}_{\hat{N}} & =  \dot\Theta - \frac{ \cot \Theta \dot{\mathcal{P}}}{\mathcal{P}}  \\
    \mathcal{K}_{\hat{T} \hat{T}}  & = \frac{\sqrt{\mu}}{2\sqrt{2}}\left(-\dot\Theta^2 + \frac{\dot{\mathcal{P}}^2}{\mathcal{P}^2 + 2 a^2 \mu}\right) 
  \\
   \mathcal{K}_{\hat{T} \hat{T} \hat{N}}  &=  \frac{\mu a^2 \dot \Theta^3}{4\mathcal{P}} + \frac{a^2 \mu \cot\Theta \dot{\mathcal{P}}^3}{2\mathcal{P}^2(\mathcal{P}^2  + 2a^2\mu)}  - \frac{\mu a^2 \dot \mathcal{P}^2 \dot\Theta}{\mathcal{P}(\mathcal{P}^2 + 2 a^2\mu)} \; . 
 \end{eqnarray} Using the now familiar strategy of a series expansion to establish boundary conditions, along with a
numerical integration, some representative MOTSodesics are shown in Figures \ref{fig:ManyLoops}-\ref{fig:InnerUnstable}. 

Figure \ref{fig:ManyLoops} shows looping MOTSs for a very low value of the rotation parameter 
$a = 0.001 \sqrt{\mu}$. At this level of rotation Myers-Perry is essentially indistinguishable from five-dimensional 
Schwarzschild-Tangherlini, although note that we are using $\tilde{r}$ instead of $r$ as a radial coordinate
here so the curves appear a little different from those in Figure \ref{fig:MOTS5DSchwarz}.
Figure \ref{fig:MPa0p1} is a little more interesting as the rotation parameter is increased to 
$a =0.1 \sqrt{\mu}$. Then we can no longer find many examples of the Schwarzschild-type looping MOTSs. In 
this case there are only three. At the four-loop level, the inner horizon and singularity interfere
and the MOTSosdesic is either repelled to infinity or ends up in the singularity. Very similar behaviour
was seen for Reissner-Nordstr\"om in \cite{Hennigar:2021ogw}, where with increasing charge parameter 
the maximum number of loops eventually decreased to zero. Figure \ref{fig:MPa0p5} demonstrates how this
happens for Myers-Perry by $a = 0.5 \sqrt{\mu}$. Note too that the details of this behaviour are quite
different from what we saw for Kerr: there, the singularity in the time slice was extended and 
many more MOTSodesics intersected with it. By contrast, for these equal-rotation Myers-Perry solutions, the entire $\mathcal{P}$ surface is singular,  (as opposed to a ring, as in Kerr) and so their behaviour is more like Reissner-Nordstr\"om.

The similarity with Reissner-Nordstr\"om continues when we examine the inner horizon. Briefly, in 
\cite{Booth:2021sow,Hennigar:2021ogw} it was noted that the (self-adjoint) 
stability operator on the inner horizon of 
Reissner-Nordstr\"om is unstable having one or more negative eigenvalues (or a single vanishing eigenvalue
for the extremal case). In that case the horizon is spherical and so the eigenfunctions are the spherical 
harmonics $Y_{l, m}$. Then it turns out that nearby MOTSosdesics weave back and forth through the 
inner horizon  with a number of intersections corresponding to the number of $Y_{l, 0}$ with negative
eigenvalues. Their number increases from $0$ (at extremality) to become arbitrarily large as the 
charge $q \rightarrow 0$.

We have not yet analyzed the stability operator for the Myers-Perry solutions, but in general adding 
rotation removes the self-adjoint property and hence the eigenvalues are no longer guaranteed to be
real. This can been seen directly in~\cite{Kunduri:2019mfn}, where the spectrum of the stability operator was computed explicitly for the event horizon cross-sections of the five-dimensional equal-spinning Myers-Perry black hole. That said, in Figures \ref{fig:weird1}-\ref{fig:InnerUnstable} we see broadly similar properties to
those observed for Reissner-Nordstr\"om. Figure \ref{fig:weird1} and Figure \ref{fig:weird2} appear, respectively, to correspond to the cases with one or two negative eigenvalues, while Figure \ref{fig:InnerUnstable} corresponds to four, five and six. We will return to a more detailed numerical study of the spectrum of the stability operator for these surfaces in future work.

\section{Conclusions}

Our primary technical objective in this work has been to extend the MOTSodesic characterization of marginally outer trapped surfaces to more general situations than those covered in the original construction. In particular, we have extended the derivations of~\cite{Booth:2021sow} to arbitrary dimensions and to cases in which the induced spatial metric does not take the form of a warped product. Relaxing this latter assumption allows for the MOTSodesic approach to be extended to a variety of rotating black hole solutions, in various dimensions. Our final results --- and the relevant equations --- of this analysis are summarized in section~\ref{RotBH}.

To illustrate the results, we have studied MOTSodesics and their properties in a variety of spacetimes, including the static and rotating BTZ black holes, higher-dimensional analogues of the Schwarzschild solution, the Kerr solution and an example of its higher-dimensional Myers-Perry analogues. Our analysis of these black holes has not been exhaustive; rather we have focused on identifying features that are common between these solutions and previously studied exact solutions, while also highlighting some examples of notable differences. 

When studying the static BTZ black hole in Boyer-Lindquist (non-horizon penetrating) coordinates, the MOTSodesic equations reduce to the problem of finding the space-like geodesics on the surfaces of constant time. This is a consequence of the vanishing of the extrinsic curvature in the Boyer-Lindquist coordinates, and also the fact that the timeslice is two-dimensional. In this case, the MOTSodesic equations can be solved exactly. We find that the MOTSodesics can wrap around the horizon an infinite number of times. In both the static case and also the rotating case, in coordinates systems that penetrate the horizon and those that do not, we find no examples of closed MOTSs besides those that correspond to the event or inner horizon. This can be understood as a consequence of the differences in dimensionality: in $(2+1)$ dimensions, there is no axis-repulsion terms in the MOTSodesic equations. Such terms appear to be crucial for obtaining exotic, closed interior MOTSs in higher-dimensional spacetimes.

We have also considered the Schwarzschild solution in five dimensions. In this case, we have demonstrated the existence of a family of closed, self-intersecting MOTSs contained in the black hole interior. This is consistent with what has been previously seen in four-dimensions~\cite{Booth:2020qhb}.

Lastly, we have studied MOTSodesics in rotating black hole spacetimes, namely the Kerr black hole and the five-dimensional Myers-Perry black hole with equal angular momentum. In both cases, we again find closed, self-intersecting MOTSs in the black hole interior. However, in these cases the number of such surfaces appears to depend on the value of the black hole angular momentum: there are more self-intersecting MOTSs when the angular momentum is small, and less when it is large. These observations are qualitatively consistent with previous studies of interior MOTSs for black holes that possess an inner horizon~\cite{Hennigar:2021ogw}.

The case of the higher-dimensional Myers-Perry black hole with equal angular momentum displayed remarkable similarities with Reissner-Nordstrom~\cite{Hennigar:2021ogw}. Here we have identified, in addition to the standard sequence of self-intersecting MOTSs, additional closed MOTSs with a variety of shapes, some containing self-intersections and some not. We expect that, like Reissner-Nordstrom, a number of these surfaces will be connected to one another as the spin parameter is varied, and we have presented some evidence for this scenario here. This provides a simple arena for understanding the spectrum of the stability operator during bifurcations/annihilations of MOTSs away from the spherically symmetric examples studied in~\cite{Hennigar:2021ogw}. Namely, the spectrum will now in general be complex and it remains to be seen how the various features observed in~\cite{Pook-Kolb:2021gsh, Hennigar:2021ogw} carry over (or if they do). 

The four-dimensional Kerr spacetime was found to depart most significantly from spacetimes previously studied. In large part, these differences can be attributed to the more complicated nature of the singularity, with MOTSodesics now able to pass through the ring singularity.

There remains a number of natural directions for further study on these topics. One obvious future direction could involve the implementation of the more general methods developed here into packages available for locating horizons in numerical simulations~\cite{pook_kolb_daniel_2021_4687700}. It is also natural to wonder whether the methods developed here, and previously, could be applied with benefit to the extremal surfaces relevant in computations of entanglement entropy in the context of holography.

With MOTSodesic methods now available in higher dimensions, one could also check whether MOTSs with novel topology can be exist in the interiors of various more exotic higher-dimensional black objects, akin to the toroidal MOTS discovered in~\cite{Pook-Kolb:2021jpd}. Of course, it remains also to more fully characterize the interior MOTSs of rotating black holes, such as Kerr, as this may shed light on the structure of apparent horizons in mergers of rotating black holes~\cite{Booth:2020qhb}.

Related to the general theory presented here, there remain questions of whether or not the full set of properties for MOTSodesics described in~\cite{Booth:2021sow} hold also under these more general considerations. Foremost here is the connection between MOTSodesics and the stability operator. In previous work it was found that the number of axisymmetric eigenfunctions of the stability operator that have  negative eigenvalues  is counted by the number of times a nearby MOTSodesic intersects with the MOTS of interest. Extending this result to the present case deserves some consideration, as, for example, in the presence of rotation, the eigenvalues of the stability operator need not be real (see, for example, the calculation of the spectrum for the stability operator on the event horizon performed perturbatively for Kerr in \cite{Bussey:2020whg} and exactly for equal-angular momenta Myers-Perry in \cite{Kunduri:2019mfn}).

\ack 

The work of RAH received the support of a fellowship from ``la Caixa” Foundation (ID 100010434) and from the European Union’s Horizon 2020 research and innovation programme under the Marie Skłodowska-Curie grant agreement No 847648 under fellowship code LCF/BQ/PI21/11830027. IB and KTBC were supported by Natural Science and Engineering Research Council of Canada Discovery Grant 2018-04873. HK was supported by Natural Science and Engineering Research Council of Canada Discovery Grant 2018-04887. SM was supported by both of these NSERC Discovery Grants. Memorial University (St.\ John's campus) is situated on the ancestral homelands of the Beothuk. McMaster University is located on the traditional territories of the Mississauga and Haudenosaunee nations, and within the lands protected by the ``Dish with One Spoon'' wampum agreement.

\appendix

\section{Integrating the MOTSodesic equations from the $z$-axis}
\label{sec:app}

All pairs of MOTSodesic equations encountered in this paper are very easily solved and plotted using basic differential 
equation commands in standard mathematics packages: for example \texttt{dsolve(\dots, numeric)} and \texttt{odeplot} in Maple \cite{maple}
or \texttt{NDsolve[\dots]} and \texttt{ParametricPlot} in Mathematica \cite{mathematica}. The results presented in this paper were generated using these packages. 

The only complication comes in implementing initial conditions. For the four- and five-dimensional spacetimes the MOTSodesic curvature
$\kappa$ includes  $\cot \theta$ terms which diverge as $\theta \rightarrow  0$. However these are always paired with $\dot{P}$-terms which simultaneously vanish in the same limit. Solutions are well-defined  but the initial conditions can't be implemented directly on the
$z$-axis. The solution is to solve for a series solution of the form
\begin{eqnarray}
P(s) & = P_0 + P_1 s + \frac{1}{2!} P_2 s^2 + \frac{1}{3!} P_3 s^3 + \frac{1}{4!} P_4 s^4 + \dots \\
\Theta(s) & =  \Theta_0 + \Theta_1 s + \frac{1}{2!} \Theta_2 s^2 + \frac{1}{3!} \Theta_3 s^3 + \frac{1}{4!} \Theta_4 s^4 + \dots  \nonumber
\end{eqnarray}
and then implement initial conditions at a small value of $s$. 

Starting with $P_0$ finite and $\Theta_0=0$ (expanding around the $z$-axis) we can then substitute these expansions into the MOTSodesic
equations and solve order-by-order to obtain all coefficients in terms of $P_0$. This is computationally straightforward though the resulting 
expressions are messy. As an example, for Kerr in Doran coordinates to fourth order in $s$ we find:
\begin{eqnarray}
P(s) = & P_0 + \left( \frac{P_0}{2(P_0^2+a^2)}\right) \left(1 - \sqrt{\frac{2mP_0}{P_0^2+a^2}} \right) s^2 \\
& + \Bigg( \frac{\sqrt{2P_{0} m^3} \left(13 P_{0}^{4}-6 a^{2} P_{0}^{2}-3 a^{4}\right) }{48 \left(a^{2}+P_{0}^{2}\right)^{\frac{9}{2}}}\nonumber \\
& \phantom{XX} +\frac{m\left(-43 P_{0}^{4}+36 a^{2} P_{0}^{2}+ 3 a^{4}\right) }{48 \left(a^{2}+P_{0}^{2}\right)^{4}} \nonumber \\
&  \phantom{XX} + \frac{ \sqrt{2m P_0^3}\, \left(7 P_{0}^{2}-11 a^{2}\right) }{16 \left(a^{2}+P_{0}^{2}\right)^{\frac{7}{2}}}+\frac{P_{0} \left(3 a^{2}-P_{0}^{2}\right)}{8 \left(a^{2}+P_{0}^{2}\right)^{3}}
 \Bigg) s^4 +  O (s^6)  \nonumber
\end{eqnarray}
and
\begin{eqnarray}
\Theta(s) = & \frac{s}{\sqrt{P_0^2+a^2}} \\
 & + \left(  -\frac{mP_{0}^{3} }{3 \left(a^{2}+P_{0}^{2}\right)^{\frac{7}{2}}}+ \frac{\sqrt{2m P_0^5}}{2 \left(a^{2}+P_{0}^{2}\right)^{3}}-\frac{2 P_{0}^{2}-a^{2}}{6 \left(a^{2}+P_{0}^{2}\right)^{\frac{5}{2}}} \right) s^3 \nonumber \\
 & \phantom{XX} + O (s^5) \nonumber \; . 
\end{eqnarray}

Note that $P_1=0$. It turns out that for all of the four and five-dimensional metrics, we must have $P_1=0$ in order for the MOTSodesics to be well-defined. That is, the MOTSodesics must intersect the $z$-axis at a right angle. Intuitively, a very close approach to the $z$-axis results in a very tight curvature around the axis which diverges as the radius goes to zero. In general $k_u$ is finite in this limit, so in order to keep $\theta_\ell =0$ there must be also be a very tight curvature away from the $z$-axis. Thus there is an effective strong (and diverging as $\theta \rightarrow 0$) repulsion of MOTSodesics from the $z$-axis: the $\cot \theta$ term. This can 
only be avoided by a perpendicular intersection for which the approach to the $z$-axis is within the tangent plane of the MOTS and so there
is no normal curvature in that direction. 

Alternatively, by symmetry arguments one can argue that since a MOTSodesic rotates into an axisymmetric surface then smoothness at $s=0$ 
requires that all odd-order $P_n$ must vanish as must all even order $\Theta_n$. These arguments are borne out in the calculations. 

Once the series expansion is found, initial conditions for MOTSodesics can be generated by choosing a $P_o$ along with a small $s_o$ 
(we typically use $s_o=0.0001$). Then the series are used to calculate $P(s_o)$, $\Theta (s_o)$ (accurate to order $s_o^4$) and 
$P'(s_o)$, $\Theta'(s_o)$ accurate to order $s_o^3$. These can then be used as initial conditions in the differential equation solvers. 
In this paper we then found the MOTSs by the shooting method: essentially informed trial-and-error looking for MOTSodesics that 
return to the $z$-axis at a right angle. For more details along with more sophisticated methods that one would use if attempting to 
exhaustively enumerate the possibilities see \cite{Booth:2021sow,Hennigar:2021ogw}. \\

\bibliographystyle{iopart-num}
\bibliography{Gravities}

\end{document}